\documentclass[10pt]{article}

\usepackage[margin=.75in]{geometry}

\usepackage{amsmath,amssymb}
\usepackage{graphicx}

\usepackage{color}
\definecolor{indigo}{RGB}{0,0,120}
\usepackage[colorlinks=true, linkcolor=indigo, citecolor=blue, urlcolor=indigo]{hyperref}

\def\imply{\Rightarrow}
\newcommand{\pt}{\noindent {$\bullet$~}}

\newcommand{\tl}[1]{\tilde{#1}}
\newcommand{\dd}[2]{\frac {\partial #1}{\partial #2}}
\newcommand{\deldel}[2]{\frac {\delta #1}{\delta #2}}
\newcommand{\pdr}{\partial}
\newcommand{\DD}[2]{\frac {d #1}{d #2}}
\newcommand{\grad}{{\bf \nabla}}
\newcommand{\diver}[1]{{\bf \nabla \cdot {#1}}}

\newcommand{\beq}{\begin{equation}}
\newcommand{\eeq}{\end{equation}}
\newcommand{\beqs}{\begin{eqnarray}}
\newcommand{\eeqs}{\end{eqnarray}}

\newcommand{\half}{\frac{1}{2}}
\newcommand{\ov}[1]{\frac{1}{#1}}

\def\al{\alpha} 		
\def\del{\delta}	
\def\eps{\epsilon} 
\def\la{\lambda}		
\def\sig{\sigma}
\def\tht{\theta}
\def\om{\omega}		
\def\Om{\Omega}
\newcommand{\G}{{\Gamma}}

\newcommand{\bfS}{{\bf S}}
\newcommand{\bfZ}{{\bf Z}}
\newcommand{\bfP}{{\bf P}}
\newcommand{\bfQ}{{\bf Q}}
\newcommand{\bfa}{{\bf a}}
\newcommand{\bfb}{{\bf b}}
\newcommand{\bff}{{\bf f}}
\newcommand{\bfg}{{\bf g}}
\newcommand{\bfh}{{\bf h}}
\newcommand{\bfu}{{\bf u}}
\newcommand{\bfv}{{\bf v}}
\newcommand{\bfw}{{\bf w}}
\newcommand{\bfx}{{\bf x}}
\newcommand{\bfy}{{\bf y}}
\newcommand{\bfz}{{\bf z}}
\newcommand{\bfj}{{\bf j}}
\newcommand{\bfM}{{\bf M}}
\newcommand{\bfk}{{\bf k}}
\newcommand{\bfr}{{\bf r}}
\newcommand{\bfs}{{\bf s}}

\newcommand{\bfA}{{\bf A}}
\newcommand{\bfE}{{\bf E}}
\newcommand{\bfB}{{\bf B}}
\newcommand{\bfC}{{\bf C}}
\newcommand{\bfD}{{\bf D}}
\newcommand{\bfl}{{\bf l}}

\newcommand{\bfL}{{\bf L}}
\newcommand{\bfT}{{\bf T}}

\newcount\colveccount  
\newcommand*\colvec[1]{\global\colveccount#1  \begin{pmatrix} \colvecnext} \def\colvecnext#1{#1 \global\advance\colveccount-1
        \ifnum\colveccount>0 \\ \expandafter\colvecnext
        \else \end{pmatrix} \fi}

\begin{document}


\title{\normalsize 
\hfill {\tt arXiv:1510.01606} \\
\vskip 0.1mm \LARGE
Conservative regularization of compressible flow and ideal magnetohydrodynamics\\
}
\author{{\sc Govind S. Krishnaswami$^{1}$, Sonakshi Sachdev$^{1}$ and Anantanarayanan Thyagaraja$^{2}$}
\\ \\ \small
$^{1}$ Physics Department, Chennai Mathematical Institute,  SIPCOT IT Park, Siruseri 603103, India\\ \small
$^{2}$Astrophysics Group, University of Bristol, Bristol, BS8 1TL, UK
\\ \\ \small
 Email: {\tt govind@cmi.ac.in, sonakshi@cmi.ac.in, athyagaraja@gmail.com}}

\date{12 Nov, 2016}

\maketitle

\begin{abstract} \normalsize

Ideal systems like MHD and Euler flow may develop singularities in vorticity ($\bfw = \grad \times \bfv$). Viscosity and resistivity provide dissipative regularizations of the singularities. In this paper we propose a minimal, local, conservative, nonlinear, dispersive regularization of compressible flow and ideal MHD, in analogy with the KdV regularization of the 1D Hopf or kinematic wave equation. This work extends and significantly generalizes earlier work on incompressible Euler and ideal MHD. It involves a density-dependent cut-off length $\lambda \propto 1/\sqrt{\rho}$ which is like a position-dependent mean free path. In MHD, $\la$ can be taken to be of order the electron collisionless skin depth $c/\omega_{pe}$. The regularizing `twirl' term is $- \la^2 \bfw \times (\grad \times \bfw)$. Such a non-linear dispersive term could be important in high speed flows with vorticity and arise in an expansion of kinetic equations in Knudsen number. A magnetic analogue of the twirl term $-\bfB \times (\grad \times \bfB)/\rho\mu_0$, arises as the Lorentz force in ideal MHD. Our regularization preserves the symmetries of the original systems, and with appropriate boundary conditions, leads to associated conservation laws. Energy and enstrophy are subject to a priori bounds determined by initial data in contrast to the unregularized systems. A Hamiltonian and Poisson bracket formulation is developed and applied to generalize the constitutive relation to bound higher moments of vorticity and its curl. A `swirl' velocity field is identified, and shown to transport $\bfw/\rho$ and $\bfB/\rho$, generalizing the Kelvin-Helmholtz and Alfv\'en theorems. The steady regularized equations are used to model a rotating vortex, MHD pinch, a plane vortex sheet, channel flow, plane flow and propagating spherical and cylindrical vortices; solutions are more regular than the corresponding Eulerian ones. The proposed regularization could facilitate numerical simulations of fluid/MHD equations and provide a consistent statistical mechanics of vortices/current filaments in 3D, without blowup of enstrophy. Implications for detailed analyses of fluid and plasma dynamic systems arising from our work are briefly discussed.

\end{abstract}






\footnotesize

\tableofcontents

\normalsize

\section{Introduction}

This work continues and seeks to extend the program of ``regularization'' of three-dimensional, conservative, continuum dynamical systems exemplified by the classic, incompressible Euler equations of fluid dynamics and ideal, incompressible magneto-hydrodynamics (MHD)\cite{thyagaraja,AT1}. Well-known examples of one-dimensional (1D) models of conservative systems like the Korteweg-de Vries equation (KdV) and the nonlinear Schr\"{o}dinger equation (NLS) show that effective analysis and computation are greatly facilitated when the dynamics imply bounded motions rather than the development of singularities which prevent a proper understanding of the system dynamics and statistical mechanics. A standard example is provided by the so called ``kinematic wave'' or Hopf equation ($u_t + u u_x = 0$) which has singular behaviour in a finite time for certain initial data. This equation admits a dissipative regularization in the well-known Burgers equation ($u_t + u u_x = \nu u_{xx}$), and thereby provides an excellent, exactly soluble [via the Cole-Hopf transformation] model of random arrays of one-dimensional shocks, traffic flows and the like. A much deeper dispersive regularization is provided by the KdV equation ($u_t -6 u u_x + u_{xxx} = 0$) which has been extensively discussed \cite{Miura,davidson} as the paradigmatic, conservatively regularized extension of the KWE with applications in many fields (e.g. solitons and integrable systems, shallow water waves, ion acoustic waves, long internal ocean waves and blood pressure waves). It is this latter example that provides the motivation for the results presented in the works cited and in the present paper.

Three-dimensional (3D) fluid dynamics fundamentally involves ``vortex stretching'', a process which in the standard Euler equations leads (as indicated in the classic work of Taylor and Green \cite{taylor-green} on Navier-Stokes (NS) with very low viscosity, see also \cite{Frisch,KRSreenivasan-onsager}) to unbounded growth of the fluid enstrophy [enstrophy density is the square of local vorticity]. Enstrophy may also diverge in the presence of singular structures such as vortex sheets, with discontinuous tangential velocity. This is analogous to the loss of single-valuedness of $u$ and development of singularities in derivatives of $u$ in the KWE. Like the KWE, the Euler system has a standard, dissipative regularization in the NS equations, on which almost all of modern fluid dynamics rests. However, until the appearance of \cite{thyagaraja} no minimal local regularization of the incompressible Euler equations comparable with KdV or NLS satisfying essential Galilean, parity and time-reversal symmetries, valid for arbitrary initial data, leading to bounded enstrophy and positive-definite conserved energy and admitting an adjoint variational principle was published\footnote{There are other interesting conservative regularizations of the 3D Euler equations, motivated partly by numerical schemes or involving averaging procedures, such as the Euler-$\alpha$ and vortex blob regularizations \cite{vortex-blob,Euler-alpha}. The incompressible Euler-$\alpha$ equations are the geodesic equations for the H1 metric on the group of volume preserving diffeomorphisms of the flow domain. They correspond to the energy functional $\rho \int \left(\half \bfv^2 + \half \al^2 (\pdr_i v_j)^2 \right) d\bfr = \rho \int \left(\half \bfv^2 + \half \al^2 \bfw^2 \right) d\bfr$ for $\grad \cdot \bfv = 0$, with $\al$ a regularizing length. However, the resulting Euler-$\alpha$ equation of motion for $\bfv$ is highly non-local as it involves the advecting velocity $(1 - \al^2 \grad^2)^{-1} \bfv$.}. 

The regularized incompressible Euler equations are
	\beq
	\dd{\bfv}{t} + (\bfv \cdot \grad) \bfv = -\frac{\grad p}{\rho} - \la^2 \bfw \times (\grad \times \bfw) \quad \text{and} \quad \grad \cdot \bfv = 0.
	\label{e:R-Euler-incompress}
	\eeq
$\la$ is a constant with dimensions of length, which acts as a short-distance regulator. Since the `twirl' term $- \la^2 \bfw \times (\grad \times \bfw)$ is quadratic in velocities, it should be significant in high-speed, i.e., compressible flows. In this paper we present a relatively `natural' extension to compressible neutral fluid flow and to compressible, ideal MHD. A key ingredient in this compressible extension is for $\la$ to be generalized to a field satisfying the constitutive relation $\la^2 \rho = $ constant. This relation might be expected from the following vortical-magnetic analogy. Indeed, the twirl force $- \la^2 \rho \, \bfw \times (\grad \times \bfw)$ can be thought of as a vortical counterpart of the magnetic Lorentz force $\bfj \times \bfB = - \bfB \times (\grad \times \bfB)/\mu_0$ familiar from non-relativistic MHD, with $\la^2 \rho$ replacing the constant $1/\mu_0$. The constitutive relation $\la^2 \rho = $ constant, can be interpreted in terms of the mean free path and/or the inter-particle distance $a \propto n^{-1/3}$ where $n$ is the number density of molecules in the medium. Thus if $L$ is a macroscopic length-scale of the system and we take $\la^2/L^2 \propto a^3/L^3$ then $\la^2 \rho$ will be a constant as required in the model. $\la$ is a microscopic length, which leads to a bounded enstrophy without altering the macro- and meso-scale conservative dynamics of ideal compressible flow. Although introduced as a formal regularizer, it is conceivable that such a twirl term could arise in a Chapman-Enskog-like expansion of kinetic equations in the Knudsen number.

Unlike in 1D, it is possible to show that there is no KdV-like regularizer linear in velocity that preserves Eulerian symmetries. The twirl term is a minimal (in the sense of effective local field theory) non-linear dispersive regularization of the ideal equations leading to bounded enstrophy and conservation laws. Indeed, as with the NS regularization of Euler, (\ref{e:R-Euler-incompress}) increases the spatial order by unity. On the other hand, Ladyzhenskaya's `hyperviscosity' regularization\cite{ladyzhenskaya-1} of the Euler [and Navier-Stokes] equation involves the fourth order term $\eps (\grad^2)^2 \bfv$ with $\eps$ constant. In \cite{ladyzhenskaya-2} she also considers a non-linear regularization term $\nu_3 \grad^2 \bfv$ where the viscosity coefficient $\nu_3$ depends on the sum of squares of the components of the rate of strain tensor. Both these dissipative regularizations serve to balance, in principle, the non-linear vortex-stretching mechanism of 3D inviscid flow. Our conservative non-linear twirl term is similarly responsible for controlling the growth of vorticity at short distances of order $\la$.

More generally, our compressible regularized MHD (R-MHD) equations are given by $\grad \cdot \bfB = 0$,
	\beq
	\dd{\rho}{t} + \grad \cdot (\rho \bfv) = 0, \quad \dd{\bfv}{t} + (\bfv \cdot \grad) \bfv = -\frac{\grad p}{\rho} + \frac{\bfj \times \bfB}{\rho}- \la^2 \bfw \times (\grad \times \bfw) \quad \text{and} \quad \dd{\bfB}{t} = \grad \times(\bfv \times \bfB - \la^2 \bfB \times (\grad \times \bfw)).
	\eeq
These R-MHD equations are seen to include both vortical $\la^2 \bfw \times (\grad \times \bfw)$ and magnetic $\la^2 \bfB \times (\grad \times \bfw)$ twirl regularizers. As before, the regularizing length $\la$ must satisfy the constitutive relation $\la^2 \rho = $ constant. We note that in plasma physics there are natural length-scales which are inversely proportional to the square-root of the number density. For example, the electron collisionless skin-depth $\del = c/\om_{pe} \propto 1/\sqrt{n_e}$. Thus if $\la \approx \del$ then $\la^2 \rho$ will indeed be a constant. In any event, it is well-known that ideal MHD is not valid at length scales of order $\del$. Another example is provided by the electron Debye length $\la_D = \sqrt{2 T_e \eps_0/n_e e^2}$ in an isothermal plasma. If we take $\la/\la_D$ constant, then we obtain the postulated constitutive relation. Thus, having a cut-off of this kind will not affect any major consequence of ideal MHD on meso- and macro-scales and yet provide a finite upper bound to the enstrophy of the system and a valid statistical mechanics.

Inclusion of these twirl regularizations should lead to more controlled numerical simulations of Euler, NS and MHD equations without finite time blowups of enstrophy. In particular, these regularized models are capable of handling three-dimensional tangled vortex line and sheet interactions in engineering and geophysical fluid flows, as well as corresponding current filament and sheet dynamics which occur in astrophysics (e.g. as in solar prominences and coronal mass ejections, pulsar accretion disks and associated turbulent jets, and on a galactic scale, jets driven by active galactic nuclei) and in strongly nonlinear phenomena such as edge localised modes in tokamaks. There is no known way of studying many of these phenomena at very low collisionality  [i.e. at very high, experimentally relevant Reynolds, Mach and Lundquist numbers] with un-regularized continuum models. Thus, we  note that recent theories \cite{henneberg-cowley-wilson,chandra-thyagaraja,lashmore-mccarthy-thyagaraja,thyagaraja-valovic-knight} of the nonlinear evolution of ideal and visco-resistive plasma turbulence in a variety of fusion-relevant devices (and many geophysical situations) can be numerically investigated in a practical way using our regularization.

In \cite{thyagaraja} the question of a hamiltonian formulation for (incompressible) R-Euler equations was raised. This has been addressed and solved both for compressible and incompressible R-Euler and R-MHD in the present work, making use of the elegant Poisson structures \cite{arnold} for compressible flow due to Morrison and Greene \cite{morrison-and-greene,morrison-review,morrison-aip} anticipated in Landau's \cite{landau} quantum theory of superfluids [cf. Equations (1.7,1.8)] and developed by London \cite{london}. Although not essential to the theory of regularization, this formalism shows that the extended systems formally share the Hamiltonian, non-canonical Poisson structures of the original, singular conservative dynamics. The existence of a positive definite Hamiltonian and bounded enstrophy should facilitate the formulation of a valid statistical mechanics of 3D vortex tubes, extending the work of Onsager on 2D line vortices. The same remark also applies to the 2D statistical mechanics of line current filaments developed by Edwards and Taylor \cite{edwards-taylor} and many other authors in ideal MHD theory. Furthermore, we indicate several new and remarkable results and present some simple but representative applications of the regularized Euler (R-Euler) equations.

Thus, the development of regularized compressible flow and MHD presented in this paper (minimal extension of ideal equations with Hamiltonian-PB structure, conservation laws, bounded enstrophy, identification of appropriate boundary conditions, applications etc.) brings these 3D models a step closer to what KdV achieves for 1D flows.

This article is organized as follows: we begin in \S \ref{s:formulation} by giving the equations of regularized compressible flow and their extension to compressible MHD. Criteria for the choice of regularization term and its physical interpretation are provided. Local conservation laws for `swirl' energy, helicity, linear and angular momenta are derived in \S \ref{s:cons-laws} followed by boundary conditions for the R-Euler equations in \S\ref{s:BC}. The corresponding results for R-MHD may be found in \S \ref{s:R-MHD-cons-laws-alfven}. Regularized versions of the Kelvin-Helmholtz and Alfv\'en theorems on freezing-in of vorticity and magnetic field into the `swirl' velocity are derived in \S \ref{s:Kelvin-circulation-Kelvin-Helmholtz-swirl-vel}. Integral invariants associated with closed curves, surfaces and volumes moving with the `swirl' velocity field are discussed in \S \ref{s:integral-inv-v-star}.
Poisson brackets for compressible and incompressible R-Euler and R-MHD are introduced in \S \ref{s:pb-for-fluid} and \S \ref{s:PB-for-R-MHD}. The regularized equations are shown to be Hamilton's equations for the swirl energy. The Poisson algebra of conserved quantities is obtained paying special attention to boundary conditions. Some properties of the Poisson brackets and a novel proof of the Jacobi identity are given in appendix \ref{a:PB-properties}. The Poisson bracket formulation is used in  \S \ref{s:other-const-laws-and-regs} to identify new regularization terms (involving new constitutive relations) that guarantee bounded higher moments of vorticity and its curl while retaining the symmetries of the ideal equations. \S \ref{s:examples} contains several applications to steady flows. The regularized equations are used to model a rotating columnar vortex and MHD pinch, channel flow, plane flow, a plane vortex sheet and propagating spherical and cylindrical vortices. These examples elucidate many interesting physical consequences. They show that our conservatively regularized flows are indeed more regular than the corresponding Eulerian solutions. In \S \ref{s:discussion} we conclude by placing our conservative regularization of ideal Euler flow and MHD in a wider physical context and discuss several open questions. A condensed version of some of these results may be found in \cite{govind-sonakshi-thyagaraja-pop}.

\section{Formulation of regularized compressible flow and MHD}
\label{s:formulation}

For compressible, barotropic flow with mass density $\rho$ and velocity field $\bfv$, the continuity and Euler equations are
	\beq
	\dd{\rho}{t} + \grad \cdot (\rho {\bf v}) = 0 \quad \text{and} \quad
	\dd{\bf v}{t} + \left( {\bf v} \cdot \grad \right) {\bf v} = - \frac{\grad  p}{\rho}.
	\eeq
The pressure $p$ is related to $\rho$ through a constitutive relation in barotropic flow. Let us introduce the stagnation pressure $\sigma$ and specific enthalpy $h$ for adiabatic flow of an ideal gas (or specific Gibbs free energy for isothermal flow) through the equation
	\beq
	\sigma = \left( \frac{\gamma}{\gamma -1} \right) \frac{p}{\rho} + \half {\bf v}^2 \equiv h + \half \bfv^2 \quad \text{where} \quad
	\frac{p}{\rho^\gamma} = \text{constant} \quad \text{with} \;\; \gamma = C_p/C_v.
	\eeq
Then using the identity $\half \grad {\bf v}^2 = {\bf v} \times \left( \grad \times {\bf v} \right) + \left({\bf v} \cdot \grad \right) {\bf v}$, the Euler equation may be written in terms of vorticity $\bfw = \grad \times \bfv$;
	\beq
	\dd{\bf v}{t} + {\bf w} \times {\bf v} = - \grad \sigma.
	\label{e:Euler-eqn}
	\eeq
In \cite{thyagaraja} a `twirl' regularization term $-\la^2 \bfT$ was introduced into the incompressible $(\grad \cdot \bfv = 0)$ Euler equations
	\beq
	\frac{D \bfv}{Dt} \equiv \dd{\bf v}{t} + \left( {\bf v} \cdot \grad \right) {\bf v} = - \frac{\grad p}{\rho} - \la^2 {\bf w} \times (\grad \times {\bf w}) \quad \text{with} \quad \bfT = \bfw \times (\grad \times \bfw).
	\label{e:R-Euler-eqn}
	\eeq
Here $D/Dt$ is the material derivative. The twirl term is a singular perturbation, making R-Euler $2^{\rm nd}$ order in space derivatives of $\bfv$ while remaining $1^{\rm st}$ order in time. The regularizing vector may be written ${\bf T} = {\bf w} \times \grad (\grad \cdot {\bf v}) - {\bf w} \times \grad^2 {\bf v}$. For incompressible flow it becomes ${\bf T} = - {\bf w} \times \grad^2 {\bf v}$. The parameter $\la$ with dimensions of length is a constant for incompressible flow. We will see that $\la$ acts as a short-distance regulator that prevents the enstrophy $\int \bfw^2 \: d\bfr$ from diverging. Unlike a lattice or other cut-off R-Euler ensures bounded enstrophy while retaining locality and all the space-time symmetries and conservation laws of the Euler equation. The sign of $\bfT$ ensures that the conserved energy $E^*$ obtained below (\ref{e:cons-egy-incompress}) is positive definite. The twirl acceleration is clearly absent in irrotational or constant vorticity flows. Since $\bfT$ involves derivatives of $\bfw$, it kicks in when vorticity develops large gradients and thereby prevents unbounded growth of enstrophy. As discussed below, $\bfT$ is chosen to have as few spatial derivatives and non-linearities as possible. A linear term in $\bfv$ (as in KdV) preserving the symmetries of the Euler equation does not exist. The twirl term $- \la^2 \bfT$ is a conservative analogue of the viscous dissipation term $\nu \grad^2 \bfv$ in the incompressible NS equations
	\beq
	\dd{\bfv}{t} + (\bfv \cdot \grad) \bfv = - \frac{\grad p}{\rho} + \nu \grad^2 \bfv, \quad \grad \cdot \bfv = 0.
	\eeq
Kinematic viscosity $\nu$ and the regulator $\la$ play similar roles. The momentum diffusive time scale in NS is set by $\nu k^2$ where $k$ is the wave number of a mode. On the other hand in the non-linear twirl term of R-Euler, the dispersion time-scale of momentum is set by $\la^2 k^2 |\bfw|$. So for high vorticity and short wavelength modes, the twirl effect would be more efficient in controlling enstrophy than pure viscous diffusion. 

It is instructive to compare incompressible Euler, R-Euler and NS under rescaling of coordinates and velocities ($\bfr = L \bfr'$, $\bfv = U \bfv'$ so that $t = (L/U) t'$). The incompressible Euler equations for vorticity
	\beq
	\dd{\bfw}{t} + \grad \times (\bfw \times \bfv) = 0 \quad \text{and} \quad \grad \cdot \bfv = 0,
	\eeq
are invariant under such rescalings. The NS equation is {\em not} invariant under independent rescalings of $\bfr$ and $\bfv$ unless $LU = 1$:
	\beq
	\dd{\bfw'}{t'} + \grad' \times (\bfw' \times \bfv') = \left( \frac{\nu}{L U} \right) \grad'^2 \bfw'.
	\eeq
As is well-known, flows with the same Reynolds number ${\cal R} = LU/\nu$ are similar. Interestingly, the R-Euler equation $\pdr \bfw/\pdr t + \grad \times (\bfw \times \bfv) = - \la^2 \grad \times (\bfw \times (\grad \times \bfw))$ is invariant under rescaling of time alone: $\bfr = \bfr', t = t'/U, \bfv = U \bfv'$ but not under independent rescalings of time and space. With both viscous and twirl regularizations present, under the rescaling $\bfr =  L \bfr', \bfv = U \bfv'$, we get
	\beq
	\dd{\bfw'}{t'} + \grad' \times (\bfw' \times \bfv') = \frac{\nu}{LU} \grad'^2 \bfw' - \frac{\la^2}{L^2} \grad' \times (\bfw' \times (\grad' \times \bfw')).
	\eeq
We may also compare the relative sizes of the dissipative viscous and conservative twirl stresses in vorticity equations. Under the usual rescaling $\bfr = L \bfr', \bfv = U \bfv'$ ($t = (L/U) t'$, $\bfw = (U/L) \bfw'$) and $|\grad'| = k$, $F_{visc} \sim (\nu/L^2) k^2 \om$ whereas $F_{twirl} \sim (\la^2 U/L^3) k^2 \om^2$ where $\om$ is the magnitude of the non-dimensional vorticity. Then $F_{twirl}/F_{visc} \sim {\cal R} \om (\la/L)^2$. This shows that at any given Reynolds number ${\cal R} = LU/\nu$ and however small $\la/L$ is taken, at sufficiently large vorticity the twirl force will always be larger than the viscous force.

Since $\bfT$ is quadratic in $\bfw$ (or $\bfv$), it should be important in high-vorticity or high-speed flows. Thus it is natural to seek a generalization of the twirl regularization to compressible flows. Consider adiabatic flow of an ideal compressible fluid whose pressure and density are related by $(p/p_0) = (\rho/\rho_0)^\gamma$. The {\it compressible} R-Euler equations are
	\beq
	\dd{\rho}{t} + \grad \cdot (\rho \bfv) = 0 \quad \text{and} \quad
	\dd{\bf v}{t} + \left( {\bf v} \cdot \grad \right) {\bf v} = - \frac{\gamma}{\gamma - 1} \grad \left(\frac{p}{\rho} \right) - \la^2 {\bf w} \times (\grad \times {\bf w}).
	\label{e:continuity-R-Euler}
	\eeq
For compressible flows we find that $\la(\bfr,t)$ and $\rho(\bfr,t)$ must satisfy a constitutive relation taking the form,
	\beq
	\la^2 \rho = \text{constant} = \la_0^2  \rho_0,
	\label{e:constitutive-relation}
	\eeq
to ensure that a positive-definite conserved energy exists for an arbitrary flow [more general constitutive relations are possible, see \S \ref{s:other-const-laws-and-regs}]. The constant $\la_0^2 \rho_0$ depends on the fluid and not the specific flow. We also note that the introduction of the twirl force entails a modification of the stress tensor $S_{ij} = p \del_{ij}$ appearing in the ideal Euler equation $\rho (Dv_i/Dt) = - \pdr_j S_{ij}$. The regularized stress tensor is $S_{ij} = p \del_{ij} + \la^2 \rho \left( \frac{w^2}{2} \del_{ij} - w_i w_j \right)$. 

As before, we write the R-Euler equation as
	\beq
	\dd{\bf v}{t} + {\bf w} \times {\bf v} = - \grad \sigma - \la^2 {\bf w} \times (\grad \times {\bf w}).
	\label{e:reg-Euler-3D}
	\eeq
Here ${\bf w} \times {\bf v}$ is the `vorticity acceleration' and $- \la^2 \bfw \times (\grad \times \bfw)$ is the twirl acceleration while $\grad \sigma$ includes acceleration due to pressure gradients. The regularization term increases the spatial order of the Euler equation by one (since $\bfw=\grad \times \bfv$), just as $\nu \grad^2 \bfv$ does in going from Euler to NS. However the boundary conditions required by the above conservative regularization involve the first spatial derivatives of $\bfv$, unlike the no-slip condition of NS. Furthermore, the regularizing viscous stress in NS is linear in $\bfv$ as opposed to the quadratically non-linear twirl stress. The twirl term involves three derivatives and should be important at high wave numbers, as is the dispersive $u_{xxx}$ term in KdV. The R-Euler equation is invariant under parity (all terms reverse sign) and under time-reversal (all terms retain their signs). It is well-known that NS is not invariant under time-reversal, since it includes viscous dissipation. Moreover, we shall see that R-Euler possesses local conservation laws for energy, flow helicity, linear and angular momenta, in common with the Euler system.

The R-Euler equation takes a compact form in terms of the `swirl' velocity field $\bfv_* = \bfv + \la^2 \grad \times \bfw$:
	\beq
	\dd{\bfv}{t} + \bfw \times \bfv_* = - \grad \sigma.
	\label{e:R-Euler-v*}
	\eeq
Here $\bfw \times \bfv_*$ is a regularized version of the Eulerian vorticity acceleration $\bfw \times \bfv$. The swirl velocity $\bfv_*$ plays an important role in the regularized theory, as will be demonstrated. In fact, the continuity equation can be written with $\bfv_*$ replacing $\bfv$:
	\beq
	\dd{\rho}{t} + \grad \cdot (\rho \bfv_*) = 0.
	\label{e:v*-continuity-eqn}
	\eeq
This is a consequence of the constitutive relation  (\ref{e:constitutive-relation}) which implies $\grad \cdot (\rho \bfv_*) =  \grad \cdot (\rho \bfv + \la^2 \rho (\grad \times \bfw)) = \grad \cdot (\rho \bfv)$. Taking the curl of the R-Euler momentum balance equation we get the R-vorticity equation:
	\beq
	{\bf w}_t + \grad \times ( {\bf w} \times {\bf v}) = - \grad \times \left( \la^2 {\bf w} \times (\grad \times {\bf w}) \right) 
	\qquad 
	\text{or} \qquad 
	\bfw_t + \grad \times (\bfw \times \bfv_*) = 0.
	\label{e:R-vorticity-eqn-compressible}
	\eeq
The incompressible regularized evolution equations possess a positive definite integral invariant [with suitable boundary data]:
	\beq
	\DD{E^*}{t} = \DD{}{t} \left( \int_V \left[ \half \rho {\bf v}^2 + \half \la^2 \rho {\bf w}^2 \right] \: d^3r \right) = 0.
	\label{e:cons-egy-incompress}
	\eeq
For compressible flow, $E^*$ is {\em not} conserved if $\la$ is a constant length. On the other hand, we do find a conserved energy if we include compressional potential energy and also let the field $\la(\bfr,t)$ be a dynamical length governed by the constitutive relation $\la^2 \rho = \la_0^2 \rho_0 = $ constant (\ref{e:constitutive-relation}). As a consequence, $\la$ is not an independent propagating field like $\bfv$ or $\rho$, its evolution is determined by that of $\rho$. Here $\la_0$ is some constant short-distance cut-off (e.g. a mean-free path at mean density) and $\rho_0$ is a constant mass density (e.g. the mean density). $\la$ is smaller where the fluid is denser and larger where it is rarer. This is reasonable if we think of $\la$ as a position-dependent mean-free-path. However, it is only the combination $\la_0^2 \:\rho_0$ that appears in the equations. So compressible R-Euler involves only one new dimensional parameter, say $\la_0$. A dimensionless measure of the cutoff $n \la^3 = \la_0^3 \, n_0^{3/2} \,n^{-1/2}$ may be obtained by introducing the number density $n = \rho/m$ where $m$ is the molecular mass. It is clearly smaller in denser regions and larger in rarified regions. As noted in the introduction, if we take $(\la/L)^2 \propto a^3/L^3$ where $a \propto n^{-1/3}$ and $L$ are inter-particle spacing and macroscopic system size, then $\la^2 \rho$ would be a constant. The conservation of $E^*$ implies an a priori bound on enstrophy; no such bound is available for Eulerian flows, where enstrophy could diverge due to vortex stretching \cite{Frisch, KRSreenivasan-onsager}. Note that boundedness of enstrophy under R-Euler evolution may still permit $\bfw$ to develop discontinuities or mild divergences for certain initial conditions.

The KdV and R-Euler equations are conservative regularizations in one and three dimensions. The dimensional reduction of R-Euler provides a possible regularization of ideal flows in $2$ dimensions. However, for incompressible 2d flow, the twirl term becomes a gradient and does not affect the evolution of vorticity (see \S \ref{s:plane-flow}). This is to be expected as {\it incompressible} 2d Euler flows do not require regularization: there is no vortex stretching, enstrophy and all moments of $\bfw^2$ are conserved. On the other hand, the twirl term leads to a new and non-trivial regularization of compressible flow in 2d (see \S \ref{s:plane-flow}).


It is possible to show that the twirl term is unique among regularization terms that are at most quadratic in $\bfv$ with at most $3$ spatial derivatives subject to the following physical requirements (1) it must preserve Eulerian symmetries and (2) admit a Hamiltonian formulation with the standard Landau Poisson brackets and continuity equation. A proof of this uniqueness result will be given in a future paper.


In the light of possible astrophysical applications, we briefly note two important generalisations of the R-Euler system. Suppose a conservative body force ${\bf F}= - \rho \grad V$ is operative, where the potential $V$ arises for instance from gravity. Then (\ref{e:reg-Euler-3D}) has the additional term $- \grad V$, signifying acceleration due to the body force. Evidently, we may now set,
	\beq
	\sigma = \left( \frac{\gamma}{\gamma -1} \right) \frac{p}{\rho} + \half {\bf v}^2 + V \equiv h + \half \bfv^2 
	\eeq
where the new enthalpy includes a contribution from potential energy. The conservation laws of the next Section generalize upon including the potential energy of the body force.

A much less trivial extension will also be briefly indicated: in compressible ideal MHD the body force is the magnetic Lorentz force ${\bf j \times B}$, which has to be related to the fluid motion through Maxwell's equations for a quasi-neutral, compressible, ideal fluid. The governing equations for mass density $\rho$, magnetic field $\bfB$ and velocity $\bfv$ take the following forms:
	\beq
	\dd{\rho}{t} + \grad \cdot (\rho \bfv) = 0, \quad
		\dd{\bfv}{t} + (\bfv \cdot \grad ) \bfv = - \ov{\rho} \grad p + \frac{\bfj \times \bfB}{\rho}
	\quad \text{and} \quad
	\dd{\bfB}{t} = \grad \times (\bfv \times \bfB)\label{e:unreg-ideal-MHD}.
	\eeq
The electric body force cancels out when one adds the momentum equations for electrons and ions. Thus one  arrives at the above momentum equation for the center of mass velocity $\bfv$ of the electrons and ions in the quasi-neutral plasma treated as a single fluid. In non-relativistic plasmas, the displacement current term in Ampere's law can be neglected, allowing us to express the electric current as the curl of the magnetic field: $\mu_0 \bfj = \grad \times \bfB$. In particular, $\bfj$ is not an independent dynamical variable, its evolution is determined by that of $\bfB$. So the magnetic body force may be written as $ (\grad \times \bfB) \times \bfB/\rho \mu_0$. In MHD, the constitutive equation relating the electric and magnetic fields to the fluid motion is the ideal Ohm's law: $\bfE + (\bfv \times \bfB)=0$, which leads to the above expression for Faraday's law.

The regularized compressible MHD (R-MHD) equations follow from arguments similar to those presented for neutral compressible flows. The continuity equation, $\rho_t + \grad \cdot (\rho \bfv) = 0$ is unchanged. As noted, it may be written in terms of swirl velocity: $\rho_t + \grad \cdot (\rho \bfv_*) = 0$. As in regularized fluid theory, we introduce the twirl acceleration on the RHS of the momentum equation, where $\la$ is again subject to (\ref{e:constitutive-relation}):
	\beq
	\dd{\bfv}{t} + (\bfv \cdot \grad) \bfv = -\ov{\rho}\grad p + \frac{\bfj \times \bfB}{\rho} - \la^2 \bfw \times (\grad \times \bfw) = -\ov{\rho}\grad p - \frac{\bfB \times(\grad \times\bfB)}{\mu_0\rho} - \la^2 \bfw \times (\grad \times \bfw).
	\label{e:R-MHD-Euler-v}
	\eeq
The twirl regularization term is the vortical analogue of the magnetic Lorentz force term with $1/\mu_0$ replaced with $\la^2 \rho$. This is also evident in the R-MHD stress tensor $S_{ij} = p \del_{ij} + \la^2 \rho \left( \half w^2 \del_{ij} - w_i w_j \right) +  \left( \half B^2 \del_{ij} - B_i B_j \right)/\mu_0$ appearing in the momentum equation $\rho (Dv_i/Dt) = - \pdr_j S_{ij}$. Equation (\ref{e:R-MHD-Euler-v}) can be obtained from the unregularized equation (\ref{e:Euler-eqn}) by replacing $\bfv$ with $\bfv_*$ in the vortex acceleration term:
	\beq
	\dd{\bfv}{t} + \bfw \times \bfv_* = -\ov{\rho}\grad p - \half \grad \bfv^2 + \frac{\bfj \times \bfB}{\rho}.
	\label{e:R-MHD-Euler-v*}
	\eeq
Similarly, the regularized Faraday law in R-MHD is obtained by replacing $\bfv$ by $\bfv_*$ in (\ref{e:unreg-ideal-MHD}) i.e.,
	\beq
	\dd{\bfB}{t} = \grad \times (\bfv_* \times \bfB) = \grad \times \left(\bfv \times \bfB - \la^2 \bfB \times (\grad \times \bfw) \right).
	\label{e:R-MHD-Faraday}
	\eeq
The regularization term in Faraday's law is the curl of the `magnetic' twirl $- \la^2 \bfB \times (\grad \times \bfw)$ term in analogy with the `vortical' twirl term $-\la^2 \bfw \times (\grad \times \bfw)$. The regularized Faraday equation is $3^{\rm rd}$ order in space derivatives of $\bfv$ and first order in $\bfB$. From (\ref{e:R-MHD-Faraday}), we deduce that the potentials ($\bfA, \phi$) in any gauge must satisfy
	\beq
	\pdr_t \bfA = \bfv_* \times \bfB - \grad \phi.
	\label{e:A-evolution-R-MHD}
	\eeq
It turns out that compressible R-MHD possesses conservation laws similar to those deduced in \cite{thyagaraja} for incompressible R-MHD, see \S \ref{s:R-MHD-cons-laws-alfven}. One can readily include conservative body forces like gravity into R-MHD. The inclusion of regularization terms arising from electron inertia and Hall effect\cite{thyagaraja} and extension to the two-fluid plasma system will be presented in a future work.

\section{Conservation laws for  regularized compressible flow and MHD}

\subsection{Conservation laws for regularized compressible fluid flow}
\label{s:cons-laws}

{\noindent \bf Swirl Energy Conservation:} Under compressible R-Euler evolution, the ``swirl'' energy density and flux vector
	\beq
	{\cal E}^* = \left[ \frac{\rho \bfv^2}{2}+ U(\rho) + \frac{\la^2 \rho \bfw^2}{2} \right] 
	\quad \text{and} \quad
		\bff = \rho \sigma \bfv + \la^2 \rho (\bfw \times \bfv_*) \times \bfw.
	\label{e:swirl-egy-density-current}
	\eeq
satisfy the local conservation law $\dd{{\cal E}^*}{t} + \grad \cdot \bff = 0$. Here $U(\rho) = p/(\gamma - 1)$ is the compressional potential energy for adiabatic flow. Given suitable boundary conditions [BCs, discussed below], the system obeys a global energy conservation law:
	\beq
	\frac{d E^*}{dt} = 0 \quad \text{where} \quad E^* = \int \left[ \frac{\rho \bfv^2}{2}+ U(\rho) + \frac{\la^2 \rho \bfw^2}{2} \right] \: d\bfr.
	\label{e:swirl-energy-R-Euler}
	\eeq

{\noindent \bf Flow Helicity Conservation:} The R-Euler equations possess a local conservation law for helicity density $\bfv \cdot \bfw$ and its flux $\bff_{\cal K}$:
	\beq
	\pdr_t (\bfv \cdot \bfw)
	+ \grad \cdot \left( \sig \bfw + (\bfw \times \bfv_*) \times \bfv \right)  = 0.
	\label{e:helicity-current-conservation}
	\eeq
This local conservation law implies global conservation of helicity ${\cal K} = \int \bfv \cdot \bfw \, d\bfr$, provided $\bff_{\cal K} \cdot \hat n = 0$ on the boundary $\pdr V$ of the flow domain $V$. Here $\hat n$ is the unit outward-pointing normal vector on the surface $\pdr V$.

{\noindent \bf Momentum Conservation:} Flow momentum is ${\bf P} = \int \rho \bfv \; d\bfr$. Momentum density ${\cal P}_i = \rho v_i$ and the stress tensor $\Pi_{ij}$ satisfy
	\beq
	\dd{{\cal P}_i}{t} + \pdr_j \Pi_{ij} = 0 \quad \text{where} \quad
	\Pi_{ij} = \Pi_{ji} = \rho v_i v_j + p \del_{ij} + \rho \la^2 \left(\half \bfw^2 \del_{ij} - w_i w_j \right).
	\label{e:momentum-current-tensor}
	\eeq
For $\bfP$ to be globally conserved, we expect to need a translation-invariant flow domain $V$. If $V = \mathbb{R}^3$, decaying boundary conditions ($\bfv \to 0$, $\rho \to$ constant) ensure $d{\bf P}/dt = 0$. Periodic BCs in a cuboid also ensure global conservation of $\bf P$.

{\noindent \bf Angular Momentum Conservation:} For regularized compressible flow, we define the angular momentum density as $\vec {\cal L} = \rho {\bf r} \times \bfv$. We find that the angular momentum satisfies the local conservation law:
	\beq
	\dd{{\cal L}_i}{t} + \pdr_l \Lambda_{il} = 0 \quad \text{where} \quad
	\Lambda_{il} = \eps_{ijk} r_j \Pi_{kl}.
	\label{e:ang-mom-current-conservation}
	\eeq
$\Lambda_{il}$ is the angular momentum flux tensor. For $\bfL = \int \vec {\cal L} \: d^3 r$ to be globally conserved, the system must be rotationally invariant. For instance, decaying BC in an infinite domain would guarantee conservation of $\vec {\cal L}$. We also note that in symmetric domains [axisymmetric torus or circular cylinder] corresponding components of angular momentum or linear momentum associated with the symmetry may also be conserved. The situation here is similar to typical Eulerian systems.

\subsection{Boundary Conditions}
\label{s:BC}

In the flow domain $\mathbb{R}^3$, it is natural to impose decaying BCs ($\bfv \to 0$ and $\rho \to$ constant as $|\bfr| \to \infty$) to ensure that total energy $E^*$ is finite and conserved. For flow in a cuboid, periodic BCs ensure finiteness and conservation of energy. For flow in a bounded domain $V$, demanding global conservation of energy leads to another natural set of BCs. Now $d E^*/dt = - \int_{\pdr V} \bff \cdot \hat n \: dS$ where $\bff$ is the energy current (\ref{e:swirl-egy-density-current}) and $\pdr V$ the boundary surface. $\bff \cdot \hat n = 0$ if the following conditions hold:
	\beq
	\bfv \cdot \hat n = 0 \quad
	\text{and} \quad
	\bfw \times \hat n = 0.
	\label{e:BCs-for-energy-cons}
	\eeq
These BCs are, for instance, satisfied at the top and bottom of a bucket of rigidly rotating fluid. The BC $\bfv \cdot \hat n = 0$ also ensures global conservation of mass as $\DD{}{t} \int \rho d\bfr = - \int \rho \bfv \cdot \hat n \: dS$. Since the R-Euler equation is $2^{\rm nd}$ order in spatial derivatives of $\bfv$, it is consistent to impose conditions on both $\bfv$ and its $1^{\rm st}$ derivatives. These boundary conditions imply that the twirl acceleration is tangential to the boundary surface $\bfT \cdot \hat n = (\bfw \times \la^2 (\grad \times \bfw)) \cdot \hat n = (\hat n \times \bfw) \cdot (\la^2 \grad \times \bfw) = 0$. It is interesting to note that the BCs ensuring helicity conservation (see \S \ref{s:direct-proofs}) are `orthogonal' to those for energy conservation
	\beq
	\bfv \times \hat n = 0 \quad 
	\text{and} \quad
	\bfw \cdot \hat n = 0 \quad \imply \quad \bff_{\cal K} \cdot \hat n = 0.
	\eeq
So helicity and energy cannot both be globally conserved simultaneously with these BCs [in bounded domains]. However, periodic or decaying BC would ensure simultaneous conservation of both. Similarly, neither angular momentum nor linear momentum is conserved in a finite flow domain with the BCs that ensure energy conservation. However, with sufficiently rapidly decaying BCs, energy, momentum, angular momentum and helicity can all be conserved simultaneously.

\subsection{Direct proofs of the Conservation Laws}
\label{s:direct-proofs}

We derive the stated conservation relations for R-Euler flows from the equations of motion (\ref{e:continuity-R-Euler},\ref{e:reg-Euler-3D},\ref{e:R-vorticity-eqn-compressible}) and the imposed BC's. Later these conservation laws will also be obtained using Poisson brackets.

{\noindent \bf Swirl energy conservation:} To prove the local conservation law for ${\cal E}^*$ (\ref{e:swirl-egy-density-current}) we begin by computing the time derivative of each term in the energy density.
\beqs
	\dd{}{t} \left(\half \rho {\bf v}^2 \right) &=& \half {\bf v}^2 \dd{\rho}{t} + \rho {\bf v} \cdot \dd{\bf v}{t}
	= - \half {\bf v}^2 \grad \cdot (\rho {\bf v}) - \rho {\bf v} \cdot \grad \sigma - \la^2 \rho {\bf v} \cdot {\bf T},
	\cr
	\dd{}{t} \left( \frac{p}{\gamma - 1} \right) &=& \frac{p_o}{\gamma - 1} \dd{}{t} \left(\frac{\rho}{\rho_o} \right)^\gamma
		= -\frac{\gamma}{\gamma -1} \left( \frac{p}{\rho} \right) \grad \cdot (\rho {\bf v}),
	\cr
	\dd{}{t} \left( \half \rho \la^2 {\bf w}^2 \right) &=& \rho_0 \la_0^2 {\bf w} \cdot \dd{\bf w}{t} = \rho \la^2 {\bf w} \cdot
	\left[ \grad \times ({\bf v} \times {\bf w}) - \grad \times (\la^2 {\bf T}) \right].
	\eeqs
It follows that:
	\beq
	\dd{{\cal E}^*}{t} = - \sig \grad \cdot (\rho {\bf v}) - \rho {\bf v} \cdot \grad \sigma
	- \rho \la^2 \left[ {\bf v} \cdot {\bf T} - {\bf w} \cdot
	\grad \times ({\bf v} \times {\bf w}) \right]
	- \rho \la^2 {\bf w} \cdot \grad \times (\la^2 {\bf T}).
	\eeq
Since $\la$ is a free parameter, the coefficient of each power of $\la$ must be shown to be a divergence. It follows from straightforward but somewhat lengthy algebra [which we omit for brevity] that this is indeed the case, leading to a local conservation equation $\pdr{{\cal E}^*}/\pdr t +  \grad \cdot \bff = 0$ with the energy flux vector $\bff$ given in (\ref{e:swirl-egy-density-current}). It should be noted that this local conservation law crucially depends on the constitutive relation (\ref{e:constitutive-relation}). The conservation of $E^* = \int {\cal E}^* \: d\bfr$ follows from Gauss' divergence theorem and our choice of boundary conditions ($\bfv \cdot \hat n = 0$ and $\bfw \times \hat n = 0$), which follow from writing
	\beq
	\bff \cdot \hat n =	\rho \sigma \bfv \cdot \hat n + \la^2 \rho (\bfw \times \bfv_*) \cdot (\bfw \times \hat n).
	\eeq
 
{\noindent \bf Flow helicity conservation:} To obtain the local conservation law for $\bfv \cdot \bfw$, we use the regularized equations (\ref{e:reg-Euler-3D},\ref{e:R-vorticity-eqn-compressible}) to write
	\beq
	\bfw \cdot \bfv_t = - \bfw \cdot ( \grad \sigma - \la^2 \bfT)  \quad
	\text{and} \quad
	\bfv \cdot \bfw_t = \bfv \cdot (\grad \times (\bfv \times \bfw) - \grad \times (\la^2 \bfT)).
	\eeq
Now $\bfv \cdot (\grad \times (\bfv \times \bfw)) = - \grad \cdot (\bfv \times (\bfv \times \bfw))$ since $(\bfv \times \bfw) \cdot \bfw = 0$. Similarly, $\bfv \cdot (\grad \times (\la^2 \bfT)) = \grad \cdot (\la^2 \bfT \times \bfv)$ since $\bfT \cdot \bfw = 0$. Combining these two, the time derivative of flow helicity density is a divergence $\grad \cdot \bff_{\cal K}$,
	\beq
	\pdr_t (\bfv \cdot \bfw) = \bfw \cdot \bfv_t + \bfv \cdot \bfw_t
	= - \bfw \cdot \grad \sigma
	- \grad \cdot ( \bfv \times (\bfv \times \bfw) ) - \grad \cdot (\la^2 \bfT \times \bfv)
	= \grad \cdot \left(\sig \bfw + (\bfw \times \bfv_*) \times \bfv \right),
	\eeq
as $\bfw$ is solenoidal. Writing
	\beq
	\bff_{\cal K} \cdot \hat n
	= \sig \bfw \cdot \hat n + (\bfw \times \bfv_*) \cdot (\bfv \times \hat n),
	\eeq
we infer BCs $\bfw \cdot \hat n = 0$ and
$\bfv \times \hat n = 0$ that ensure global helicity conservation [decaying BCs would of course also work].

{\noindent \bf Linear and angular momentum conservation:} The proof of local conservation of momentum density $\rho \bfv$ uses the continuity and R-Euler equations:
	\beq
	\rho \dd{v_i}{t} = - \rho v_j \pdr_j v_i - \pdr_i p - \rho \la^2 T_i
	\quad \text{and} \quad
	v_i \dd{\rho}{t} = - v_i \pdr_j (\rho v_j).
	\eeq
By the constitutive relation, $\la^2 \rho$ is a constant, so
	\beq
	\dd{{\cal P}_i}{t} = - \pdr_j (\rho v_i v_j) - \pdr_i p - \rho \la^2 T_i = - \pdr_j \left[ \rho v_i v_j + p \del_{ij} + \rho \la^2 \left(\half \bfw^2 \del_{ij} - w_i w_j \right) \right] \equiv - \pdr_j \Pi_{ij}.
	\eeq
Thus, we have local conservation of momentum $\pdr {{\cal P}_i}/\pdr t + \pdr_j \Pi_{ij} = 0$. The time derivative of angular momentum density $\vec {\cal L} = \rho {\bf r} \times \bfv$ is calculated using the local conservation law for momentum density and the symmetry of $\Pi_{ij}$:
	\beq
	\dd{{\cal L}_i}{t} = \eps_{ijk} r_j \dd{(\rho v_k)}{t} 
	= - \eps_{ijk} r_j \pdr_l \Pi_{kl} = - \pdr_l \left( \eps_{ijk} r_j \Pi_{kl} \right) = - \pdr_l \Lambda_{il}.
	\eeq
So angular momentum satisfies $\pdr {\cal L}_i/ \pdr t + \pdr_l \Lambda_{il} = 0$ where $\Lambda_{il}$ is the angular momentum flux tensor  (\ref{e:ang-mom-current-conservation}).

\subsection{Conservation Laws for R-MHD and boundary conditions}
\label{s:R-MHD-cons-laws-alfven}

{\noindent \bf Swirl energy conservation}: In R-MHD, we obtain the following local energy conservation law:
		\beqs
	      \dd{{\cal E}_{\rm mhd}^*}{t} &+& \grad \cdot \bff_{\rm mhd} = 0 \;\; \rm{where} \quad
	      {\cal E}_{\rm mhd}^*=\left( \frac{\rho(x) \bfv^2(x)}{2}+ U(\rho) + \frac{\la^2 \rho \bfw^2(x)}{2} + \frac{\bfB^2(x)}{2 \mu_0} \right)  \quad\rm{and} \cr
		 \bff_{\rm mhd} 
		     & = & \left( \rho \sigma \bfv + \la^2 \rho (\bfw \times \bfv_*) \times \bfw \right) 
		     + \ov{\mu_0} \left[ \bfB \times  (\bfv_* \times \bfB) + \la^2 \left( \bfw \times ((\grad \times \bfB) \times \bfB) \right)  \right]
	\eeqs
is the energy flux vector and $E^*_{\rm mhd} = \int_V{\cal E}^*_{\rm mhd}\: d^3r$ is the the total `swirl' energy of barotropic compressible R-MHD. 

{\noindent \sc Proof:} The time derivative of the swirl energy density is calculated using the evolution equations (\ref{e:R-MHD-Euler-v},\ref{e:R-MHD-Faraday}) for $\bfv, \bfw, \bfB$ and $\rho$: 
\beqs
	\dd{}{t} \left(\half \rho {\bf v}^2 \right) &=& \half {\bf v}^2 \dd{\rho}{t} + \rho {\bf v} \cdot \dd{\bf v}{t}
	= - \half {\bf v}^2 \grad \cdot (\rho {\bf v}) - \rho {\bf v} \cdot \grad \sigma - \la^2 \rho {\bf v} \cdot {\bf T} + \bfv \cdot (\bfj \times \bfB),
	\cr
	\dd{}{t} \left( \frac{p}{\gamma-1} \right) &=& \frac{p_o}{\gamma-1} \dd{}{t} \left(\frac{\rho}{\rho_o} \right)^\gamma
		= -\frac{\gamma}{\gamma -1} \left( \frac{p}{\rho} \right) \grad \cdot (\rho {\bf v}),
	\cr
	\dd{}{t} \left( \half \rho \la^2 {\bf w}^2 \right) &=& \rho_0 \la_0^2 {\bf w} \cdot \dd{\bf w}{t} = \rho \la^2 {\bf w} \cdot
	\left( \grad \times ({\bf v} \times {\bf w}) - \grad \times (\la^2 {\bf T}) + \ov{\rho} \grad \times (\bfj \times \bfB) \right) \cr
	\dd{}{t} \left( \frac{\bfB^2}{2\mu_0} \right) &=& \ov{\mu_0} \bfB \cdot \dd{\bfB}{t} = \ov{\mu_0} \bfB \cdot \left( \grad \times (\bfv_* \times \bfB) \right).
	\eeqs
Therefore the time derivative of energy density is :
	\beqs
	\dd{{\cal E}_{\rm mhd}^*}{t} &=& - \sig \grad \cdot (\rho {\bf v}) - \rho {\bf v} \cdot \grad \sigma
	- \rho \la^2 \left[ {\bf v} \cdot {\bf T} - {\bf w} \cdot
	\grad \times ({\bf v} \times {\bf w}) \right]
	- \rho \la^2 {\bf w} \cdot \grad \times (\la^2 {\bf T})
	\cr 
	&& + {\mu_0}^{-1} \left[ \bfv \cdot ((\grad \times \bfB) \times \bfB) + \la^2 \bfw \cdot \grad \times ((\grad \times \bfB) \times \bfB) + \bfB \cdot \left(\grad \times ((\bfv + \la^2 \grad \times \bfw) \times \bfB ) \right) \right].
	\eeqs
The first line containing terms independent of $\bfB$ has already been expressed as the divergence of the R-Euler fluid energy current $\bff = \rho \sigma \bfv + \la^2 \rho (\bfw \times \bfv_*) \times \bfw$. Now we split the terms containing $\bfB$ into those of order $\la^0$ and those quadratic in $\la$ and express each as a divergence using the vector identity $\grad \cdot (\bfA \times \bfB) = \bfB \cdot \grad \times \bfA - \bfA \cdot \grad \times \bfB$:
	\beqs
\la^0: &&	\hspace{-.6cm}	\bfB \cdot ( \grad \times (\bfv \times \bfB)) + \bfv \cdot ((\grad \times \bfB) \times \bfB) = \grad \cdot \left[ (\bfv \times \bfB) \times \bfB \right] \cr
\la^2: &&	\hspace{-.6cm}
\bfw \cdot \grad \times \left( (\grad \times \bfB) \times \bfB \right) + \bfB \cdot \grad \times \left( (\grad \times \bfw) \times \bfB \right) = - \grad \cdot \left[ \bfw \times ((\grad \times \bfB) \times \bfB) + \bfB \times ((\grad \times \bfw) \times \bfB) \right].
	\eeqs
Thus we obtain the abovementioned conserved energy current density for regularized compressible MHD. Boundary conditions on the surface $\pdr V$ of the flow domain $V$ that ensure  global conservation of $E^*_{\rm mhd}$ are \beq
	\bfv \cdot \hat n = 0, \quad
	\bfw \times \hat n = 0, \quad 
	(\grad \times \bfw) \cdot \hat n = 0 	\quad \text{and} \quad
	\bfB \cdot \hat n = 0.
	\label{e:R-MHD-energy-BCs}
	\eeq
The R-MHD equations of motion (\ref{e:R-MHD-Euler-v},\ref{e:R-MHD-Faraday}) are $3^{\rm rd}$ order in $\bfv$ and $1^{\rm st}$ order in $\bfB$. So we may impose BCs on $\bfB$, $\bfv$, the $1^{\rm st}$ and $2^{\rm nd}$ derivatives of $\bfv$. It also follows from (\ref{e:R-MHD-energy-BCs}) that $\bfB \cdot \bfw = 0$ and $\bfv_* \cdot \hat n = 0$ on the boundary. These BCs follow from writing
	\beqs
	\bff_{\rm mhd} \cdot \hat n &=& \rho \sigma \bfv \cdot \hat n + \la^2 \rho (\bfw \times \bfv_*) \cdot (\bfw \times \hat n) \cr
	&& + \,{\mu_0}^{-1} \left[\bfB^2 (\bfv_* \cdot \hat n) - (\bfv_* \cdot \bfB) \bfB \cdot \hat n
	+ \la^2 \left\{  (\bfw \cdot \bfB) (\grad \times \bfB \cdot \hat n) - (\grad \times \bfB \cdot \bfw)  (\bfB \cdot \hat n)	\right\} \right].
	\eeqs
{\noindent \bf Magnetic helicity conservation}: We define magnetic helicity as ${\cal K}_B = \int_V \bfA \cdot \bfB d\bfr$. This is the magnetic analogue of flow helicity ${\cal K} = \int_V \bfv \cdot \bfw d\bfr$ where we make the replacements $\bfv \to \bfA, \bfw \to \bfB$. Despite appearances, ${\cal K}_B$ is gauge-invariant for decaying boundary conditions or if $\bfB$ is tangential to $\pdr V$. For, under a gauge transformation $\bfA \to \bfA + \grad \tht$,
	\beq
	{\cal K}_B \to {\cal K}_B + \int_V \bfB \cdot \grad \tht d^3 r = K + \int_V \grad \cdot (\tht \bfB) \: d^3r = K + \int_{\pdr V} \tht \bfB \cdot \hat n \: dS.
	\eeq
Magnetic helicity density is locally conserved in any gauge with potentials $(\bfA,\phi)$
	\beq
	\dd{(\bfA \cdot \bfB)}{t} + \grad \cdot (\bfA \times (\bfv_* \times \bfB) + \bfB \phi) = 0.
	\eeq
{\noindent \sc Proof}: Using (\ref{e:R-MHD-Faraday}, \ref{e:A-evolution-R-MHD}) the time derivative of $\bfA \cdot \bfB$ is
	\beq
	\dd{(\bfA \cdot \bfB)}{t} = \bfA \cdot \dd{\bfB}{t} + \bfB \cdot \dd{\bfA}{t} = \bfA \cdot \grad \times (\bfv_* \times \bfB) + \bfB \cdot (\bfv_* \times \bfB - \grad\phi).
	\eeq
The second term is zero. Using the vector identity $\grad \cdot (\bfA \times \bfD) = \bfD \cdot \grad \times \bfA - \bfA \cdot \grad \times \bfD$ and $\grad \cdot \bfB$ = 0 we may write
	\beqs
	\dd{(\bfA \cdot \bfB)}{t} &=& \bfA \cdot \grad \times (\bfv_* \times \bfB) - \bfB \cdot \grad \phi = - \grad \cdot (\bfA \times (\bfv_* \times \bfB)+\bfB\phi) + (\bfv_* \times \bfB) \cdot (\grad \times \bfA) \cr
	&=& - \grad \cdot (\bfA \times (\bfv_* \times \bfB) + \bfB\phi) .
		\eeqs
Thus we get the local conservation law for magnetic helicity density as stated above.
	$\bfA \times (\bfv_* \times \bfB) + \bfB \phi$ is the flux of magnetic helicity\footnote{In the laboratory gauge used in the Poisson brackets of \S \ref{s:PB-for-R-MHD}, $\phi = \bfv_* \cdot \bfA$ so the magnetic helicity current is $(\bfA \cdot \bfB) \bfv_*$ in this gauge.}. Global conservation of ${\cal K}_B$ requires the flux of magnetic helicty across the boundary surface to be zero. This is guaranteed by the conditions $\bfB \cdot \hat n = 0$, $\bfv \cdot \hat n = 0$ and $(\grad \times \bfw) \cdot \hat n = 0$. This is because 
	\beq
	\left( \bf A \times (\bfv_* \times \bfB) \right) \cdot \hat n = (\bfv_* \cdot \hat n) (\bfA \cdot \bfB) - (\bfv_* \cdot \bfA) (\bfB \cdot \hat n)
	= (\bfA \cdot \bfB) \left( \bfv \cdot \hat n + \la^2 (\grad \times \bfw) \cdot \hat n \right) - (\bfv_* \cdot \bfA) (\bfB \cdot \hat n).
	\eeq
Note that for conservation of ${\cal K}_B$ it suffices that both $\bfB$ and $\bfv_*$ be tangential to $\pdr V$. The BC $\bfB \cdot \hat n = 0$ also guarantees gauge-invariance of ${\cal K}_B$. Moreover, unlike for flow helicity, the BCs that guarantee $E^*$ conservation also ensure conservation of ${\cal K}_B$ (though not vice versa). In an infinite domain energy and magnetic helicity are conserved if $\bfv , \bfB \to 0$ and $\rho \to $ constant as $\bfr \to \infty$. For a finite flow domain, we may also impose periodic BC for energy and magnetic helicity conservation. 


{\noindent \bf Cross helicity conservation:} Cross helicity $\int \bfv \cdot \bfB d\bfr$ measuring the degree of linkage of vortex and magnetic field lines is locally conserved in R-MHD:
	\beq
	\pdr_t (\bfv \cdot \bfB) + \grad \cdot (\sigma\bfB + \bfv \times (\bfv_* \times \bfB)) = 0.
	\eeq
The cross helicity current may be obtained from the magnetic helicity current by replacing $\phi \to \sigma$ and $\bfA \to \bfv$. To see this, we express $\pdr_t (\bfv \cdot \bfB)$ as a divergence 
	\beqs
	\pdr_t (\bfv \cdot \bfB) &=& \bfB \cdot \bfv_t + \bfv \cdot \bfB_t = \bfB \cdot ( - \grad \sigma + \bfv_* \times \bfw) + \bfv \cdot (\grad \times (\bfv_* \times \bfB)) \cr
	&=& - \bfB \cdot \grad \sigma + \bfB \cdot \bfv_* \times \bfw + \bfv_* \times \bfB \cdot \bfw +\grad \cdot ((\bfv_* \times \bfB) \times \bfv) = -\grad \cdot (\sigma\bfB + \bfv \times (\bfv_* \times \bfB)). 
	\eeqs
Boundary conditions that lead to global cross helicity conservation are $\bfv_* \cdot \hat n =0$ and $\bfB \cdot \hat n = 0$. 


{\noindent \bf Locally conserved linear and angular momenta:} The momentum density ${\cal P}_i = \rho v_i$ and stress tensor $\Pi_{ij}$ satisfy a local conservation law
	\beq
	\dd{{\cal P}_i}{t} + \pdr_j \Pi_{ij} = 0, \quad \text{where} \quad \Pi_{ij} = \rho v_i v_j + p \del_{ij} + \la^2 \rho \left( \half w^2 \del_{ij} - w_i w_j \right) + \ov{\mu_0} \left( \half B^2 \del_{ij} - B_i B_j \right).
	\label{e:mom-cons-r-mhd}
	\eeq
$\bfB$ and $\bfw$ enter $\Pi_{ij}$ in the same manner since the twirl force ($-\la^2 \rho \bfw \times (\grad \times \bfw)$) and magnetic Lorentz force ($- \ov{\mu_0} \bfB \times (\grad \times \bfB)$) are of the same form. The proof is as follows
	\beq
	\dd{{\cal P}_i}{t} = v_i \dd{\rho}{t} + \rho \dd{v_i}{t} = - \pdr_j \left( \rho v_i v_j + p \del_{ij} + \la^2 \rho \left( \half w^2 \del_{ij} - w_i w_j \right) \right) + \ov{\mu_0} ((\grad \times \bfB) \times \bfB)_i.
	\eeq
The first term is known from the conservation of momentum in R-Euler flow and the second comes from the magnetic force. The magnetic force term can be expressed as a divergence leading to the above-mentioned result:
	\beq
	((\grad \times \bfB) \times \bfB)_i = - \half \pdr_i B^2 + B_j \pdr_j B_i = - \pdr_j \left( \half B^2 \del_{ij} - B_i B_j \right).
	\eeq
We define angular momentum density in R-MHD as $\vec {\cal L} = \rho \bfr \times \bfv$. Using the local conservation of $\rho \bfv$ we find that $\vec {\cal L}$ too is locally conserved in R-MHD:
	\beq
	\dd{{\cal L}_i}{t} = \eps_{ijk} r_j \dd{\rho v_k}{t} = - \pdr_l \left( \eps_{ijk} r_j \Pi_{kl} \right) = - \pdr_l \Lambda_{il}.
	\label{e:ang-mom-cons-r-mhd}
	\eeq
Linear momentum $\int {\cal P}_i \, d\bfr$ and angular momentum $\int {\cal L}_i \, d\bfr$ are globally conserved for appropriate boundary conditions (e.g. decaying BC in an infinite domain or periodic BC in a cuboid for linear momentum).

\subsection{Regularized Kelvin-Helmholtz and Alfv\'{e}n freezing-in theorems and swirl velocity}
\label{s:Kelvin-circulation-Kelvin-Helmholtz-swirl-vel}

{\noindent \bf Regularized Kelvin-Helmholtz freezing-in theorem}: For incompressible ideal flow, it is well known that vorticity is frozen into the velocity field: $\bfw_t + \bfv \cdot \grad \bfw - \bfw \cdot \grad \bfv = 0$ or $\bfw_t + {\cal L}_\bfv \bfw = 0$. Here ${\cal L}_\bfv \bfw$ is the Lie derivative of $\bfw$ along $\bfv$, which is also the commutator of vector fields $[\bfv,\bfw]$. Kelvin's and Helmholtz's theorems on vorticity follow from the freezing of $\bfw$ into $\bfv$. This result has an extension to the compressible,  regularized theory. We show that $\bfw/\rho$ is frozen into the swirl velocity $\bfv_* = \bfv + \la^2 \grad \times \bfw$ (\ref{e:v*-continuity-eqn}). The R-vorticity equation (\ref{e:R-vorticity-eqn-compressible}) can be written as
	\beqs
	\dd{\bfw}{t} + \grad \times (\bfw \times \bfv_*) = 0
	\quad \imply \quad  \dd{\left\{(\bfw/\rho) \rho\right\}}{t} + \grad \times \left(\rho \frac{\bfw}{\rho} \times  \bfv_* \right) &=& 0 \cr
	\imply \quad 
	\rho \dd{(\bfw / \rho)}{t} + \frac{\bfw}{\rho} \dd{\rho}{t} + \bfw (\grad \cdot \bfv_*) - \bfv_* (\grad \cdot \bfw) + (\bfv_* \cdot \grad) (\rho (\bfw/\rho))
	- (\rho (\bfw/\rho) \cdot \grad) \bfv_* &=& 0.
	\eeqs
We use the continuity equation (\ref{e:v*-continuity-eqn}) to write $\rho_t = - \rho \grad \cdot \bfv_* - \bfv_* \cdot \grad \rho$. The last term is one that appears in the Lie derivative ${\cal L}_{\bfv_*}(\bfw/\rho)$ and the penultimate term also contributes to ${\cal L}_{\bfv_*}(\bfw/\rho)$ upon using the Leibnitz rule. Thus
	\beq
	\rho \dd{(\bfw / \rho)}{t} - \frac{\bfw}{\rho} (\bfv_* \cdot \grad) \rho - \bfw \grad \cdot \bfv_* + \bfw \grad \cdot \bfv_* + \rho \bfv_* \cdot \grad \left(\frac{\bfw}{\rho} \right) + \frac{\bfw}{\rho} (\bfv_* \cdot \grad) \rho - \left[\rho \left(\frac{\bfw}{\rho} \right) \cdot \grad \right] \bfv_* = 0.
	\eeq
So dividing by $\rho$ we obtain the freezing-in of $\bfw/\rho$ into $\bfv_*$:
	\beq
	\dd{(\bfw/\rho)}{t} + (\bfv_* \cdot \grad) (\bfw/\rho)  - ((\bfw/\rho) \cdot \grad) \bfv_* = 0 \quad
	\text{or} \quad
	\dd{(\bfw/\rho)}{t} + {\cal L}_{\bfv_*} (\bfw/\rho) = 0.
	\label{e:freezing-in-w-by-rho-into-v*}
	\eeq
Indeed, it is well-known in Eulerian compressible, barotropic flow [$\la \rightarrow 0$] that $\bfw/\rho$ is frozen into $\bfv$.

{\noindent \bf Regularized Alfv\'{e}n's Theorem:} $\bfB/\rho$ is frozen into the swirl velocity $\bfv_*$ (\ref{e:v*-continuity-eqn}), i.e., it is Lie dragged along $(\pdr_t, \bfv_*)$:
	\beq
	\dd{}{t}\left(\frac{\bfB}{\rho}\right) + {\cal L}_{\bfv_*} \frac{\bfB}{\rho} = 
	\dd{}{t}\left(\frac{\bfB}{\rho}\right) + (\bfv_* \cdot \grad) \frac{\bfB}{\rho} - \left(\frac{\bfB}{\rho} \cdot \grad \right) \bfv_* = 0.
	\label{e:Bbyrho-frozen-in-v*}
	\eeq
{\sc Proof:} Multiplying and dividing by $\rho$ in the regularized Faraday's law (\ref{e:R-MHD-Faraday}) and using Leibnitz rule we get:
	\beq
	\pdr_t \bfB =  \pdr_t \left(\rho \frac{\bfB}{\rho} \right) 
	= \frac{\bfB}{\rho} \dd{\rho}{t} + \rho \pdr_t \frac{\bfB}{\rho} = \grad \times \left(\rho \bfv_* \times \frac{\bfB}{\rho} \right) \quad \imply \quad 
	\rho \dd{}{t}\left(\frac{\bfB}{\rho}\right)
	= \grad \times \left(\rho \bfv_* \times \frac{\bfB}{\rho} \right) - \frac{\bfB}{\rho} \dd{\rho}{t}.
	\eeq
Using the continuity equation expressed in terms of $\bfv_*$ (\ref{e:v*-continuity-eqn}), this simplifies to
	\beqs
	\rho \dd{}{t}\left(\frac{\bfB}{\rho}\right)
	&=& \grad \times \left(\rho \bfv_* \times \frac{\bfB}{\rho} \right) + \frac{\bfB}{\rho} \grad \cdot (\rho \bfv_*) \cr
	&=& \rho \bfv_*\grad \cdot \left(\frac{\bfB}{\rho}\right) - \frac{\bfB}{\rho} \grad \cdot (\rho \bfv_*) + \left(\frac{\bfB}{\rho}\right) \cdot \grad (\rho \bfv_*) - \rho \bfv_* \cdot \grad  \left(\frac{\bfB}{\rho}\right)
	+ \frac{\bfB}{\rho} \grad \cdot (\rho \bfv_*)\cr
	&=& \rho \bfv_* \left(\bfB \cdot \grad \ov{\rho}\right) + \bfv_* (\grad \cdot \bfB) 
	+ \rho \left(\frac{\bfB}{\rho}\right) \cdot \grad \bfv_* + \bfv_* \left(\frac{\bfB}{\rho}\right) \cdot \grad \rho
	- (\rho \bfv_* \cdot \grad ) \left(\frac{\bfB}{\rho}\right) \cr
	&=& \bfB \cdot \grad \bfv_*
	- (\rho \bfv_* \cdot \grad ) \left(\frac{\bfB}{\rho}\right).
	\eeqs
where we used the Leibnitz rule and  $\grad \cdot \bfB = 0$. Thus we get the above-mentioned result.


{\noindent \bf Swirl energy in terms of swirl velocity:} It is useful to note that the conserved swirl energy $E^*$ (in both R-Euler and R-MHD) can be expressed compactly in terms of $\bfv_*$ (for appropriate BC):
	\beq
	E^* = \int_V \left[ \frac{\rho \bfv^2}{2}+ U(\rho) + \frac{\la^2 \rho \bfw^2}{2} + \frac{\bfB^2}{2 \mu_0} \right] \: d\bfr  \; = \; 	\int_V \left(\ov{2}\rho(x)\bfv_*(x)\cdot \bfv(x) + U(\rho) + \frac{\bfB^2}{2 \mu_0} \right) \: d\bfr
	\equiv E_{\bfv_*}^*.
	\eeq
So up to a boundary term, $\bfv \cdot \bfv_*$ accounts for both kinetic and enstrophic energies. To see this, we begin by substituting for  $\bfv_* = \bfv + \la^2 \grad \times \bfw $ in $E_{\bfv_*}^*$ and use the divergence of a cross product to get
	\beqs
	E_{\bfv_*}^* &=& \int_V \left( \frac{\rho \bfv^2}{2} + \frac{\la^2 \rho}{2} (\grad \times \bfw)\cdot \bfv + U(\rho) + \frac{\bfB^2}{2 \mu_0} \right) \: d\bfr \cr
	&=& \; \int_V \left( \frac{\rho \bfv^2}{2} + \frac{\la^2 \rho}{2} \bfw^2 + U(\rho) + \frac{\bfB^2}{2 \mu_0} + \frac{\la^2 \rho}{2} \grad \cdot(\bfw \times \bfv)\right) \: d\bfr \cr
	&=& \int_V \left( \ov{2}\rho \bfv^2 + \ov{2}\la^2 \rho \bfw^2 + U(\rho) + \frac{\bfB^2}{2 \mu_0}\right)\: d\bfr + \half \la^2 \rho \int_{\pdr V} (\bfw \times \bfv) \cdot \hat n \: dS.
	\eeqs
The boundary term vanishes if $\bfv \times \hat n = 0$ or $\bfw \times \hat n = 0$. In both R-Euler and R-MHD, the BCs for $E^*$ conservation include $\bfw \times \hat n = 0$. So it is possible to express $E^*$ in terms of $\bfv_*$ with the same BCs that lead to $E^*$ conservation. Moreover, in R-Euler the BCs that guarantee conservation of flow helicity include $\bfv \times \hat n = 0$. So in R-Euler it is possible to express $E^*$ in terms of $\bfv_*$ with the BCs that lead to either $E^*$ or flow helicity conservation.


{\noindent \bf Time evolution of $\bfv_*$:} In compressible R-Euler flow, the evolution equation for $\bfv_*$ is  
	\beq
	\bfv_{*t} + \bfw \times \bfv_* + \grad \sigma = \frac{\la^2}{\rho} \grad \cdot (\rho \bfv_*) \grad \times \bfw - \la^2 \grad \times \left( \grad \times (\bfw \times \bfv_*) \right).
	\eeq
\normalsize
Here $\sigma = h + \half (\bfv_* - \la^2 \grad \times \bfw)^2$ and $\bfw$ satisfies (\ref{e:R-vorticity-eqn-compressible}). This is a {\em local} formulation of R-Euler in terms of $\bfv_*$, $\rho$ and $\bfw$. In R-MHD, for $\sigma$ as above, the evolution eqaution for $\bfv_*$ becomes
	\beq
	\bfv_{*t} + \bfw \times \bfv_* + \grad \sigma = \frac{\la^2}{\rho} \grad \cdot (\rho \bfv_*) \grad \times \bfw - \la^2 \grad \times \left( \grad \times (\bfw \times \bfv_*) \right) + \frac{\bfj \times \bfB}{\rho} + \la^2 \grad \times \left(\grad \times \left(\frac{\bfj \times \bfB}{\rho}\right)\right).
	\eeq

\section{Integral invariants associated to swirl velocity in R-Euler and R-MHD}
\label{s:integral-inv-v-star}
\subsection{Swirl Kelvin Theorem: Circulation around a contour moving with $\bfv_*$ is conserved}

We show here that the circulation $\G$ of $\bfv$ around a closed contour $C^*_t$ (that moves with $\bfv_*$) is independent of time. This is a regularized version of the Kelvin circulation theorem.
	\beq
		\frac{d\G}{dt} = \frac{d}{dt} \oint_{C^*_t} \bfv \cdot d\bfl = \frac{d}{dt} \int_{S^*_t} \bfw \cdot d\bfS = 0.
	\label{e:swirl-kelvin-theorem}
	\eeq
Here $S^*_t$ is any surface moving with $\bfv_*$ spanning $C^*_t$. Note that the circulation is that of $\bfv$ while the advecting velocity is $\bfv_*$.
{\flushleft \sc Proof:} When the time derivative is taken inside the integral sign to act on Eulerian quantities transported by $\bfv_*$, we introduce the operator $D_t^* \equiv \frac{D^*}{Dt} = \pdr_t + \bfv_* \cdot \grad$:
	\beq
	\DD{}{t} \oint_{C^*_t} \bfv \cdot d\bfl 
	= \oint_{C^*_t} \frac{D^* \bfv}{Dt} \cdot d\bfl + \oint_{C^*_t} \bfv \cdot \frac{D^* d\bfl}{Dt}.
	\eeq
Since $d \bfl$ is a line element that moves with $\bfv_*$, $\frac{D^* d\bfl}{Dt} = d \frac{D^* \bfl}{Dt} = d \bfv_*$. To see this we make use of the flow map from the fixed initial coordinates $\bfx_0$ to the coordinates $\bfx$ at time $t$.
	\beq
      dx_i = \frac{\partial x_i}{\partial x_{0 j}}dx_{0 j} \quad \imply \quad
      \frac{d^{*}}{dt}(dx_i)=\frac{\partial}{\partial x_{0 j}}\left(\frac{d^{*}x_i}{d t}\right)dx_{0 j}=\frac{\partial v_{*i}}{\partial x_{0 j}}dx_{0 j}=\frac{\partial v_{* i}}{\partial x_k} dx_k = dv_{* i}.
      \eeq
Thus
	\beq
	\frac{d\G}{dt} = \oint_{C^*_t} \left( \dd{\bfv}{t} + \bfv_* \cdot \grad \bfv \right) \cdot d\bfl + \oint_{C^*_t} \bfv \cdot d\bfv_*.
	\eeq
Using the R-Euler equation $\bfv_t = - \bfw \times \bfv_* - \grad \sigma$ and the vector identity $\bfv_* \cdot \grad \bfv =   \grad \bfv \cdot \bfv_*- \bfv_* \times (\grad \times \bfv)$ where $(\grad \bfv \cdot \bfv_*)_i = v_{*j} \pdr_i v_j$ we get
	\beq
	\frac{d\G}{dt} = \oint_{C^*_t} \grad \bfv \cdot \bfv_* \cdot d\bfl + \oint_{C^*_t} \bfv \cdot d \bfv_* - \oint_{C^*_t} \grad \sig \cdot d\bfl.
	\eeq
$\grad \sig$ integrates to zero around a closed contour. Finally, using $\bfv \cdot d \bfv_* = v_j \pdr_i v_{*j} dl^i$ and $\grad \bfv \cdot \bfv_* \cdot d\bfl = \bfv_{*j} \pdr_i v_j dl^i$ we get
	\beq
	\DD{\G}{t} = \oint_{C^*_t} \pdr_i (\bfv_* \cdot \bfv) dl^i = \oint_{C^*_t} d (\bfv_* \cdot \bfv) = 0.
	\eeq
The final equality of (\ref{e:swirl-kelvin-theorem}) follows from Stokes' theorem $\G = \int_{S^*_t} (\grad \times \bfv) \cdot d\bfS$.

\subsection{Swirl Alfv\'en theorem on conservation of magnetic flux}

We show that the line integral $\Phi = \oint_{C_t^*} \bfA \cdot d\bfl$ over a closed contour $C^*_t$ moving with $\bfv_*$ is a constant of the motion.

{\flushleft \sc Proof:} Using the equation of motion for $\bfA$: $\dd{\bfA}{t} = \bfv_* \times \bfB - \grad \phi$ we can write 
	\beqs
	\DD{}{t}\oint_{C^*_t} \bfA \cdot d\bfl &=& \oint_{C^*_t} \frac{D^*\bfA}{Dt}  \cdot d\bfl  + \oint_{C^*_t} \bfA \cdot d\bfv_* = \oint_{C^*_t} \left( \bfv_* \times \bfB - \grad \phi + \bfv_* \cdot \grad \bfA \right) \cdot d\bfl + \oint_{C^*_t} \bfA \cdot d\bfv_* \cr
	&=& \oint_{C^*_t} \left( v_{*j} \pdr_i A_j dl^i + A_i\pdr_j v_{*i}dl^j \right) = \oint_{C^*_t} \grad(\bfv_* \cdot \bfA) \cdot d\bfl = 0.
	\eeqs
We used the identity 
$(\bfv_* \times \bfB + \bfv_* \cdot \grad \bfA)_i = v_{*j} \; \pdr_i A_j$ and wrote $(d\bfv_*)_i = \pdr_j v_{*i} dl^j$ as in our proof of the swirl Kelvin theorem. Now if $S^*$ is any surface spanning the contour $C^*$ and $\bfB = \grad \times \bfA$ is the magnetic field, from Stokes' theorem we see that $\Phi = \int_{S^*} \bfB \cdot d\bfS$ is a constant of the motion. This is the regularized version of Alfv\'en's frozen-in flux theorem.

\subsection{Surfaces of vortex and magnetic flux tubes move with $\bfv_*$}

Given any smooth function $S(\bfr,t)$ we may consider its level surfaces at a given instant of time. We define an evolution of such a surface through an equation for $S({\bf r},t)$:
	\beq
	\frac{\partial S}{\partial t}+{\bf v}_{*}.\nabla S = D^{*}_{t}S = 0 \quad \text{where the operator} \quad D^{*}_{t}\equiv \frac{\partial }{\partial t}+{\bf v}_{*}.\grad.
	\label{e:material-surface-adv-v_*}
	\eeq
It follows that level surfaces of $S$ are advected by ${\bf v}_{*}$. Suppose the equation $({\bf w}/\rho) \cdot \grad S = 0$ holds at $t=0$, it implies that $\bfw$ is tangential to the level surfaces of $S$ at $t = 0$. For $\bfw$ to remain tangential to the level surfaces of $S$ at all times, $D^*_t(\frac{\bfw}{\rho} \cdot \grad S)$ must vanish. This is indeed so as a consequence of the freezing of $\bfw/\rho$ into $\bfv_*$ (\ref{e:freezing-in-w-by-rho-into-v*}) and the advection of $S$ by $\bfv_*$:
               \beq
D^{*}_{t}\left[\left(\frac{\bfw}{\rho}\right)\cdot\nabla S\right]
	         =   \left(\frac{\bfw}{\rho}\right).\nabla {\bf v}_{*} \cdot\nabla S + \left(\frac{\bfw}{\rho}\right)\cdot D^{*}_{t} \nabla S  \\
	         =   \left(\frac{\bfw}{\rho}\right).\nabla {\bf v}_{*}\cdot \nabla S + \left(\frac{\bfw}{\rho}\right)\cdot \left[-\nabla \left({\bf v_{*}\cdot \nabla S}\right)+{\bf v_{*}}\cdot \nabla \nabla S\right] = 0.
	\eeq
In particular, the surface of a vortex tube is advected by $\bfv_*$ (and {\em not} by $\bfv$). As in the case of vorticity, $\bfB/\rho$ is frozen into $\bfv_*$ by virtue of (\ref{e:Bbyrho-frozen-in-v*}). Thus magnetic flux tubes, like vortex tubes are transported by $\bfv_*$ .

\subsection{Curves advected by $\bfv_*$}
 Consider the level surfaces of two functions, $\alpha({\bf x},t)$ and $\beta({\bf x},t)$, advected 
by ${\bf v}_{*}$:
	\beq
	\al_t + \bfv_*\cdot \grad \alpha = 0 \quad \text{and} \quad \beta_t + \bfv_* \cdot \grad \beta = 0.
	\label{e:alpha-beta-advected-by-v*}
	\eeq
If $\alpha$ and $\beta$ are not functions of each other, the curve defined by the [solenoidal] direction vector, ${\bf Z}=\grad \alpha \times \grad \beta$ is a space curve, varying with time. We show that this space curve moves with $\bfv_*$, i.e. that ${\bf Z}/\rho$ is `frozen' into $\bfv_*$:
    \beq
    \bfZ_t = \grad \alpha_t \times \grad \beta + \grad \alpha \times \grad \beta_t .
    \eeq
From (\ref{e:alpha-beta-advected-by-v*}) and the identity $\grad a \times \grad b = \grad\times (a\grad b)$, we get:
	\beq
	\bfZ_t = - \grad(\bfv_* \cdot \grad \alpha) \times \grad \beta + \grad(\bfv_* \cdot \grad \beta)\times \grad \alpha
	= \grad \times \left[ (\bfv_* \cdot \grad \beta) \: \grad \al - (\bfv_* \cdot \grad \alpha) \: \grad \beta \right]
	= \grad \times (\bfv_* \times \bfZ).
	\eeq  
A solenoidal field satisfying $\bfZ_t = \grad \times (\bfv_* \times \bfZ)$ is termed a `Helmholtz' field associated to $\bfv_*$ \cite{thyagaraja-IITM}. Combining this with the continuity equation, we find that $\bfZ/\rho$ is frozen into $\bfv_*$:
	\beq
    \frac{\partial}{\partial t}\left(\frac{{\bf Z}}{\rho}\right)+\bfv_* \cdot \grad \left(\frac{{\bf Z}}{\rho}\right) = \frac{D^*}{Dt}\left(\frac{{\bf Z}}{\rho}\right) = \left(\frac{{\bf Z}}{\rho}\right)\cdot\grad \bfv_*.
    \eeq
Not every Helmholtz field is expressible as $\bfZ = \grad \al \times \grad \beta$ for a pair of functions advected by $\bfv_*$. We will show in \S \ref{s:helmholtz-field-g-helicity-g-tube} that such a Helmholtz field has zero `$\bfZ$-helicity', unlike Helmholtz fields like vorticity and magnetic field which lead to generally non-trivial flow and magnetic helicity.

\subsection{Analogue of Reynolds' transport theorem for volumes advected by $\bfv_*$}

There is useful version of Reynolds' transport theorem for volumes advected by the swirl velocity $\bfv_*$. Suppose $f(\bfx,t)$ is a scalar function associated with a volume $V^*$ moving with $\bfv_*$, then
	\beq
  \frac{d}{dt}\int_{V^*_t} f d\bfx =   \int_{V^*_t} D^{*}_{t}\left(\frac{f}{\rho}\right)\rho d \bfx.
  \label{e:reynolds-transport-theorem}
	\eeq
It is useful to develop briefly the ``Lagrangian'' theory underlying Reynolds' transport theorem. Let ${\bf x}(t)$ be the location of a ``fluid particle'' being transported by the swirl velocity ${\bf v}_{*}({\bf x},t)$. By definition $\partial^0 \bfx/\partial t = {\bf v}_{*}({\bf x},t)$ where the `Lagrangian' time derivative is taken holding the initial position $\bfx_0$ fixed unlike the `local' Eulerian time derivative. If ${\bf v}_{*}({\bf x},t)$ is known, integration gives, ${\bf x}={\bf x}({\bf x}_{0},t)$, so that at any instant the fluid position is a function of $t$ and initial location ${\bf x}_{0}$. The Jacobian, $J=\frac{\partial (x,y,z)}{\partial (x_{0},y_{0},z_{0})}$ relates the volume elements in the two coordinates $\bfx_0$ and $\bfx$ : $Jd\bfx_{0} = d\bfx$. It is a standard result \cite{chorin-marsden} that:
	\beq
 	\ov{J} \dd{^0 J}{t} = \grad \cdot {\bf v_*}
  \eeq
where ${\bf v}_{*}$ is the advecting velocity and the RHS is the standard Eulerian divergence taken at ${\bf x}$ at the instant $t$. Using the continuity equation :$D^*_t \rho = -\rho \grad \cdot \bfv_*$ we get
	\beq
	\grad \cdot \bfv_* = -\ov{\rho} D^*_t \rho = \ov{J} \dd{^0 J}{t} \quad \imply \quad D^*_t (\rho J) =0.
	\eeq
In fact, $\rho J = \rho_0$ where $\rho_0 = \rho(\bfx, t=0)$ as $J(t=0) = 1$. Now if $f({\bf x},t)$ is a scalar function associated with a volume $V$ moving with $\bfv_*$ we have
	\beq
  \frac{d}{dt}\int_{V^*_t} f d\bfx 
  = \frac{d}{dt}\int_{V^*_{0}}fJ \, d\bfx_0 
  = \int_{V^*_{0}} D^*_t\left(\frac{f}{\rho} \rho J\right)d\bfx_{0} 
  = \int_{V^*_{0}} D^*_t \left(\frac{f}{\rho} \right)\rho J d\bfx_{0} 
  = \int_{V^*_t} D^{*}_{t}\left(\frac {f}{\rho}\right)\rho \,d\bfx.
	\eeq
We have used $D^*_t (\rho J) = 0$, $D^*_t d\bfx_0 = 0$ and $Jd\bfx_0 = d\bfx$.

\subsection{Conservation of mass in a volume moving with $\bfv_*$}

Suppose a volume $V_t^*$ moves with $\bfv_*$. The mass of fluid within such a volume is independent of time. From (\ref{e:reynolds-transport-theorem}),
	\beq
	\frac{d}{dt} \int_{V_t^*} \rho \: d\bfx 
	= \int_{V^*_t} \rho D^*_t\left(\frac{\rho}{\rho} \right) \: d\bfx =	0
	\eeq

\subsection{Conservation of flow helicity in a closed vortex tube}

As we have noted, vortex tubes move with $\bfv_*$. Here we show that the flow helicity ${\cal K}$ associated with such a tube enclosing a volume $V^*_t$ is independent of time:
	\beq
	\frac{d{\cal K}}{dt} = \frac{d}{dt}\int_{V^*_t} \bfw \cdot \bfv \: d\bfx = 0.
	\eeq
{\flushleft \sc Proof} : Applying (\ref{e:reynolds-transport-theorem}) to ${\cal K} $ and using the freezing in condition $D_t^*(\bfw/\rho) = (\bfw/\rho) \cdot \grad \bfv_*$ and equation of motion (\ref{e:R-Euler-v*}) we get 
	\beq
	\dot {\cal K} = \int_{V_t^*} D^{*}_{t}
	\left(\frac{\bfw}{\rho} \cdot {\bfv} \right) \, \rho d\bfx
	= \int_{V^*_t} \left[D^{*}_{t}\left(\frac{\bf w}{\rho}\right)\cdot{\bf v}+\left(\frac{\bf w}{\rho}\right)\cdot D^{*}_{t} ({\bf v})\right] \rho d\bfx                 
	= \int_{V^*_t} {\bf w}\cdot \left[ \grad\bfv_* \cdot \bfv + {\bf v}_{*}\cdot {\bf \grad v}+{\bf v}_{*}\times {\bf w} - \grad \sigma \right] d\bfx.
  \eeq
The middle two terms combine (${\bf v}_{*}\cdot {\bf \grad v}+{\bf v}_{*}\times {\bf w} = \grad \bfv \cdot \bfv_*$) to give
	\beq
	\frac{d {\cal K}}{dt} = \int_{V^*_t} {\bf w}\cdot[\grad\bfv_* \cdot \bfv + \grad\bfv \cdot \bfv_* - \grad \sigma ]d\bfx
		= \int_{V^*_t} \bfw \cdot \grad [\bfv \cdot \bfv_* - \sig] \,d\bfx
		= \int_{\pdr V^*_t} (\bfv \cdot \bfv_* - \sigma) \bfw \cdot \hat n \, dS = 0.
	\eeq
Here we used $\grad \cdot \bfw = 0$ and the fact that $\bfw$ is tangential to the surface (vortex tube) bounding the volume $V^*_t$.

\subsection{Conservation of magnetic helicity in a magnetic flux tube}
In R-MHD, the magnetic helicity ${\cal K}_B$ (but {\it not} flow helicity) associated with a volume $V^*_t$ bounded by a closed magnetic flux tube is independent of time:
	\beq
	\frac{d {\cal K}_B}{dt}= \frac{d}{dt}\int_{V^*_t} \bfB \cdot \bfA \,d\bfx = 0.
   	\eeq
This is a consequence of the fact that $\bfB$ is tangential to the boundary of such a volume by the freezing of $\bfB/\rho$ into $\bfv_*$.

{\flushleft \sc Proof} : As before, we apply (\ref{e:reynolds-transport-theorem}) to $d{\cal K}_B/dt$ and use the freezing-in condition $D^*_t(\bfB/\rho) = (\bfB/\rho) \cdot \grad \bfv_*$ and equation for the evolution of the vector potential (\ref{e:A-evolution-R-MHD}) to get\footnote{$\phi$ is arbitrary, it depends on the choice of gauge. In the PB formulation $\phi = \bfv_* \cdot \bfA$}
	\beqs
	\frac{d{\cal K}_B}{dt}&=& \int_{V_t^*} D^{*}_{t}\left[\left(\frac{\bfB}{\rho}\right)  {\bf \cdot A} \right] \rho \, d\bfx =  \int_{V^*_t} \left[D^{*}_{t}\left(\frac{\bfB}{\rho}\right)\cdot{\bfA}+\left(\frac{\bfB}{\rho}\right)\cdot D^{*}_{t} (\bfA)\right] \rho \,d\bfx \cr
    &=& \int_{V^*_t} {\bf B}\cdot[{\bf \grad v}_{*}\cdot{\bf A} +\bfv_* \times \bfB - \grad \phi +\bfv_* \cdot \grad \bfA]\,d\bfx   
	= \int_{V^*_t} {\bf B}\cdot \grad \left[\bfv_* \cdot \bfA - \phi \right] \, d\bfx \cr
	&=& \int_{\pdr V^*_t} \left[\bfv_* \cdot \bfA - \phi \right] {\bf B} \cdot \hat n \, d\bfx = 0.
      \eeqs
The last equality follows as $\grad \cdot \bfB = 0$ and since $\bfB$ is tangential to a surface that moves with $\bfv_*$ ($V_t^*$ is a magnetic flux tube).

\subsection{Helmholtz fields $\bfg$ and their conserved helicities in $\bfg$-tubes}
\label{s:helmholtz-field-g-helicity-g-tube}

The conservation of flow and magnetic helicity in vortex and magnetic flux tubes are special cases of a more general result. Recall that a Helmholtz field \cite{thyagaraja-IITM} is a solenoidal vector field $\bfg$ that evolves according to $\bfg_t + \grad \times (\bfg \times \bfv_*) = 0$. If $\bfg$ is a Helmholtz field, then $\bfg/\rho$ is frozen into $\bfv_*$, i.e., $D^*_t (\bfg/\rho) = (\bfg/\rho) \cdot \grad \bfv_*$. A Helmholtz field in a simply-connected region is expressible in terms of a `vector potential' ${\bf u}$:
	\beq
	\bfg = \grad \times {\bf u} \quad \text{with} \quad {\bf u}_t + \bfg \times \bfv_* + \grad \tht  = 0
	\eeq
for some scalar function $\tht(\bfx,t)$. Examples of Helmholtz fields in R-Euler and R-MHD include $\bfw$ and $\bfB$. The corresponding vector potentials are $\bfv$ and $\bfA$, with $\tht$ corresponding to the stagnation enthalpy $\sigma$ and electrostatic potential $\phi$ respectively. 

If $\bfg$ is a Helmholtz field then its flux through a surface $S^*_t$ spanning a closed contour $C^*_t$ moving with $\bfv_*$ is conserved, generalizing the Kelvin and Alfv\'en theorems:
	\beq
	\DD{}{t} \oint_{C^*_t} \bfu \cdot d\bfl = \DD{}{t} \int_{S^*_t} \bfg \cdot d\bfS = 0.
	\eeq
Given a Helmholtz field, a closed surface everywhere tangent to $\bfg$ is called a $\bfg$-tube, generalizing vortex tubes and magnetic flux tubes. The freezing of $\bfg/\rho$ into $\bfv_*$ then implies that a $\bfg$-tube moves with $\bfv_*$. Associated to a Helmholtz field $\bfg$ and its vector potential $\bf u$ is a $\bfg$-helicity density, $\bfg \cdot {\bf u}$. It follows from the transport theorem and the above equations of motion that the $\bfg$- helicity in a $\bfg$-tube is independent of time:
	\beq
	\DD{}{t} \int_{V^*_t} \bfg \cdot {\bf u} \: d\bfx = \int_{V^*_t} D^*_t \left(\frac{\bfg}{\rho} \cdot {\bf u}\right) \rho \: d\bfx = 0
	\eeq
{\flushleft \bf Note:} If $\bfZ = \grad \al \times \grad \beta$ is a Helmholtz field defined by two independent scalar functions advected by $\bfv_*$, then its vector potential is of the form $\bfu = \al \grad \beta + \grad \gamma$ where $\gamma$ is a scalar function. The corresponding $\bfZ$-helicity in a moving volume $V^*_t$, $\int_{V^*_t} \bfZ \cdot \grad \gamma \: d\bfx = \int_{\pdr V^*_t} \gamma \bfZ \cdot d\bfS - \int_{V^*_t} \gamma \grad \cdot \bfZ \: d\bfx$ is identically zero since $\bfZ$ is solenoidal and tangential to the boundary $\pdr V^*_t$.

\section{Poisson brackets for the R-Euler equations}
\label{s:pb-for-fluid}

Commutation relations among `quantized' fluid variables were proposed by Landau \cite{landau} in an attempt at a quantum theory of superfluid He-II. As a byproduct, one obtains Poisson brackets (PB) among {\it classical} fluid variables allowing a Hamiltonian formulation for compressible flow. Suppose $F$ and $G$ are two functionals of $\rho$ and $\bfv$, then their equal-time PB (see \cite{morrison-and-greene,morrison-review}) is
	\beqs
	\{ F, G \} &=& \int \left[ \frac{\bfw}{\rho} \cdot \left( \deldel{F}{\bfv} \times \deldel{G}{\bfv} \right) - \deldel{F}{\bfv} \cdot \grad G_{\rho} + \deldel{G}{\bfv} \cdot \grad F_{\rho} \right] d\bfr \cr
	&=& \int \left[ \frac{\bfw}{\rho} \cdot \left( \deldel{F}{\bfv} \times \deldel{G}{\bfv} \right) +\grad \cdot \left( \deldel{F}{\bfv} \right) G_{\rho} - \grad \cdot \left(\deldel{G}{\bfv} \right) F_{\rho} \right] d\bfr.
	\label{e:pb-between-functionals-of-rho-v}
	\eeqs
The two formulae are related by integration by parts. If $\rho$ and mass current ${\bf M} = \rho \bfv$ are taken as the basic variables, then
	\beq
	\{ F , G \} = - \int \left[ \rho \left( \deldel{F}{\bf M} \cdot \grad G_{\rho} - \deldel{G}{\bf M} \cdot \grad F_{\rho} \right) + M_i \left( \deldel{F}{\bf M} \cdot \grad \deldel{G}{M_i} - \deldel{G}{\bf M} \cdot \grad \deldel{F}{M_i} \right) \right] \: d\bfr.
	\eeq
We will show that this PB, along with our conserved swirl energy hamiltonian $E^*$ lead to the R-Euler equations. The PB is manifestly anti-symmetric and the dimension of $\{ F, G \}$ is that of $FG/\hbar$. The PB of $F[\rho,\bfv]$ with a constant (independent of $\rho$ and $\bfv$) is zero. The Leibnitz rule $\{ FG, H \} = F \{ G, H\} + \{F, H \} G$ for three functionals follows from the (\ref{e:pb-between-functionals-of-rho-v}) upon using the Leibnitz rule for functional derivatives. In other words, the PB $\{ F, G \}$ is a derivation in each entry holding the other fixed.

From (\ref{e:pb-between-functionals-of-rho-v}) we deduce the PB among basic dynamical variables subject to the constitutive relation $\la^2 \rho =$ constant:
	\beqs
	\{ \rho(\bfx), \rho(\bfy) \} = 0, \quad \{ \rho(\bfx), \la(\bfy) \} = 0, &&
	\{ v_i(\bfx), v_j(\bfy) \} = (\om_{ij}/\rho)(\bfx \; \text{or} \; \bfy) \: \del(\bfx-\bfy),
	\cr \cr
	\{ \rho(\bfx), \bfv(\bfy) \} = - \grad_\bfx \del(\bfx-\bfy) =  \frac{(\grad_\bfy - \grad_\bfx)}{2} \del(\bfx-\bfy), && \{ \la^2(\bfx), \bfv(\bfy) \} = - \frac{\la^2(\bfx)}{\rho(\bfx)} \{ \rho(\bfx), \bfv(\bfy) \}.
	\label{e:PB-among-basic-var}
	\eeqs
Here $\om_{ij} = \pdr_i v_j - \pdr_j v_i$ is the dual of vorticity, $w_i = \half \eps_{ijk} \omega_{jk}$ or $\omega_{ij} = \eps_{ijk} w_k$. (\ref{e:PB-among-basic-var}) generalises Gardner's PB $\{ u(\bfx), u(\bfy) \} = \half (\pdr_\bfy - \pdr_\bfx) \del(\bfx-\bfy)$ for KdV \cite{gardner}. The $\{ v_i , v_j \}$ is akin to the PB between canonical momenta of a charged particle in a $\bfB$ field
	\beq
	\left\{ p_i - ({e}/{c}) A_i(\bfx), p_j - ({e}/{c}) A_j(\bfx) \right\} = ({e}/{c}) F_{ij}(\bfx) \quad \text{where} \quad F_{ij} = \eps_{ijk} B_k.
	\eeq
$\bfB$ is analogous to $\bfw$ and $F_{ij}$ to $\omega_{ij}$. The Morrison-Greene PBs among functionals (\ref{e:pb-between-functionals-of-rho-v}) follow from the basic PBs (\ref{e:PB-among-basic-var}) by postulating that the PB is a derivation in either entry. For instance, denoting functional derivatives by subscripts we have:
	\beqs
	\{F[\rho], G[\bfv]\} &=& \int \frac{\del F}{\del \rho(x)}\frac{\del G}{\del v_i(y)}\{\rho(x), v_i(y)\} \:d\bfx\: d\bfy 
	= \int \frac{\del F}{\del \rho(x)}\frac{\del G}{\del v_i(y)}\pdr_{y^i} \del(x -y) \:d\bfx\: d\bfy = -\int  F_{\rho} \grad \cdot G_\bfv \: d\bfx.\cr
 \{F[\bfv], G[\bfv]\} &=& \int \frac{\del F}{\del v_i(x)}\frac{\del G}{\del v_j(y)}\{v_i(x), v_j(y)\} d\bfx \; d\bfy 
    = \int \frac{\del F}{\del v_i(x)}\frac{\del G}{\del v_j(y)}\frac{\eps_{ijk} w_k(x)}{\rho(x)}\del(x-y) d\bfx \;d\bfy \cr
    &=& \int \frac{\bfw}{\rho} \cdot (F_{\bfv} \times G_{\bfv}) d\bfx.
	\eeqs
Some useful PBs follow from (\ref{e:PB-among-basic-var}). For instance $\rho$ commutes with vorticity:
	\beqs
	(a) && \{ \rho(\bfx), \bfw(\bfy) \} = 0 = \{ \la(\bfx), \bfw(\bfy) \}, \cr
	(b) && \{ v_i(\bfx), w_j(\bfy) \} = \eps_{jkl} \pdr_{\bfy^k} \left( \rho^{-1} \, \om_{il}(\bfy) \del(\bfx-\bfy) \right) = (\del_{jk} \pdr_{\bfy^i} - \del_{ij} \pdr_{\bfy^k}) (\rho^{-1} w_k(\bfy) \del(\bfx-\bfy)), \cr
	(c) && \{ w_i(\bfx), w_j(\bfy) \} = \eps_{ikl} \eps_{jmn} \pdr_{\bfx^k} \pdr_{\bfy^m} \left( \rho^{-1} {\om_{ln}(\bfx \; \text{or} \; \bfy)} \, \del(\bfx-\bfy) \right), \cr
	(d) && \{ v_k(\bfx), \om_{ij}(\bfy) \} = \pdr_{\bfy^i} \left( \rho^{-1}\om_{k j}(\bfy) \, \del(\bfx-\bfy) \right) - (i \leftrightarrow j), \cr
	(e) && \{ (\bfv \cdot \bfw)(\bfx), \rho(y)\bfy \} = - (\bfw(\bfx) \cdot \grad_\bfx) \del(\bfx-\bfy), \cr
	(f) && \{ (\grad \cdot \bfv)(\bfx) , \rho(\bfy) \} = - \grad^2_\bfx \del(\bfx-\bfy).
	\label{e:list-of-useful-PB}
	\eeqs

Some PBs of ${\bf M} = \rho \bfv$ and $\bfv_*$ are collected in \S \ref{a:pb-mass-curr-and-v*}. Properties of PBs among linear functionals are discussed in \S \ref{a:solenoidal-irrot-pb}. The basic PBs may also be written in Fourier space, which should be useful for numerics in a periodic domain:
	\beqs
	\{ \tl \rho(\bfk), \tl \rho(\bfk') \} = 0, &&
	\{ \tl \rho(\bfk), v_j(\bfk') \} = -i k_j (2\pi)^3 \del(\bfk + \bfk'), \quad
	\{ \tl v_i(\bfk) , \tl v_j(\bfk') \} = \widetilde{\left( \frac{\om_{ij}}{\rho} \right)}(\bfk + \bfk'),
	\cr
	\text{where} \quad
	\tl \rho(\bfk) &=& \int \rho(\bfx) e^{- i \bfk \cdot \bfx} \: d\bfx, \quad
	v_i(x) = \int \tl v_i(k) e^{i \bfk \cdot \bfx} \: \frac{d \bfk}{(2\pi)^3}, \quad \text{etc.}	
	\eeqs
The Jacobi identity is $\{ \{ F[\rho,\bfv], G[\rho,\bfv] \}, H[\rho, \bfv] \} + {\rm cyclic} = 0$. Using the PB among $\rho$ and $\bfv$, it is straightforward to check the Jacobi identity in some special cases, e.g., for coordinate functionals $F = \rho(x), G = \rho(y)$ and $H = \bfv(z)$ or for two $\bfv$'s and a $\rho$. It is not so straightforward to check the Jacobi condition in general, see the discussion in \cite{morrison-aip}. In \S \ref{a:jacobi} we give an elementary proof of the Jacobi identity for three linear functionals of $\rho$ and $\bfv$. It involves a remarkable integral identity. In \S \ref{s:Jacobi-general-proof} we extend the proof to exponentials of linear functionals and use a functional Fourier transform to establish the identity for a much wider class of non-linear functionals. The Jacobi identity should also follow by interpreting these PBs as among functions on the dual of a Lie algebra, see \cite{holm-kupershmidt}. Furthermore, one formally expects the Jacobi identity to hold if we regard these PB as the semi-classical limit of commutators in Landau's quantized superfluid model.


\subsection{Equations of motion from Hamiltonian and Poisson brackets}

We show in this section that the continuity and R-Euler equations 
	\beq
	\dd{\rho}{t} + \grad \cdot (\rho \bfv) = 0, \quad
	{\rm and} \quad
	\dd{\bfv}{t} + ({\bfv} \cdot \grad) \bfv = - \grad U'(\rho) - \la^2 \bfw \times \grad \times \bfw
	\eeq
follow from Hamilton's equations $\dd{\rho}{t} = \{ \rho, H \}$ and $\dd{\bfv}{t} = \{ \bfv, H \}$ for the swirl hamiltonian
	\beq
	H = \int \left[ \frac{\rho \bfv^2}{2}+ U(\rho) + \frac{\la^2 \rho \bfw^2}{2} \right] \: d\bfr.
	\eeq
We call the $3$ terms kinetic (KE), potential (PE) and enstrophic (EE) energies. By the constitutive relation $\la^2 \rho$ is a constant. Here $U'(\rho) = h(\rho)$, e.g., for adiabatic flow $U(\rho) = p/(\gamma - 1)$ so that $U'(\rho) = h(\rho) = \gamma/(\gamma - 1) (p/\rho)$ and $\grad U'(\rho) = \grad h = \ov{\rho} \grad p$. 

For the continuity equation, we note that only KE contributes to $\{ H, \rho \}$ since $\{ \rho, \rho \} = \{ \bfw , \rho \} = 0$:
	\beq \nonumber
	\{ H, \rho(\bfy) \} 
	= - \int_V \rho(\bfx) v_i(\bfx) \pdr_{\bfx^i} \del(\bfx-\bfy) \: d\bfx
	= \int_V \pdr_i [\rho(\bfx) v_i(\bfx)] \: \del(\bfx-\bfy) \: d\bfx - \int_{\pdr V} \rho(\bfx) v_i(\bfx) n_i \, \del(\bfx-\bfy) dS
	= \grad \cdot (\rho \bfv).
	\eeq 
The boundary term vanishes as $y$ is in the interior and $x$ on the boundary ($\bfv \cdot \hat n = 0$ also ensures this).

To get the R-Euler equation, we evaluate $\{ H, \bfv \}$. The individual PBs are
	\beqs
	\{ KE, v_i \} &=& (\bfv \cdot \grad) v_i - \int_{\pdr V} \bfv^2 n_i \del(\bfx-\bfy) \: dS, \quad
	\{ PE, v_i \} = \pdr_{i} U'(\rho) - \int_{\pdr V} U'(\rho) n_i \del(\bfx-\bfy) dS \cr
	{\rm and} \quad
	\{ EE, v_i \} &=& \la^2 (\bfw \times (\grad \times \bfw))_i - \int_{\pdr V} \la^2 ((\bfw \times \hat n) \times \bfw)_i \del(\bfx-\bfy) \: dS.
	\eeqs
The boundary terms vanish as before. The equation of motion for $\bfv$ then follows:
	\beq
	\{\bfv, H\} = \frac{\pdr \bfv}{\pdr t} = - (\bfv \cdot \grad) \bfv - \grad U'(\rho) -\la^2 \bfw \times (\grad \times \bfw).
	\eeq
For this to agree with the Euler equation $U'(\rho)$ must be chosen to be the enthalpy $h(\rho)$.

\subsection{Poisson brackets among locally conserved quantities and symmetry generators}
\label{s:PB-cons-qty}

We work out the PBs among locally conserved quantities of regularized compressible flow. As one might expect, linear and angular momenta and helicity Poisson commute with the {\it swirl} hamiltonian
	\beq
	\{ P_i, H \} = \{ L_i, H \} = \{ {\cal K}, H \} = 0.
	\eeq
BC are important: we would not expect linear or angular momenta to be conserved in a finite container that breaks translation or rotation invariance. Decaying BC ($\bfv \to 0, \rho \to$ constant) in an infinite domain would guarantee the above PB. More generally, we show below that the above PB may be expressed in terms of the conserved (regularized) currents of momentum, angular momentum and helicity. So these PB vanish provided the corresponding currents have zero flux across the boundary. 
	
$\{ P_i, H \}$ can be expressed as the divergence of the momentum current $\Pi_{ij}$ using $\rho \grad U' = \grad p$ and the constitutive relation:
	\beq
	\{ P_i, H \} = - \int_V \left( \pdr_i p +\pdr_j(\rho v_i v_j) + \la^2 \rho \left(\ov{2} \pdr_i \bfw^2 - \pdr_j (w_j w_i) \right) \right) d\bfr = - \int_{\pdr V} \Pi_{ij} n_j dS.
	\eeq
This vanishes if the momentum current (\ref{e:momentum-current-tensor}) has zero flux across the boundary. Similarly, $\{L_i , H \}$ can be expressed as a boundary term after dropping some terms using antisymmetry of $\eps$:
	\beq
	\{L_i , H \} = \eps_{ijk} \int_{\pdr V} x_j \left[ -\rho v_l v_k - p \delta_{kl} - \la^2 \rho \left( \half \bfw^2 \del_{lk} - w_l w_k \right) \right] n_l \: dS = - \int_{\pdr V} \Lambda_{il} n_l \: dS.
	\eeq
This vanishes if the regularized angular momentum current (\ref{e:ang-mom-current-conservation}) has zero flux across the boundary. The PB of the $H$ with flow helicity can be expressed in terms of the regularized helicity current. Let us first consider the unregularized $H$, for which
	 \beqs \nonumber
	 \{KE + PE,{\cal K}\}
	&=& \int_V \left[ - v_j(\bfx) w_i(\bfy) \om_{ij}(\bfx) \del(\bfx-\bfy) + \rho(\bfx) v_j(\bfx) v_i(\bfy) \eps_{ilk} \pdr_{\bfy^l} \left( \frac{\om_{jk}(\bfy)}{\rho(\bfy)} \del(\bfx-\bfy) \right) \right] \: d\bfx \: d\bfy 
	\cr && - \int_V \left( \half v^2(\bfx) + U'(\rho(\bfx)) \right) \left[w_i(\bfy) \pdr_{\bfx_i} \del(\bfx-\bfy)\right ]\: d\bfx \: d\bfy 
	= \int_{\pdr V} [\bfv \times (\bfv \times \bfw) + \sigma \bfw ] \cdot \hat n \: dS.
	\eeqs
$\bfv \times (\bfv \times \bfw) + \sigma \bfw$ is the unregularized ($\la \to 0$) helicity current. Using (\ref{e:constitutive-relation}) and repeated integration by parts we get
	\beqs
	\{ EE, {\cal K} \} &=& \iint_V \left\{ \half \la^2 \rho \bfw^2 , \bfv \cdot \bfw \right\} \, d\bfx \, d\bfy = -\la^2 \rho \iint_V \left( w_i(\bfx) v_j(y) \eps_{ikl} \eps_{jmn} \pdr_{\bfx^k} \pdr_{\bfy^m} \left( \frac{\om_{nl}(\bfx)}{\rho(\bfx)} \del(\bfx-\bfy) \right) \right) \, d\bfx \, d\bfy
	\cr
	 &=& \int_{\pdr V} \la^2 (\bfT \times \bfv) \cdot \hat n \, dS - \int_{\pdr V} \int_{\pdr V} \la^2 \bfw \cdot ((\bfw \times \hat n) \times (\bfv \times \hat n)) \, dS \, dS.
	\eeqs
We conclude that $\{ H, {\cal K} \} = \int_{\pdr V} \bfj_{\cal K} \cdot \hat n - \int_{\pdr V} \int_{\pdr V} \la^2 \bfw \cdot ((\bfw \times \hat n) \times (\bfv \times \hat n)) \, dS \, dS$ where $\bfj_{\cal K}$ is the conserved helicity current (\ref{e:helicity-current-conservation}). So if we use decaying or $\bfw \cdot \hat n = 0$ and $\bfv \times \hat n = 0$ BCs, then $\bfj_{\cal K}$ has zero flux across $\pdr V$ and the double boundary term also vanishes ensuring $\{ H, {\cal K} \} = 0$. Helicity also commutes with $\bfP$ and $\bfL$ with decaying or $\bfw \cdot \hat n = 0$ and $\bfv \times \hat n = 0$ BCs
	\beq
	\{ \bfP, {\cal K} \} = \int_{\pdr V} \left[ (\bfv \times \hat n) \times \bfw + (\bfw \cdot \hat n) \bfv \right] \, dS, \quad
	\{ \bfL, {\cal K} \} = \int_{\pdr V} \bfr \times \left[ (\bfv \times \hat n) \times \bfw + (\bfw \cdot \hat n) \bfv \right] \, dS.
	\eeq
Indeed it is known that helicity is a Casimir of the Poisson algebra with decaying or $\bfw \cdot \hat n = 0$ and $\bfv \times \hat n = 0$ BCs. Using $\del {\cal K}/\del \bfv = 2 \bfw$ (assuming $\bfv \times \hat n = 0$ on $\pdr V$), we have for any functional $F$ of $\rho$ and $\bfv$,
	\beq
	\{ {\cal K}, F[\rho, \bfv] \} = 2 \int_V \left[\frac{\bfw}{\rho} \cdot \left( \bfw \times \deldel{F}{\bfv} \right) - \bfw \cdot \grad F_{\rho} \right] d\bfx = -2 \int_{\pdr V} (\bfw \cdot \hat n) F_{\rho} dS = 0.
	\eeq
The PBs among $\bfP$ and $\bfL$ are
	\beqs
	\{ L_i , L_j \} &=& \eps_{ijk} L_k 
	+ \int_{\pdr V} \rho(\bfr) [(\bfr \times \bfv)_i (\bfr \times \hat n)_j - (i \leftrightarrow j)] \, dS \cr
	{\rm and} \quad 
	\{ P_i , L_j \} &=& \eps_{ijk} P_k +\int_{\pdr V} \rho(\bfr) \left[ (\bfr \times \hat n)_j v_i - (\bfr \times \bfv)_j n_i \right] \, dS.
	\eeqs
So with, say decaying BCs, both $\bfP$ and $\bfL$ transform as vectors under rotations generated by $\bfL$. Finally, the generator of Galilean boosts is ${\bf G} = \int \bfr \rho \: d\bfr$. $\bf G$ is not conserved, its PB with swirl energy is momentum
	\beq
	\{G_i, H \} = \int \bfx_i \{ \rho(\bfx), H \} \:d\bfx = \int \bfx_i \dot \rho \: d\bfx = - \int \bfx_i \pdr_j (\rho v_j) \: d\bfx = \del_{ij} \int \rho v_j = P_i.
	\eeq
We similarly check that $\bf G$ transforms as a vector under rotations $\{ G_i, L_j \} = \eps_{ijk} G_k$ and that $\{ {\bf G}, {\cal K} \} = 0$ and $\{ G_i, G_j \} = 0$. Finally, there is a central term in $\{ G_i, P_j \} = M \del_{ij}$ where $M$ is the total mass of fluid.

\subsection{Poisson brackets for incompressible flow}

PB for incompressible flow $(\grad \cdot \bfv = 0, \rho = \text{constant})$ are given in the literature (see \S 1.5 of \cite{marsden-ratiu}). Suppose $F[\bfv], G[\bfv]$ are two functionals of $\bfv$, then the `ideal fluid bracket' is
	\beq
	\{ F, G \} = - \ov{\rho} \int \bfv \cdot \left[ \deldel{F}{\bfv} , \deldel{G}{\bfv} \right] \: d\bfr.
	\eeq
The square brackets above denote the commutator of incompressible vector fields $[\bff , \bfg] = \bff \cdot \grad \bfg - \bfg \cdot \grad \bff$. These PBs follow from the compressible PBs when we impose the conditions
	\beq
	\grad \cdot \bfv = 0, \quad 
	\grad \cdot \deldel{F}{\bfv} = 0 = \grad \cdot \deldel{G}{\bfv} \quad \text{and} \quad \rho = \text{constant}.
	\label{conditions-incompress-PB}
	\eeq
We start with the compressible PB and impose (\ref{conditions-incompress-PB}) so that the quantity in the second parentheses below vanishes, giving 	\beqs
	\{F, G \} &=& \int \frac{\bfw}{\rho} \cdot \left(\deldel{F}{\bfv} \times \deldel{G}{\bfv} \right) d\bfr 
	= \ov{\rho} \int \eps_{ijk} \eps_{ilm} \pdr_l v_m \deldel{F}{v_j} \deldel{G}{v_k} d\bfr
	= \ov{\rho} \int (\pdr_j v_k - \pdr_k v_j) \deldel{F}{v_j} \deldel{G}{v_k}  d\bfr \cr
	&=& \ov{\rho} \int \left[ v_j \deldel{G}{v_k} \pdr_k \deldel{F}{v_j} - v_k \deldel{F}{v_j} \pdr_j \deldel{G}{v_k} \right] + \left[ v_j \deldel{F}{v_j} \pdr_k \deldel{G}{v_k} - v_k \deldel{G}{v_k} \pdr_j \deldel{F}{v_j} \right] d\bfr
	= - \ov{\rho} \int \bfv \cdot \left[ F_\bfv , G_\bfv \right] d\bfr.
	\eeqs 
\subsubsection{Incompressible R-Euler from PB}

The incompressible R-Euler equation (\ref{e:R-Euler-incompress}) follows from the above PB and Hamiltonian (with $\la$ and $\rho$ constant)
	\beq
	H = \rho \int \left( \half \bfv^2 + \half \la^2 \bfw^2 \right) \: d\bfz \quad \imply \quad
		\rho \dd{v_i(y)}{t} = \rho \{ v_i(y) , H \} = - \int v_k(x) \left[ \deldel{v_i(y)}{v_j(x)} \pdr_j \deldel{H}{v_k(x)} - \deldel{H}{v_j(x)} \pdr_j \deldel{v_i(y)}{v_k(x)} \right] d\bfx.
	\eeq
Here, $\del{KE}/\del{\bfv} = \rho \bfv$
and $\del{EE}/\del{\bfv} = \la^2 \rho \grad \times \bfw$ are divergence free as required, but $\del{v_i(x)}/\del{v_j(x)} = \del_{ij} \del(x-y)$ is not. Hence we will need to take care to project the equation of motion resulting from these PBs to the incompressible subspace. We will do this after calculating the PBs.
	\beqs
	\rho \{ v_i(y) , KE \} &=& - \int v_k(x) \left[ \del_{ij} \del(x-y) \pdr_j (\rho v_k(x)) - \rho v_j(x) \pdr_j \left( \del_{ik} \del(x-y) \right) \right] 
	= - \rho \left[ v_j \pdr_i v_j + v_j \pdr_j v_i \right], \cr
	\rho \{ v_i(y), EE \} &=& - \int v_k(x) \left[ \del_{ij} \del(x-y) \pdr_j (\la^2 \rho (\grad \times \bfw)_k(x)) - \la^2 \rho (\grad \times \bfw)_j(x) \pdr_j \left( \del_{ik} \del(x-y) \right) \right]
	\cr
	&=& - \la^2 \left[ v_j \pdr_i (\grad \times \bfw)_j + \pdr_j \left( v_i (\grad \times \bfw)_j \right) \right]
	= - \la^2 \left[ \pdr_i \left( \bfv \cdot (\grad \times \bfw) \right) - (\grad \times \bfw)_j \pdr_i v_j + (\grad \times \bfw)_j \pdr_j v_i \right] \cr
	&=& - \la^2 \left[ \bfT_i + \pdr_i (\bfv \cdot (\grad \times \bfw)) \right].
	\eeqs
Thus the momentum equation is
	\beq
	\dd{\bfv}{t} + \mathbb{P} \left( \bfv \cdot \grad \bfv + \la^2 \bfT + \grad \left(\half \bfv^2 + \la^2 \bfv \cdot \grad \times \bfw \right) \right) = 0
	\quad \text{or} \quad
	\dd{\bfv}{t} + \mathbb{P} \left( \bfv \cdot \grad \bfv + \la^2 \bfT + \grad \left(\bfv \cdot \bfv_* -\half \bfv^2 \right) \right) = 0
	\eeq
where $\mathbb{P}$ is the projection to the incompressible subspace, which we can define using the Helmholtz decomposition. Given a vector field $\bfv$ we may write it as the sum of curl-free and divergence-free parts $\bfv = - \grad \phi + \grad \times \bfA$ where $\phi = {(4\pi)}^{-1} \int \frac{\grad \cdot \bfv}{|\bfr - \bfs|} d \bfs $. Then, $\mathbb{P} (\bfv) = \bfv + \grad \phi = \grad \times \bfA$. In particular, the projection of a gradient vanishes. Thus $\mathbb{P}\left(\grad \left(\bfv \cdot \bfv_* - \half \bfv^2\right) \right)= 0$ while
	\beq
	\mathbb{P}( \bfv \cdot \grad \bfv + \la^2 \bfT) =  \bfv \cdot \grad \bfv + \la^2 \bfT +\ov{\rho} \grad p \quad \text{where} \quad \frac{p(\bfr)}{\rho} = \ov{4\pi} \int \frac{\grad_s \cdot (\bfv \cdot \grad \bfv(s) + \la^2 \bfT(s))}{|\bfr - \bfs|} \: d\bfs.
	\eeq
So after projecting to the incompressible subspace we get the incompressible R-Euler equation $\bfv_t + \bfv \cdot \grad \bfv = - \grad p/\rho - \la^2 \bfT$. Note that the above definition of pressure may be written as a Poisson equation for $p$ or $\sigma$
	\beq
	\grad^2 p = - \rho \grad \cdot (\bfv \cdot \grad \bfv + \la^2 \bfT) \quad \text{or} \quad
	\grad^2 \sigma = - \grad \cdot (\bfw \times \bfv + \la^2 \bfT) = - \grad \cdot (\bfw \times \bfv_* ).
	\eeq

\section{Poisson brackets for regularized MHD}
\label{s:PB-for-R-MHD}

Poisson brackets among functionals of velocity, density and magnetic field, for ideal compressible MHD were given by Morrison and Greene in \cite{morrison-and-greene}. The PB of functionals $F,G$ of $\rho, \bfv, \bfB$ is
	\beqs
	\{ F, G \} &=& \int \left[\frac{\bfw}{\rho} \cdot \left( F_{\bfv} \times G_{\bfv} \right) - F_{\bfv} \cdot \grad G_{\rho} + G_{\bfv} \cdot \grad F_{\rho}\right] d\bfr \cr
	&& - \int \left[\frac{\bfB}{\rho} \cdot \left[ \left( F_{\bfv} \cdot \grad \right) G_{\bfB} - \left( G_{\bfv} \cdot \grad \right) F_{\bfB} \right]
	+ \frac{B_i}{\rho} \left(\deldel{F}{v_j} \pdr_i \deldel{G}{B_j} - \deldel{G}{v_j} \pdr_i \deldel{F}{B_j}  \right)\right] d\bfr.
	\label{e:pb-mhd-functionals-rho-v-B}
	\eeqs
There are other forms related to the above formula via integration by parts using $\diver{\bfB} = 0$ and appropriate BCs.

From these we get the PBs between $\rho, \bfv$ and $\bfB$. As before (\S \ref{s:pb-for-fluid}) for the fluid variables $\rho,\bfv$ and $\bfw$ we have
	\beq
	\{ \rho(x), \rho(y) \} = 0, \quad
	\{ v_i(x), v_j(y) \} = \frac{\eps_{ijk}w_k(x)}{\rho(x)} \del(x-y) , \quad \text{and} \quad
	\{ v_i(x), \rho(y) \} = - \pdr_{x^i} \del(x-y),
	\eeq
Like $\bfw$, $\bfB$ Poisson commutes with $\rho$, but unlike $\bfw$ its components commute. The PB of $\bfv$ with $\bfB$  is
	\beq
	\{ v_i(x), B_j(y) \} = \ov{\rho(x)} \left[ \del_{ij} B_k(x) \pdr_{x^k} - B_j(x) \pdr_{x^i} \right] \del(x-y)
	= \ov{\rho(x)} \eps_{ilk} \eps_{jmk} B_l(x) \pdr_{x^m} \del(x-y) .
	\label{e:pb-v-B}
	\eeq
Taking the curl of (\ref{e:pb-v-B}) we get the PB of vorticity with magnetic field: 
	\beqs
	\{ w_i(x), B_j(y) \} &=& \eps_{ilm} \pdr_{x^l} \left( \ov{\rho(x)} \left[ \del_{mj} B_k(x) \pdr_{x^k} - B_j(x) \pdr_{x^m} \right] \del(x-y) \right) \quad \text{or} \cr
	\{ B_i(x), w_j(y) \} &=& - \eps_{jlm} \pdr_{y^l} \left( \ov{\rho(y)} \left[ \del_{mi} B_k(y) \pdr_{y^k} - B_i(y) \pdr_{y^m} \right] \del(x-y) \right).
	\eeqs 
MHD PBs can also be written for functionals of $\rho, \bfM = \rho \bfv$ and $\bfB$. Denoting the commutator of vector fields in the usual way,
	\beqs
	\{ F , G \} &=& - \int \left[ \rho \left( F_{\bfM} \cdot \grad G_{\rho} - G_{\bfM} \cdot \grad F_{\rho} \right) 
	+ \bfM \cdot \left[ F_{\bfM}, G_{\bfM}\right]\right]d\bfr \cr
	&& -  \int \left[\bfB \cdot \left[ \left( F_{\bfM} \cdot \grad \right) G_{\bfB} - \left( G_{\bfM} \cdot \grad \right) F_{\bfB} + \grad \left( F_{\bfM} \right) \cdot G_{\bfB} - \grad \left( G_{\bfM} \right) \cdot F_{\bfB} \right]\right] d\bfr.
	\eeqs
We use the dyadic notation in the last term e.g. $\bfB \cdot \grad(\bfC) \cdot \bfD = B_i (\pdr_i C_j) D_j$. If $\bfA$ is the magnetic vector potential $\bfB = \grad \times \bfA$, then the PBs of functionals of $\rho, \bfM$ and $\bfA$ in the laboratory gauge (to be discussed below) is given by
	\beqs
	\{ F , G \} &=& - \int \left[ \rho \left( F_{\bfM} \cdot \grad G_{\rho} - G_{\bfM} \cdot \grad F_{\rho} \right) + \bfM \cdot \left[ F_{\bfM}, G_{\bfM} \right]\right] d\bfr \cr
	&& + \int \bfA \cdot \left[ F_{\bfM} \grad \cdot G_{\bfA} - G_{\bfM} \grad \cdot F_{\bfA} - \grad \times \left( F_{\bfM} \times G_{\bfA} - G_{\bfM} \times F_{\bfA} \right) \right] \, d\bfr.
	 \label{e:PB-RMHD-A}
	\eeqs
Thus the components of $\bfA$ commute with $\rho$ and among themselves while the PB with mass current and velocity are
	\beq
	\{ M_i(x) , A_j(y) \} = (F_{ij}(x) + A_i(x) \pdr_{y^j}) \del(x-y) \quad \text{and} \quad 
	 \{ v_i(x), A_j(y) \} = \frac{(F_{ij}(x) + A_i(x) \pdr_{y^j}) \del(x-y)}{\rho(x)}.
	\label{e:v-A-pb}
	\eeq
Here $F_{ij} = \pdr_i A_j - \pdr_j A_i = \eps_{ijk} B_k$. We check that these PBs of $\bfA$ imply the above PBs of $\bfB$. Taking the curl of $\{ \bfv(x), \bfA(y) \}$ in $y$, the second term is a curl of a gradient and vanishes and we recover (\ref{e:pb-v-B}).
The curl of (\ref{e:v-A-pb}) gives the PB between vector potential and vorticity:
	\beq
	\{ A_i(x) , w_j(y)\} = \eps_{jkl} \pdr_{y^k} \left[ \ov{\rho(y)} \left( F_{li} (y) - A_l(y) \pdr_{y^i}\right)\del(x-y) \right].
	\eeq
For {\it incompressible} ($\grad \cdot \bfv = 0$ and constant $\rho$) R-MHD, the above PBs (\ref{e:PB-RMHD-A}) in laboratory gauge reduce to the following PBs
	\beqs
	\{ F[\bfv, \bfA] , G[\bfv, \bfA] \} &=& - \ov{\rho}\int \left[ \bfv \cdot \left[ F_{\bfv}, G_{\bfv} \right]
	 + \bfA\cdot \left[ F_{\bfv} \grad \cdot G_{\bfA} - G_{\bfv} \grad \cdot F_{\bfA} - \grad \times \left( F_{\bfv} \times G_{\bfA} - G_{\bfv} \times F_{\bfA} \right) \right]\right] \, d\bfr  \cr
	 &=& -\ov{\rho}\int  \left[\bfv \cdot \left[ F_{\bfv}, G_{\bfv} \right]
	 + \bfA \cdot \left([F_{\bfA}, G_{\bfv}] -[ G_{\bfA}, F_{\bfv}] \right) \right]\, d\bfr.
	\eeqs
As for incompressible neutral fluids, functional derivatives with respect to $\bfv$ are assumed solenoidal: $\grad \cdot F_{\bfv} = 0$ and $\grad \cdot G_{\bfv} = 0$.

\subsection{R-MHD Equations of motion from Poisson brackets}

The Hamiltonian for R-MHD is the conserved swirl energy of R-Euler with the additional magnetic energy term:
	\beq
	H = \int \left[ \frac{\rho(x) \bfv^2(x)}{2}+ U(\rho) + \frac{\la^2 \rho \bfw^2(x)}{2} + \frac{\bfB^2(x)}{2 \mu_0} \right] \: d\bfr.
	\eeq
Since $\rho$ commutes with $\bfB$, $\{ H, \rho \}$ is the same in R-MHD as in R-Euler. So the continuity equation  $\pdr \rho/\pdr t = \{ \rho, H \} = - \grad \cdot (\rho \bfv)$ follows. On the other hand, the introduction of the magnetic field alters the evolution equation for $\bfv$. We show that our PB give the correct evolution equations for $\bfv$ and $\bfB$ in regularized compressible MHD.

\subsubsection{Evolution of $\bfA$ and $\bfB$ from Poisson brackets} 

Here we derive the evolution equation for $\bfA$ using PB :$\pdr \bfA/\pdr t = \{\bfA,H\}$. Let us evaluate $\{ \bfA ,KE + EE\}$. $\{\bfA, PE \}= \{ \bfA ,ME\} = 0$ since both $\rho$ and $\bfB$ commute with $\bfA$.  
	\beqs
     \{A_j(y), H \} &=& \int \left[\rho(x)v_i(x)\{A_j(y) ,  v_i(x)\} + \la^2 \rho w_i(x)\{A_j(y),w_i(x) \} \right]dx \cr
     &=& \int \left[v_i(x)\left( A_i(x)\pdr_{x^j} - F_{ij} \right) \del(x-y)
     + \la^2 \rho w_i(x)\eps_{ikl}\pdr_{x^k}\left(\rho(x)^{-1} \left( A_l(x) \pdr_{x^j} - F_{lj} (x) \right) \del(x-y) \right)\right] dx \cr
     &=& -\pdr_j(v_i A_i) - v_i F_{ij}+\la^2\eps_{ikl}\left[(\pdr_k w_i)F_{lj} + \rho \pdr_j \left((\pdr_k w_i)\left({A_l}/{\rho}\right) \right)\right] \cr
     &=& (\bfv \times \bfB)_j +\left(\la^2 (\grad \times \bfw)\times \bfB \right)_j -\pdr_j(\bfv \cdot\bfA) - \la^2 \rho \pdr_j \left({(\grad \times \bfw) \cdot \bfA }/{\rho}\right)\cr
     \imply \quad \bfA_t &=& \{\bfA, H\} = (\bfv_* \times \bfB) - \grad(\bfv_* \cdot \bfA)  \quad \text{or} \quad
     \left[ - \grad(\bfv_* \cdot \bfA) -  {\bfA}_{t} \right]  + (\bfv_* \times \bfB) = 0. \label{e:pb-A-with-H}
     \eeqs
In this calculation we omitted the boundary terms assuming suitable BCs (e.g. $\bfw \times \hat n = 0$ and $\bfA \times \hat n = 0$). We identify the electric field as $\bfE = - \pdr \bfA/\pdr t - \grad (\bfv_* \cdot \bfA)$. Thus in this `laboratory' gauge, the electrostatic potential $\phi = \bfv_* \cdot \bfA$. This would be the electrostatic potential in the lab frame for the case where the electrostatic potential is zero in a `plasma' frame moving at $\bfv_*$ (See eq. 24.39 of \cite{Fock}). In the lab frame, if $\bfv_* = 0$ at a point, then the electrostatic potential would be zero in this gauge at that point. This gauge is distinct from Coulomb gauge, indeed $\grad \cdot \bfA$ evolves according to
	\beq
	\pdr_t (\grad \cdot \bfA) = \grad \cdot (\bfv_* \times \bfB) - \grad^2 (\bfv_* \cdot \bfA).
	\eeq
Taking the curl of (\ref{e:pb-A-with-H}) we arrive at the regularized Faraday law governing evolution of $\bfB$
	\beq
	\pdr_t \bfB = \{ \bfB, H \}
	= \grad \times \left[ \bfv_* \times \bfB \right].
	\eeq
An ab initio calculation of $\{ \bfB, H \}$ from the PBs (\ref{e:pb-mhd-functionals-rho-v-B}) assuming the BCs $\bfv \cdot \hat n = 0$, $\bfB \cdot \hat n = 0$ and $\bfw \times \hat n= 0$ gives the same regularized Faraday's law .

\subsubsection{Evolution of velocity from Poisson brackets}

Here we show that $\dd{\bfv}{t} = \{ \bfv , H \}$ gives the R-Euler equation including the Lorentz force term
	\beq
	\dd{\bfv}{t} + ({\bf v} \cdot \grad) \bfv = - \grad U'(\rho) - \la^2 \bfw \times (\grad \times \bfw) + \frac{{\bf j} \times \bfB}{\rho}.
	\eeq
Recall that $H = KE + PE + EE + ME$ and the PB of $KE + PE + EE$ with velocity is the same as in R-Euler and gives rise to all but the Lorentz force term in the momentum equation. So it only remains to calculate the PB of ME with $\bfv$:
	\beqs
	\{ ME, v_i(x) \} &=& \ov{\mu_0} \int B_j(y) \{ B_j(y), v_i(x) \} dy
	= \ov{\mu_0} \int B_j(y)\ov{\rho(x)} \left[ B_j(x) \pdr_{x^i} - \del_{ij} B_k(x) \pdr_{x^k} \right] \del(x-y) \: dy \cr
	&=& \frac{B_j}{\mu_0 \rho} \pdr_{x^i} \int B_j(y) \del(x-y) dy
	- \frac{B_k}{\mu_0 \rho} \pdr_{x^k} \int B_i(y) \del(x-y) dy \cr
	&=& - \ov{\rho \mu_0} (B_k \pdr_k B_i - B_k \pdr_i B_k) = - \ov{\rho} (\bfj \times \bfB)_i.
	\eeqs
Here $\mu_0 \bfj = \grad \times \bfB$. This gives the Lorentz force term in the momentum equation.

\subsubsection{$\grad \cdot {\bf B}$ commutes with the Hamiltonian $H$}

The Maxwell equation $\grad \cdot \bfB = 0$ is consistent with our PBs since we show below that $\grad \cdot \bfB$ commutes with $H$. So if $\grad \cdot \bfB$ is initially zero, it will remain zero under hamiltonian time evolution. Now potential energy $\int U(\rho) dx$ commutes with $\grad \cdot \bfB$ since $\{ \rho, \bfB \} =0$. Magnetic energy $\int {\bfB^2}/{2 \mu_0}$ also commutes with $\grad \cdot \bfB$ since $\{B_i, B_j\} =0$. We will show now, that $\{ KE, \grad \cdot \bfB \}$ and $\{ EE, \grad \cdot \bfB \}$ vanish separately, so that the above assertion holds:
	\beqs \nonumber
	\{ KE, \grad \cdot \bfB\} &=&\pdr_{y^j} \int \rho(x) v_i(x) \{ v_i(x), B_j(y) \} \: dx = \pdr_{y^j} \int v_i(x)\left[ \del_{ij} B_k(x) \pdr_{x^k} - B_j(x) \pdr_{x^i} \right] \del(x-y) \: dx  \cr
	&=&  \pdr_i (v_j B_i) - \pdr_j \pdr_i (v_i B_j) = 0, \cr
	\{ EE,  \grad \cdot \bfB\} &=& \pdr_{y^j} \int \la^2\rho w_i(x) \{ w_i(x), B_j(y) \} \: dx \cr
	&=& \pdr_{y^j} \int \la^2 \rho w_i(x) 
	\eps_{ilm} \pdr_{x^l} \left(\ov{\rho(x)} \left[ \del_{mj} B_k(x) \pdr_{x^k} - B_j(x) \pdr_{x^m} \right] \del(x-y) \right) \: dx \cr
	&=& \pdr_{y^j} \int (\la^2 (\grad \times \bfw)_m)(x)
	  \left[ \del_{mj} B_k(x) \pdr_{x^k} - B_j(x) \pdr_{x^m} \right]\del(x-y) dx  \cr
	  &=& \pdr_j\pdr_m (\la^2 (\grad \times \bfw)_j  B_k)- \pdr_j\pdr_k(\la^2 (\grad \times \bfw)_j B_k)=0.
	\eeqs

\subsection{Poisson algebra of conserved quantities in R-MHD}
\label{s:pb-cons-qty-MHD}

Linear momentum $\bfP = \int \rho \bfv d\bfr$ commutes with itself and the R-MHD Hamiltonian $H$. To show that $\bfP$ commutes with the $H$ we need only calculate $\{ P_i , ME \}$ since it was shown to commute with $KE, PE$ and $EE$ in R-Euler with appropriate BCs:
	\beqs
	\{ P_i ,ME \} &=& \ov{\mu_0}\iint_V \rho(x) B_j(y) \left\{v_i(x) ,B_j(y)\right\}\,dx\, dy 
	= \ov{\mu_0}\iint_V B_j(y)\left[\del_{ij} B_k(x) \pdr_{x^k} - B_j(x) \pdr_{x^i}\right] \del(x - y) \,dx\, dy \cr
	&=& \ov{\mu_0}\int_V  B_j \pdr_{i} B_j \,dy +  \ov{\mu_0}\int_{\pdr V} \left[B_i\left(\bfB \cdot \hat n \right) - \bfB^2 n_i\right]\, dS
	= - \ov{\mu_0}\int_{\pdr V}  \left(\frac{\bfB^2}{2} \del_{ij} - B_i B_j  \right) n_j \,dS.
	\eeqs
Thus $\{ P_i, H \} = - \int_{\pdr V} \Pi_{ij} n_j dS$ where $\Pi_{ij}$ is the momentum current (\ref{e:mom-cons-r-mhd}). For periodic or decaying BC this flux is zero. Angular momentum $\bfL = \int \rho \bfr \times \bfv  \: d\bfr$ also commutes with $H$. Again we only compute $\{ L_i ,ME\}$: \small
	\beqs 
	\mu_0\{ L_i ,ME \} &=&  \iint_V \eps_{ijk} x_j\rho(x) B_l(y)\left\{v_k(x), B_l(y)\right\}\,dx\, dy
	=\iint_V \eps_{ijk} x_j B_l(y) \left[\del_{kl} B_m(x) \pdr_{x^m} - B_l(x) \pdr_{x^k}\right] \del(x - y) \,dx\, dy \cr
	&=&  \int_V \left[(\bfB \times \bfB)_i + \eps_{ijj} B^2 + \eps_{ijk}y_j \pdr_k \frac{B^2}{2} \right] dy +\int_{\pdr V} \eps_{ijk}y_j  \left [B_k \bfB \cdot \hat n- B^2 n_k \right]\, dS \cr
	&=& \int_{\pdr V} \eps_{ijk}y_j n_m \left [B_k B_m - \frac{B^2}{2} \del_{mk} \right]\, dS.
	\eeqs \normalsize
Thus $\{ L_i , H \} = - \int_{\pdr V} \Lambda_{ij} n_j dS$ where $\Lambda_{ij}$ is the angular momentum current (\ref{e:ang-mom-cons-r-mhd}). So $\{ \bfL, H \} = 0$ if this flux vanishes (as for decaying BCs). The angular momentum algebra $\{ L_i, L_j \} = \eps_{ijk} L_k$ is unaffected by the addition of $ME$. Magnetic helicity ${\cal K}_B = \int \bfA \cdot \bfB \, d\bfr$ commutes with the swirl Hamiltonian\footnote{In MHD flow helicity does not commute with $H$ due to the Lorentz force in the momentum equation.}. In fact, it is a Casimir of the Poisson algebra. Since $\bfA$ commutes with $\rho$ and itself and ${\cal K}_B$ is a functional of $\bfA$ alone, by (\ref{e:PB-RMHD-A}), the PB of ${\cal K}_B$ with any functional $F[\rho,\bfM,\bfA]$ is
	\beq
	\{ {\cal K}_B, F \} = \int_V \bfA \cdot \left[ F_{\bfM} \grad \cdot {\cal K}_{B,\:\bfA}  - \grad \times \left( F_{\bfM} \times {\cal K}_{B,\:\bfA} \right) \right]\, d\bfr.	
	\eeq
To proceed, we first show that ${\cal K}_{B,\:\bfA} \equiv \del {\cal K}_B/\del \bfA = 2 \bfB$ provided $\bfA$ is normal to the boundary: 
	\beqs
	\deldel{{\cal K}_B}{A_l(y)} &=& \deldel{}{A_l(y)} \int_V A_i(x) \eps_{ijk} \pdr_j A_k(x) \: d\bfx  = \int_V \eps_{ijk} \left[ \del_{il} \del(x-y) \pdr_j A_k(x) + A_i(x) \pdr_j (\del_{kl} \del(x-y)) \right] d\bfx \cr
	&=& 2B_l +	\int_{\pdr V} (\bfA \times \hat n)_l\del(x-y) \: dS.
	\eeqs
Armed with this, the PB becomes
	\beqs
	\{ {\cal K}_B, F \} &=& 2 \int_V \bfA \cdot \left[ F_{\bfM} \grad \cdot \bfB -  \grad \times \left( F_{\bfM} \times \bfB \right) \right]\, d\bfr 
	= - 2 \int_V \bfA \cdot \grad \times \left( F_{\bfM} \times \bfB \right) \:d\bfr \cr
	&=& 2\int_V \grad \cdot (\bfA \times ( F_{\bfM} \times \bfB)) \:d\bfr = 2\int_{\pdr V} \bfA \times (\bfB \times F_{\bfM}) \cdot \hat n \: d\bfr \cr
	&=& 2 \int_{\pdr V} [\bfB (F_{\bfM} \cdot \bfA) - F_{\bfM} (\bfA \cdot \bfB )] \cdot \hat n dS.
	\eeqs
Thus ${\cal K}_B$ commutes with any observable $F$ provided $\bfB \cdot \hat n = 0$, $F_{\bfM} \cdot \hat n =0$ and $\bfA \times \hat n = 0$ on the boundary $\pdr V$ of the flow domain. Taking $F = H$ and using $H_M = \ov{\rho} H_\bfv = \bfv_*$ (assuming $\bfw \times \hat n = 0$) we have
	\beq
	\{{\cal K}_B, H \} = 2\int_{\pdr V} [(\bfB \cdot \hat n)(H_{\bfM} \cdot \bfA) - (\bfA \cdot \bfB )(H_{\bfM} \cdot \hat n)]dS = 2\int_{\pdr V} [(\bfB \cdot \hat n)(\bfv_* \cdot \bfA) - (\bfA \cdot \bfB )(\bfv_* \cdot \hat n)]dS = 0.
 	\eeq
Thus magnetic helicity commutes with the Hamiltonian with decaying/periodic BCs or assuming $\bfB$ and $\bfv_*$ are tangential and $\bfw$ and $\bfA$ are normal to the boundary.

In addition to magnetic helicity, cross helicity $X = \int \bfv \cdot \bfB \:d\bfr$ is also a Casimir. To see this, we compute its PB with an arbitrary functional $G$ (assuming decaying BCs for simplicity) using (\ref{e:pb-mhd-functionals-rho-v-B}) and the functional derivatives $X_\bfv = \bfB$ and $X_\bfB = \bfv$: \small
	\beqs
	\{ X, G \} &=&  \int \left[ \frac{\bfw}{\rho} \cdot \left( \bfB \times G_{\bfv} \right) - \bfB \cdot \grad G_{\rho} 
	- \frac{\bfB}{\rho} \cdot \left[ \left( \bfB \cdot \grad \right) G_{\bfB} - \left( G_{\bfv} \cdot \grad \right) \bfv \right]
 	 + \frac{B_i}{\rho} \left(B_j \pdr_i G_{B_j} - G_{v_j} \pdr_i v_j  \right) \right] d\bfr \cr
	 &=& \int \left[ \frac{\grad \times \bfv}{\rho} \cdot \left( \bfB \times G_{\bfv} \right) + (\grad \cdot \bfB )G_{\rho} 
	+ B_j \pdr_j \left( \frac{B_i}{\rho} \right) G_{B_i}  - \pdr_j \left( \frac{B_i}{\rho} G_{v_j}\right)v_i - B_i \pdr_i \left( \frac{B_j}{\rho} \right) G_{B_j} + \pdr_i \left( \frac{B_i}{\rho} G_{v_j}\right)v_j \right] d\bfr
	\cr
	&=& \int \left[ \bfv \cdot \left( \grad \times \left(\frac{\bfB}{\rho} \times G_{\bfv}\right)\right)  - v_i \pdr_j \left( \frac{B_i}{\rho} G_{v_j}\right) + v_j \pdr_i \left( \frac{B_i}{\rho} G_{v_j}\right) \right]d\bfr = 0.
	\eeqs
\normalsize

\section{Other constitutive laws and regularizations bounding higher moments of vorticity}
\label{s:other-const-laws-and-regs}

An interesting application of our Hamiltonian and PB formulation is to the identification of other possible conservative regularizations that preserve the symmetries of the Euler equations. An interesting class of these arise by choosing new constitutive relations. Recall that the twirl regularization term $- \la^2 \bfw \times (\grad \times \bfw)$ was selected as it is the least non-linear term of lowest spatial order that preserves the symmetries of Euler. Moreover, with the constitutive relation $\la^2 \rho =$ constant, R-Euler admits a conserved swirl energy $E^*$ (\ref{e:swirl-energy-R-Euler}) which implies bounded enstrophy. R-Euler equations are Hamilton's equations for $E^*$ and the standard PBs (\ref{e:pb-between-functionals-of-rho-v}). Retaining the same Poisson brackets as before, and choosing an unaltered form for the Hamiltonian,
	\beq
	H = \int \left[\half \rho \bfv^2 + U(\rho) + \half \la^2 \rho \bfw^2 \right] \; d\bfr,
	\label{e:hamiltonian-new-constitutive-law}
	\eeq
we will now allow for more general constitutive relations, e.g., $\la_n^2 \rho = c_n \left(\bfw^2 \right)^n$ where $c_n$ is a positive constant. The virtue of this type of constitutive law is that the $(n+1)^{\rm th}$ moment of $\bfw^2$ is bounded in the flow generated by this conserved Hamiltonian\footnote{More generally $c_n$ could depend on $\rho$ without affecting the continuity equation but resulting in additional terms in the equation of motion which ensure boundedness of $\int c_n(\rho) (\bfw^2)^{n+1} \: d\bfr$.}. From Hamilton's equation for $\rho$ we see that the continuity equation is unaltered since $\rho$ commutes with itself and $\bfw$ (in fact as long as $\la$ depends only on $\rho$ and $\bfw$, the continuity equation will remain the same). However, there is a new regularization term in the equation for $\bfv$. Indeed, from (\ref{e:list-of-useful-PB}) one finds that
	\beq
	\{ \bfv , EE_n \} = \left\{ \bfv , \int \half \la_n^2 \: \rho \bfw^2 \: d\bfy \right\} 
	= c_n (n+1) \int (\bfw^2(\bfy))^n  \{ \bfv, \bfw(\bfy) \} \cdot \bfw(\bfy) \, d\bfy
	= - \frac{(n+1) c_n}{\rho} \left[ \bfw \times \left( \grad \times (\bfw^2)^n \bfw \right) \right].
	\eeq
Thus the equation of motion becomes
	\beqs
	\dd{\bfv}{t} &=& \{ \bfv, H \} = - \bfv \cdot \grad \bfv - \ov{\rho} \grad p - \frac{(n+1) c_n}{\rho} \left[ \bfw \times \left( \grad \times (\bfw^2)^n \bfw \right) \right] = - \grad \sigma - \bfw \times \bfv_{n*}, \cr
	\text{where} \quad \bfv_{n*} &=& \bfv + \ov{\rho} \grad \times ((n+1)c_n |\bfw|^{2n} \bfw)
	\label{e:mom-eqn-new-const-law}
	\eeqs
is a new swirl velocity. Clearly, $\grad \cdot (\rho \bfv) = \grad \cdot (\rho \bfv_{n*})$ so the continuity equation may be written as $\rho_t + \grad \cdot (\rho \bfv_{n*}) = 0$. Thus the form of the governing equations is unchanged; only the swirl velocity $\bfv_*$ is modified to $\bfv_{n*}$. When $n = 0$, this reduces to the R-Euler equation for which the first moment of $\bfw^2$ (enstrophy) is bounded. For $n > 0$ we get new regularization terms which are more non-linear (i.e, of degree $2n+2$ in $\bfv$) than the quadratic twirl term, though the equation {\it remains} $2^{\rm nd}$ order in space derivatives. Furthermore, $P_i, L_i$ continue to be conserved as the new constitutive relation does not break translation or rotation symmetries (it only depends on the scalar $\bfw^2$). Flow helicity is also conserved being a Casimir of the Poisson algebra. Finally, parity, time reversal and Galilean boost invariance are also preserved.

For R-MHD, the Hamiltonian (\ref{e:hamiltonian-new-constitutive-law}) is augmented by the magnetic energy ME $= \int \bfB^2/2\mu_0 \: d\bfr$. ME does not affect the continuity equation as $\{ \rho, \bfB \} = 0$ but adds the Lorentz force term to the momentum equation (\ref{e:mom-eqn-new-const-law})
	\beq
	\pdr_t \bfv = - \bfw \times \bfv_{n *} + \frac{\bfj \times \bfB}{\rho}.
	\eeq
The R-Faraday law (\ref{e:R-MHD-Faraday}) is modified by the new constitutive relation since $\bfB$ does not commute with vorticity.  Remarkably the R-Faraday equation takes the same form as (\ref{e:R-MHD-Faraday}) with $\bfv_* \mapsto \bfv_{n*}$: $\bfB_t = \grad \times (\bfv_{n*} \times \bfB)$. Indeed,
	\beqs
	\{B_i(\bfx), EE_n \} &=& \int c_n (n+1) (\bfw)^{2n} w_j(\bfy) \{B_i (\bfx) , w_j(y)\} d\bfy \cr
	&=& \int c_n (n+1) (\bfw)^{2n} w_j(\bfy)\eps_{jlm} \pdr_{y^l}\left( \ov{\rho}(B_i \pdr_m - \del_{mi}B_k\pdr_k)\right)\del(\bfx -\bfy)\,d\bfy \cr
	&=&\int \ov{\rho}c_n(n+1) \left(\grad \times (\bfw)^{2n} \bfw\right)_m \left(B_i \pdr_m - \del_{mi}B_k\pdr_k\right)\del(\bfx -\bfy)\,d\bfy \cr
	&=& c_n(n+1) \left(\left(\grad \times (\bfw)^{2n} \bfw\right) \cdot \grad \left(\frac{B_i}{\rho}\right)  - \bfB \cdot \grad \left(\frac{(\grad \times (\bfw)^{2n} \bfw)_i }{\rho}\right)\right) \cr
	&=& \grad \times \left( \ov{\rho} \grad \times ((n+1)c_n |\bfw|^{2n} \bfw) \times \bfB \right)
	\eeqs
where we have used a vector identity for $\grad \times (\bfC \times \bfD)$ taking $\bfC = \grad \times ((n+1)c_n |\bfw|^{2n} \bfw)$ and $\bfD = \bfB /\rho$. Thus $\pdr_t \bfB = \{ \bfB , H \} = \grad \times (\bfv_{n*} \times \bfB)$. It is remarkable that the PB formalism enables us to obtain, with the help of suitable constitutive relations, regularized flows with bounded higher moments of vorticity.

\subsection{Regularizations that bound higher moments of $\grad \times \bfw$}

We use the PB formalism to derive new regularized equations for which we have an a priori bound on the $L^2$ norm of the curl of vorticity (just as we had a bound on the $L^2$ norm of vorticity earlier). This is achieved by considering the Hamiltonian
	\beq
	H = \int \left[ \half \rho \bfv^2 + U(\rho) + \frac{\bfB^2}{2 \mu_0} + \half d_1 (\grad \times \bfw)^2 \right] \: d\bfr
	\label{e:hamiltonian-with-powers-of-curl-w}
	\eeq
where $d_1$ is a positive constant. By dimensional analysis, $d_1$ may be expressed in terms of a dynamical short-distance cut off $\la(\bfr,t)$ that satisfies the constitutive relation $\la^4 \rho = d_1$. The continuity equation $\rho_t = \{ \rho, H \}= - \grad \cdot (\rho \bfv) = 0$ is unchanged from that in ideal MHD since $\{ \rho, \bfw \} = 0$. The evolution equation for $\bfv$ is of fourth order in space derivatives of $\bfv$ and turns out to be expressible in the familiar form (\ref{e:R-MHD-Euler-v*}) where $\bfv_* = \bfv + \la^4 \grad \times (\grad \times (\grad \times \bfw))$ is a new swirl velocity field. To see this we compute $\{ \bfv, H \}$. It suffices to consider only the PB with new term in $H$ (\ref{e:hamiltonian-with-powers-of-curl-w}):
	\beqs
	\left\{ v_i(x) , \int \frac{d_1}{2} (\grad \times \bfw)^2 \: dy \right\}
	&=& d_1 \int \left(\grad \times (\grad \times \bfw) \right)_m \left( \del_{km} \pdr_{y^i} - \del_{im} \pdr_{y^k} \right) \frac{w_k(y)}{\rho(y)} \del(x-y) \: d\bfy \cr
	&=& - \frac{d_1 w_k(x)}{\rho(x)} \left[ \pdr_i \left(\grad \times (\grad \times \bfw) \right)_k - \pdr_k \left(\grad \times (\grad \times \bfw) \right)_i  \right] \cr
	&=& - \la^4 \left[ \bfw \times \left(\grad \times (\grad \times (\grad \times \bfw)) \right) \right]_i.
	\eeqs
Similarly, Faraday's law of ideal MHD gets modified, but takes the same form $\bfB_t = \grad \times (\bfv_* \times \bfB)$ as in R-MHD when expressed in terms of $\bfv_*$. To see this we compute the PB with the regularization term in $H$ (\ref{e:hamiltonian-with-powers-of-curl-w}):
	\beqs
	\left\{ B_i(x) , \int \frac{d_1}{2} (\grad \times \bfw)^2 \: dy \right\}
	&=& d_1 \int (\grad \times \bfw)_j \{ B_i(x) , (\grad \times \bfw)_j \} \: d\bfy 
= d_1 \int (\grad \times \bfw)_j \eps_{jlm} \pdr_{y^l} \{ B_i(x), w_m(y) \} \: d\bfy \cr
	&=& d_1 \int (\grad \times (\grad \times \bfw))_m \eps_{mnp} \pdr_{y^n} \left[ \ov{\rho(y)} \left( B_i(y) \pdr_{y^p} - \del_{ip} B_k(y) \pdr_{y^k} \right) \del(x-y) \right] \: d\bfy
	\cr
	&=& - d_1 \: \pdr_p \left[ \rho^{-1} \left(S_p B_i - S_i B_p \right) \right]
	= d_1 \left( \frac{\bfB}{\rho} \cdot \grad S_i + S_i \grad \cdot \frac{\bfB}{\rho} - \bfS \cdot \grad \left(\frac{B_i}{\rho} \right) \right) \cr
	\imply \quad \left\{ \bfB , \int \frac{d_1}{2} (\grad \times \bfw)^2 \: dy \right\}
&=& \grad \times \left(\la^4 \bfS \times \bfB \right) = \grad \times \left( (\bfv_* - \bfv) \times \bfB \right).
	\eeqs
Here we defined ${\bf S} = \grad \times (\grad \times (\grad \times \bfw))$. Including the usual contribution from KE, we get the regularized Faraday law $\bfB_t = \grad \times (\bfv_* \times \bfB)$. The freezing-in and integral theorems automatically generalize to this case with the above swirl velocity $\bfv_*$.

We can generalize to a model where the $(2m)^{\rm th}$ moment of $\grad \times \bfw$ is bounded by considering the Hamiltonian
	\beq
	H = \int \left[ \half \rho \bfv^2 + U(\rho) + \frac{\bfB^2}{2 \mu_0} + \ov{2} d_m (\grad \times \bfw)^{2m} \right] \: d\bfr \equiv H_{MHD} + H_m.
	\label{e:hamiltonian-higher-power-of-curl-w}	
	\eeq
The constant $d_m$ must have dimensions of $(M/L^3) L^{2m+2} T^{2m-2}$. To express it in terms of the dynamical short distance cut-off $\la$ and density $\rho$ we introduce a reference {\it constant} speed $c$: $d_m = \la^{4m} \rho c^{2 - 2m}$. The regularized equations take the same form as above when expressed in terms of an appropriate swirl velocity 
	$\bfv_{m*} = \bfv + m \la^{4m} c^{2 - 2m} \grad \times \left(\grad \times \left( \left(\grad \times \bfw \right)^{2m-2} \grad \times \bfw \right) \right).$ The new term in $H$ does not change the continuity equation. By the constitutive relation $\la^{4m} \rho c^{2-2m} = d_m$, a constant,  $\grad \cdot (\rho \bfv_{m*}) = \grad \cdot (\rho \bfv)$ which means the continuity equation can also be expressed as $\rho_t = -\grad \cdot (\rho \bfv_{m*})$. To verify the regularized Euler and Faraday laws, it suffices to compute the PBs of $\bfv$ and $\bfB$ with the regularizing term $H_m$ in (\ref{e:hamiltonian-higher-power-of-curl-w}):
	\beqs
	\left\{ v_i(x) , H_m \right\}
	&=& m d_m \int (\grad \times \bfw)^{2m-2} (\grad \times \bfw)_j \{ v_i(x) , (\grad \times \bfw)_j \} \: d\bfy \cr
	&=& m d_m \int (\grad \times \bfw)^{2m-2} (\grad \times \bfw)_j \eps_{jlm} \pdr_{y^l} \{ v_i(x), w_m(y) \} \: d\bfy \cr
	&=& m d_m \int \left(\grad \times ((\grad \times \bfw)^{2m-2} \grad \times \bfw) \right)_m \left( \del_{km} \pdr_{y^i} - \del_{im} \pdr_{y^k} \right) \frac{w_k(y)}{\rho(y)} \del(x-y) \: d\bfy \cr
	&=& - m \la^{4m} c^{2-2m}  \left[ \bfw \times \left(\grad \times (\grad \times ((\grad \times \bfw)^{2m-2} \grad \times \bfw)) \right) \right]_i.
	\eeqs
Similarly, if we define ${\bf S} = \grad \times (\grad \times ((\grad \times \bfw)^{2m-2} \grad \times \bfw))$, then:
	\beqs
	\left\{ B_i(x) , H_m \right\}
	&=& m d_m \int (\grad \times \bfw)^{2m-2} (\grad \times \bfw)_j \{ B_i(x) , (\grad \times \bfw)_j \} \: d\bfy \cr
	&=& m d_m \int (\grad \times \bfw)^{2m-2} (\grad \times \bfw)_j \eps_{jlm} \pdr_{y^l} \{ B_i(x), w_m(y) \} \: d\bfy \cr
	&=& - m d_m \: \pdr_p \left[ \rho^{-1} \left(S_p B_i - S_i B_p \right) \right]
	= m d_m \left( \frac{\bfB}{\rho} \cdot \grad S_i + S_i \grad \cdot \frac{\bfB}{\rho} - \bfS \cdot \grad \left(\frac{B_i}{\rho} \right) \right) \cr
	\imply \quad \{\bfB, H_m\} &=& \grad \times \left(m \la^{4m} c^{2-2m} \bfS \times \bfB \right) = \grad \times \left( (\bfv_{m*} - \bfv) \times \bfB \right).
	\eeqs
Thus use of the PBs enables us to identify new regularization terms in the momentum equation that ensure bounded higher moments of $\grad \times \bfw$, without altering the continuity equation. It is remarkable that the regularized momentum and Faraday equations involve a common swirl velocity field $\bfv_{m*}$ into which both $\bfw/\rho$ and $\bfB/\rho$ are frozen.

\section{Some solutions of regularized flow equations}
\label{s:examples}

\subsection{Compressible flow model for rotating vortex}
\label{s:modeling-vortex}

In this section we model a {\em steady} tornado [cylindrically symmetric rotating columnar vortex with axis along $z$] using the compressible R-Euler equations. The unregularized Euler equations do not involve derivatives of vorticity, and admit solutions where the vorticity can be discountinuous or even divergent (e.g. at the edge of the tornado). On the other hand, the R-Euler equations involve the first derivative of $\bfw$ and can be expected to smooth out large gradients in vorticity on a length scale of order $\la$ while ensuring bounded enstrophy.

Given appropriate initial profiles for $\bfv$ and $\rho$, the R-Euler equations should uniquely determine $\rho$ and $\bfv$ at later times. However, unlike the initial value problem, the {\em steady} R-Euler equations are under-determined (just like the steady Euler equations). As a consequence of this under-determinacy, the system may reach different steady states depending on the initial conditions. This is unlike dissipative systems (e.g. Navier-Stokes) which typically have a unique steady solution irrespective of initial conditions (except when there are bifurcations to multiple steady states allowed by the boundary conditions).

In our rotating vortex model, the density $\rho$ and pressure $p$ depend only on the distance from the central axis while $\bfv$ is purely azimuthal ($\bfv = v_\phi(r) \: \hat \phi$) and vorticity vertical $\bfw = w_z(r) \, \hat z$. In the steady state there is a single equation for the two unknowns $w_z$ and $\rho$, so we can determine the density profile given a suitable vorticity field. In the vortex core of radius $a$, we assume the fluid rotates at approximately constant angular velocity $\Omega$. Far from the core, $\bfw \to 0$. In a boundary layer of width $\ll a$, the $\bfw$ smoothly interpolates between its core and exterior values. As a consequence of the regularization term, we find that this decrease in vorticity is related to a corresponding increase in density (from a rare core to a denser periphery). By contrast, the unregularized Euler equations (i.e. $\la \to 0$) allow $\bfw$ to have unrestricted discontinuities across the layer while $\rho$ is continuous.

\subsubsection{Steady state regularized equations in cylindrical geometry}

Our infinitely long columnar vortex rotates about the $z$-axis and is assumed to be rotationally and translationally invariant about its axis. Hence $\bfv \cdot \hat z = 0$. We seek steady solutions of the R-Euler system. The continuity equation $\grad \cdot (\rho \bf v) = 0$ becomes $\pdr_x(\rho v_x) + \pdr_y(\rho v_y) = 0$. The incompressible 2-d vector field $\rho \bfv$ can be expressed in terms of a scalar stream function $\rho \bfv = - \grad \times (\psi \hat z)$. Axisymmetry dictates that $\psi$ is a function of the cylindrical coordinate $r$ alone. It follows that $\bfv$ is purely azimuthal: $v_\phi = \psi'(r)/\rho$ (primes denote differentiation in $r$) and the continuity equation is identically satisfied. The steady state R-Euler equation is
	\beq
	\bfw \times \bfv = - \grad \sigma - \la^2 \bfw \times (\grad \times \bfw) \quad \text{where} \quad \sigma = h + \half  \bfv^2,
	\eeq
and $h$ is the specific enthalpy/Gibbs free energy for adiabatic/isothermal flow. Vorticity is vertical ($w_z = r^{-1} (r v_\phi)'$) while its curl is azimuthal $(\grad \times \bfw)_\phi = - w_z'(r)$. Thus the vorticity $\bfw \times \bfv$ and twirl accelerations both point radially:
	\beq
	(\bfw \times \bfv)_r = -w_z v_\phi \quad
	\text{and} \quad
	(\bfw \times (\grad \times \bfw))_r = w_z \dd{w_z}{r}.
	\eeq
Hence $\grad \sigma$ must also be radial and $h$ and $\rho$ functions of $r$ alone. Thus the steady R-Euler equations reduce to a single $1^{\rm st}$ order nonlinear ODE for $\rho(r)$ given $v_\phi(r)$ or $w_z(r)$. To solve it we need an equation of state relating $p$ to $\rho$.
	\beq
	w_z v_\phi = \dd{}{r} \left(h + \half v_\phi^2 \right) + \frac{\la^2}{2} \dd{w_z^2}{r} \quad \text{or} \quad
	\frac{v_\phi^2}{r} = \dd{h}{r} + \frac{\la^2}{2} \dd{w_z^2}{r}.
	\label{e:steady-state-eqn-for-vortex}
	\eeq

\subsubsection{Vortex model with rigidly rotating fluid core}

\begin{figure}[h]
\begin{center}
 \includegraphics[width = 4cm]{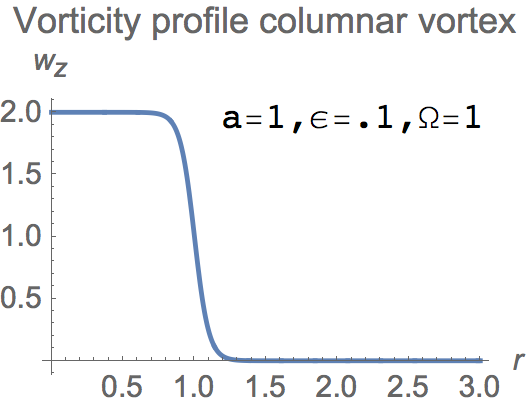}
 \hspace{1cm}
 \includegraphics[width = 4cm]{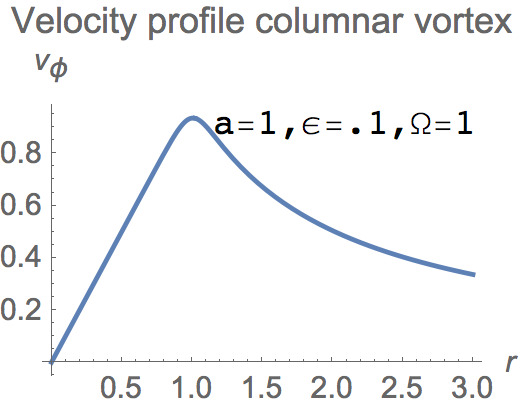}
 \hspace{1cm} 
 \includegraphics[width = 4cm]{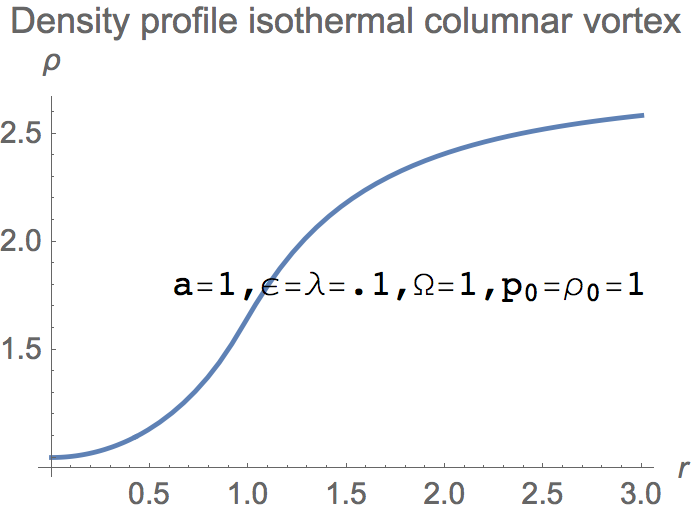}
 \caption{\footnotesize Vorticity $w_z(r)$ and velocity $v_\phi(r)$ for rotating vortex of core radius $a$ and angular velocity $\Om$. $\rho(r)$ for isothermal flow increases outwards from core and reaches an asymptotic value. The regularization relates the drop in $w_z$ to an increase in $\rho$ in a layer of thickness $\approx \eps = \lambda$ around $r = a$.}
\label{f:tornado-figs}
 \end{center}
\end{figure}

As a simple model for a rotating vortex of core radius $a$, we consider the vorticity distribution (see Fig. \ref{f:tornado-figs})
	\beq
	w_z(r) = 2 \Om \left[1 - \tanh \left(\frac{r-a}{\epsilon} \right) \right] \: \left[1+\tanh \left(a/\epsilon \right) \right]^{-1}.
	\label{e:vorticity-profile-tornado}
	\eeq
Over a transition layer of width $\approx 2 \eps \ll a$, the vorticity drops rapidly from $\approx 2 \Om$ to $\approx 0$. In the vortex core $r \ll a - \eps$, the flow corresponds to rigid body rotation at the constant angular velocity $\Omega \hat z$, apart from higher order corrections in $\eps$. Thus in the core, the vorticity is roughly twice the angular velocity and $\bfv = \Om \hat z \times \bfr \approx \Om r \hat \phi$. In the exterior region, for $r \gg a + \epsilon$ the vorticity tends to zero exponentially. The velocity is obtained by integration subject to the BC $v_{\phi}(0)=0$. \footnotesize
	\beq
	v_\phi(r) = \Omega \frac{\eps^2
   \left[\text{Li}_2\left(-e^{\frac{2 (a-r)}{\eps}}\right)-\text{Li}_2\left(-e^{\frac{2 a}{\eps}}\right)\right] 
   + 2 \left[\eps (a-r) \log
   \left(e^{\frac{2(a-r)}{\eps}}+1\right) 
   + a  \left( r + \eps \log\frac{\cosh(a/\eps)}{\cosh((a-r)/\eps)}\right) 
   - a \eps \log\left(e^{\frac{2a}{\eps}} + 1 \right)\right]}
   {r \left(\tanh \frac{a}{\eps} + 1 \right)}.
	\label{e:vel-profile-tornado-exact}
	\eeq
\normalsize
The velocity profile (Fig.\ref{f:tornado-figs}) rises nearly linearly with $r/a$ in the core [rigid body motion] and drops off as $\sim 1/r$ at large distances like a typical, irrotational potential vortex. In the transition layer $a-\eps \lesssim r \lesssim a+\eps$ the radial derivative of the velocity varies rapidly.
The density can be obtained by integrating the steady R-Euler equation. We do this below in the simpler case of isothermal flow where the equation for $\rho$ is linear since $p \propto \rho$. The adiabatic case ($p/p_0 = (\rho/\rho_0)^\gamma$) is similar, but the steady state equation (\ref{e:steady-state-eqn-for-vortex}) is a non-linear first order ODE for density: $\rho v_\phi^2/r = \gamma  \frac{p_0}{\rho_0^\gamma} \rho^{\gamma -1} \rho'(r) + \rho \la^2 (w_z^2)'/2$.

\subsubsection{Vortex with isothermal flow}

For isothermal flow the ideal gas equation of state $p V = n k_B T$ implies the pressure-density relation $p = (p_0/\rho_0) \rho$. The specific `Gibbs free energy' $h$ is obtained from the condition $\grad h = \ov{\rho} \grad p$,
	\beq
	\grad h = \frac{p_0}{\rho_0} \frac{\grad \rho}{\rho} = \frac{p_0}{\rho_0} \grad \log\frac{\rho}{\rho_0} \quad \imply \quad h = \frac{p_0}{\rho_0} \log \frac{\rho}{\rho_0}.
	\eeq
The flow is assumed purely `hydrodynamic': internal energy changes due to density variations are ignored; entropy and the internal energy equation do not play a role. The steady equation (\ref{e:steady-state-eqn-for-vortex}) in the isothermal case is
	\beq
	\frac{v_\phi^2}{r} = \frac{p_0}{\rho_0} \frac{\rho'(r)}{\rho(r)} + \frac{\la^2}{2} \dd{w_z^2}{r} \;\; \imply \;\;
	\frac{p_0}{\rho_0} \rho'(r) - \frac{v_\phi^2}{r} \rho(r) = - \frac{\la_0^2 \rho_0}{2} (w_z^2)'.
	\label{e:isothermal-steady-state-vortex-eqn}
	\eeq
This is a first order linear inhomogeneous ODE for $\rho$ with variable coefficients in the standard form\footnote{Putting $q'/q = B/A$ the equation becomes $(q \rho)' = qf/A$ whence $\rho = \ov{q} [\rho(0) q(0) + \int_0^r \frac{qf}{A} \, ds]$.}
	\beq
	A(r) \rho' + B(r) \rho = f(r), \quad \text{where} \quad
	A = \frac{p_0}{\rho_0}, \quad B = - \frac{v_\phi^2}{r} \quad \text{and} \quad f(r) = - \frac{\la_0^2 \rho_0}{2} (w_z^2)'.
	\eeq
It is convenient to take the reference values $\rho_0, \la_0, p_0$ to be at $r = 0$. The solution for $\rho(r)$ is
	\beqs
	\rho &=& \frac{\rho_0 q(0)}{q(r)} \left[ 1 + \int_0^r \frac{q(s) f(s)}{q(0) \rho(0) A} \, ds \right]
	= \frac{\rho_0 q(0)}{q(r)} \left[ 1 - \frac{\Om^2 \la_0^2 \rho_0}{2 p_0} \int_0^r \frac{q(s)}{q(0)} \left(\frac{w_z^2}{\Om^2} \right)' \, ds \right], \cr
	\text{where} \quad \frac{q(r)}{q(0)} &=& \exp \left[- \frac{\rho_0}{p_0} \int_0^r \frac{v_\phi^2}{s} \, ds \right].
	\label{e:soln-linear-inhom-1st-Order-ode}
	\eeqs
$q(r)$ is a positive monotonically (exponentially) decreasing function of $r$ and we can take $q(0)=1$ without loss of generality. The integrations are done numerically and the resulting density is plotted in Fig \ref{f:tornado-figs}. $\rho$ is monotonically increasing from $\rho(0)$ to an asymptotic value $\rho(\infty)$ (material has been `ejected' from the core). The above formula shows that one effect of the regularization is to decrease the density relative to its Eulerian value (especially outside the core). To get more insight into the role of the regularization we solve the steady equation approximately in the core, transition and exterior regions separately.


\pt {\bf Vortex Core $0 < r \lesssim a_- = a-\eps$}: In this region $w_z(r) \approx w_z(0) = 2\Omega$. The corresponding velocity $v_\phi(r) = r w_z(0)/2 = r\Omega$ grows linearly as for a rigidly rotating fluid. Since $\bfw$ is roughly constant, the regularization term may be ignored and (\ref{e:isothermal-steady-state-vortex-eqn}) becomes $\rho'(r) = \rho(r) \:(\rho_0/p_0) \Om^2 r$. The density grows exponentially inside the vortex core:
	\beq
	\rho(r) \approx \rho(0) \exp \left( \frac{\rho_0 \Om^2 r^2}{2 p_0} \right) 
	\approx \rho(a_-) \exp \left( \frac{\rho_0 \Om^2 (r^2 - a_-^2)}{2 p_0} \right) \quad \text{for} \;\; r \lesssim a_-.
	\eeq


\pt {\bf Outside the vortex $r \gtrsim a_+ = a+\eps$:} Here $w_z(r) \approx 0$ so the velocity decays as $v_{\phi}(r) = a_+ v_\phi(a_+)/r$. Again, ignoring the regularization term, the steady state density is determined by (\ref{e:isothermal-steady-state-vortex-eqn}):
	\beq
	\frac{\rho'(r)}{\rho(r)} = \frac{\rho_0 a_+^2 v_\phi(a_+)^2}{p_0} \ov{r^3}.
	\eeq
$\rho(r)$ monotonically increases from its value at the outer edge $\rho(a_+)$ to an asymptotic value $\rho(\infty)$ 
	\beq
	\rho(r)	= \rho(a_+) \exp \left( \frac{\rho_0 v_\phi(a_+)^2 \left(r^2 - a_+^2 \right)}{2 p_0 r^2} \right)
	= \rho(\infty) \exp \left(- \frac{\rho_0 v_\phi(a_+)^2 a_+^2}{2 p_0 r^2} \right) \quad \text{for} \;\; r \gtrsim a_+.
	\eeq
Even in this approximation, $\rho$ in the exterior depends on the regularization via $v_\phi(a_+)$.


\pt {\bf Transition layer $a_- \lesssim r \lesssim a_+$:} Here $w_z(r)$ (\ref{e:vorticity-profile-tornado}) rapidly falls from $w_z(0)$ to $0$. $v_\phi(r)$ is given by (\ref{e:vel-profile-tornado-exact}). $\rho$ is determined by 
	\beq
	\frac{\rho v_\phi^2}{r} = \frac{p_0}{\rho_0} \rho'(r) + \frac{\la^2 \rho}{2} \dd{w_z^2}{r}.
	\label{e:steady-R-Euler-for-rot-vortex}
	\eeq
To find the density we integrate this equation from $a_-$ to $r < a_+$ using the relation $\la^2 \rho = $ constant:
	\beq
	\int_{a_-}^r \frac{\rho v_\phi^2}{r} dr' = \frac{p_0}{\rho_0} \left[ \rho(r) - \rho(a_-) \right] + \frac{\la^2 \rho}{2} \left( w_z^2(r) - w_z^2(a_-) \right).	
	\eeq
Since the layer is thin ($\eps \ll a$) and $\rho$, $v_\phi$ are continuous across the layer, we may ignore the LHS. Thus the rapid decrease in $w_z$ must be compensated by a corresponding increase in $\rho$ across the layer
	\beq
	(p_0/\rho_0) \left[ \rho(r) - \rho(a_-) \right] \approx -  (\la^2 \rho/2) \left[ w_z^2(r) - w_z^2(a_-) \right].
	\label{e:density-vorticity-balance-across-layer}	
	\eeq
The increase in $\rho$ is not as rapid as the fall in $w_z$ since the latter is multiplied $\la^2$. For our vorticity profile (\ref{e:vorticity-profile-tornado}), taking $w_z(a_-) \approx w_z(0) = 2 \Om$, we get $\rho(r)$ in the transition layer
	\beq
	 \rho(r) \approx \rho(a_-) + \frac{2 (\Om \la_0)^2 \rho_0^2}{p_0} \left[1 - \frac{(1 - \tanh((r-a)/\eps))^2}{(1 + \tanh(a/\eps))^2} \right].	
	\eeq
In particular, $\rho(a_+)$ exceeds $\rho(a_-)$ by an amount determined by the regularization
	\beq
	\rho(a_+) \approx \rho(a_-) + \frac{2 (\Om \la_0)^2 \rho_0^2}{p_0} \left[1 - \frac{[1 - \tanh(1)]^2}{(1 + \tanh(a/\eps))^2} \right] \approx \rho(a_-) + 2 M^2  \rho_0 \quad \text{for} \quad \eps \ll a.
	\eeq
We see that for $\eps \ll a$ (vortex edge thin compared to core size), the twirl force causes an increase in density across the boundary layer by an amount controlled by the `twirl Mach number' $M = \la_0 \Om/c_s$ where $c_s = \sqrt{p_0/\rho_0}$ is the isothermal sound speed.

The steady R-Euler equation (\ref{e:steady-R-Euler-for-rot-vortex}) for the vortex is similar to Schr\"odinger's stationary equation for a non-relativistic quantum particle in a 1d delta potential: $E \psi(x) = - g \del(x) \psi(x) - (\hbar^2/2m)\: \psi''(x)$. $E \psi$ is like $\rho v_\phi^2/r$ on the LHS of (\ref{e:steady-R-Euler-for-rot-vortex}). The potential $- g \del(x) \psi(x)$ and kinetic $-(\hbar^2/2m) \psi''(x)$ terms mimic the pressure $(p_0/\rho_0) \rho'$ and twirl $\la^2 \rho \:(w_z^2)'/2$ terms respectively. The kinetic and twirl terms are both singular perturbations. The free particle regions $x < 0$ and $x > 0$ are like the interior and exterior of the vortex. The bound-state wave function is $\psi(x) = A \exp(- \kappa |x|)$ with $\kappa = \sqrt{-2mE}/\hbar$, so $\psi'$ has a jump discontinuity at $x=0$. The boundary layer is like the point $x=0$ where the delta potential is supported. Just as we integrated R-Euler across the transition layer, we integrate Schr\"odinger in a neighbourhood of $x=0$ to get $\psi'(\eps) - \psi'(-\eps) = - (2mg/\hbar^2)\: \psi(0)$. The discontinuity in $\psi'$ is determined by $\psi(0)$, just as the increase in $\rho$ across the layer is fixed by the corresponding drop in $w_z$ (\ref{e:density-vorticity-balance-across-layer}). Finally, $\la > 0$ regularizes Euler flow just as $\hbar > 0$  regularizes the classical theory, ensuring $E_{\rm gs} = - m g^2/2 \hbar^2$ is bounded below.

\subsection{A steady columnar vortex in conjunction with an MHD pinch}

A similar analysis in R-MHD involves specifying in addition to the above $w_z(r)$, a vertical (axial) current $j_z(r)$. The (solenoidal) azimuthal $B_\phi(r)$ associated with it is determined from $\mu_0 \bfj = \grad \times \bfB$, i.e. by integrating $\mu_0 j_z = {r}^{-1} (r B_\phi(r))'$. Assuming $r B_\phi(r)$ vanishes along the axis, $B_\phi(r) = {r}^{-1} \int_0^r \mu_0 s j_z(s) ds$. As in R-Euler above, the steady continuity equation $\grad \cdot (\rho(r) v_\phi(r)) \equiv 0$ is identically satisfied. The steady R-Faraday equation $\grad \times (\bfv_* \times \bfB) = 0$ is also identically satisfied since both $\bfv_* = (v_\phi - \la^2 w_z') \hat \phi$ and $\bfB$ are parallel. Thus the electric field is zero. In R-MHD, the steady momentum equation (\ref{e:R-MHD-Euler-v}) only has a non-trivial radial component. Under isothermal conditions ($p/p_0 =  \rho/\rho_0$) it becomes
	\beq
	\frac{p_0}{\rho_0} \rho' - \frac{v_\phi^2}{r} \rho = - \half \la_0^2 \rho_0 (w_z^2)' - \frac{B_\phi}{\mu_0 r} (r B_\phi)'.
	\label{e:mhd-pinch-vortex-ODE-for-rho}
	\eeq
In (\ref{e:mhd-pinch-vortex-ODE-for-rho}) the inhomogeneous term on the RHS is modified by the Lorentz force relative to (\ref{e:isothermal-steady-state-vortex-eqn}). The latter is always radially inwards (`pinching') whereas the twirl term is outwards for radially decreasing vorticity. Furthermore, the twirl term could be small for $\la_0 \ll a$. Thus the radial density variation in this magnetized columnar pinch could differ from R-Euler. For any given current and vorticity profiles (\ref{e:mhd-pinch-vortex-ODE-for-rho}) can be integrated to find $\rho(r)$ as we did in (\ref{e:soln-linear-inhom-1st-Order-ode}).

Another case of interest in R-MHD is a magnetized columnar vortex with an axial skin current. Thus we assume $j_z(r)$ is localized between $a - c/\om_{pe}$ and $a + c/\om_{pe}$ where $c/\om_{pe}$ is the electron collisionless skin depth and $\la \approx c/\om_{pe}$. In this case, in the interior $r < a_-$ we have the previous (tornado) interior solution with $B_\phi = 0$. In the exterior solution, $B_\phi(r) \approx \mu_0 I/2\pi r$ for $r \geq a_+$. The effect of the Lorentz force in the skin is seen from (\ref{e:mhd-pinch-vortex-ODE-for-rho}) to be opposite to that of the twirl term. The exclusion of the magnetic field within the vortex is reminiscent of the Meissner effect in superconductivity. Axial magnetic fields (screw pinch) and flows with the same symmetries (i.e., purely radial dependence) may be readily incorporated in the framework presented since the momentum equation remains purely radial and the continuity and R-Faraday laws are identically satisfied.

\subsection{Simple model for channel flow using regularized equations}
\label{s:channel-flow}

We consider flow along an infinitely long (in the $x$ direction) and infinitely wide (in the $z$ direction) channel. The channel extends from $y=0$ to a height of $y=a$. We seek a steady state solution of the regularized equations with velocity field $\bfv = (u(y),0,0)$ and density $\rho$ a function of $y$ alone. i.e., velocity and density vary with height but are translation invariant along the length and breadth of the channel. The steady state continuity equation $\grad \cdot (\rho \bfv) = 0$ is identically satisfied since $\pdr_x(\rho(y) u(y)) = 0$. For our velocity field the advection term in the momentum equation $\bfv \cdot \grad \bfv$ is identically zero and \footnote{Note that subscripts denote derivatives.}
	\beq
	\bfw = - u_y \hat z, \quad
	\bfw \times \bfv = - u u_y \hat y, \quad
	\grad \times \bfw = - u_{yy} \hat x
	\quad \text{and} \quad
	\bfT = \bfw \times (\grad \times \bfw) = u_y u_{yy} \hat y.
	\eeq
 So only the $\hat y$ component of the momentum equation survives:
	\beq
	\la^2 u_y u_{yy} = - \pdr_y{h(\rho(y))}.
	\label{e:reg-euler-channel-flow}
	\eeq
In other words, the steady state equations are underdetermined, we have a single second order non-linear ODE for both $u(y)$ and $\rho(y)$. So given $u(y)$ and a suitable boundary value, say $\rho(0)$, we may determine the density profile. In particular, in the unregularized theory ($\la = 0$), the Euler equation  simply states that density must be a constant since $\pdr_y h(\rho) = 0$. As a consequence, the unregularized velocity $u$ can be an arbitrary function of $y$ (satisfying appropriate boundary conditions). So the regularization introduces a non-trivial dependence of $\rho(y)$ on $u(y)$.

\noindent {\em Remark on energy conservation:} For steady flow, local conservation of energy becomes $\grad \cdot \bff = 0$. $\grad \cdot \bff \equiv 0$ for channel flow since the energy current points along $\hat x$ but depends only on $y$:
	\beq
	\bff = \rho \sigma \bfv + \la^2 \rho ((\bfw \times \bfv) \times \bfw) + \la^4 \rho \bfT \times \bfw \;
	= \; \left[ \rho \left(h + \half u^2 \right) u + \la^2 \rho u u_y^2 - \la^4 \rho u_y^2 u_{yy} \right] \hat x.
	\eeq
Furthermore, the energy flux across the upper and lower walls of the channel vanish ($\bff \cdot \hat z = 0$). So energy is conserved even though our flow does {\em not} satisfy the BC $\bfw \times \hat n = 0$ that we obtained as a sufficient condition for energy conservation in \S \ref{s:cons-laws}.

\subsubsection{Isothermal channel flow}

For isothermal flow specific enthalpy is $h = (p_0/\rho_0) \log(\rho/\rho_0)$. Since $\la^2 \rho$ is a constant, the R-Euler equation (\ref{e:reg-euler-channel-flow}) becomes 
	\beq
	\frac{\la^2 \rho}{2} \dd{u_y^2}{y} = - (p_0/\rho_0) \rho_y \quad
	\text{or} \quad
	\pdr_y \left(\half \la^2 \rho u_y^2 + \frac{p_0 \rho}{\rho_0} \right) = 0.
	\label{e:R-euler-isotherm-channel-flow}
	\eeq
As $\bfw = - u_y \hat z$, this Bernoulli-like equation states constancy of the sum of enstrophic and compressional energy densities with height. The kinetic energy contribution is absent due to the assumption of a purely longitudinal velocity field that varies only with height: recall that the advection term $\bfv \cdot \grad \bfv$ is identically zero. As a consequence, this Bernoulli-like equation is very different in character from the usual one, which involves the kinetic energy of the flow and the compressional energy along streamlines. In that case, the pressure along a streamline is lower where the velocity is higher. In the present case, there is no variation of any quantity along streamlines, but only in the $y$-direction. We find that the density, and hence the pressure, is higher where the vorticity is higher! This is fundamentally a consequence of the regularizing ``twirl acceleration''.

An exact first integral of the above equation is $\half \la^2 \rho u_y^2 + (p_0/\rho_0) \rho = K$, where $K$ is an integration constant. We make use of the constitutive relation $\la^2 \rho=\la_{0}^{2}\rho_{0}$, where both $\la_0$ and $\rho_0$ are taken at the base of the channel $y=0$, and evaluate the equation there to obtain $K = \ov{2}\la_{0}^2 \rho_{0} u_y^2(0) + p_0$. For convenience, we use the reference values $p_0 = p(0)$ and $\rho_0 = \rho(0)$ to be the pressure and density at $y=0$. For instance, we consider the example of a parabolic velocity profile:
	\beq
	u(y) = 4u_{\rm max}\left[\frac{y}{a} \left(1-\frac{y}{a} \right)\right], 
	\label{e:parabolic-vel-profile}
	\eeq
where $u_{\rm max}=u(a/2)$ is the flow velocity midway up the channel, and $u(0)=0$. It follows that 
	\beq
	\dd{u}{y} = u_{y} = 4\frac{u_{\rm max}}{a}\left[1-2 \left(\frac{y}{a} \right)\right] \quad
	\text{and so} \quad u_{y}(0)=4 \frac{u_{\rm max}}{a}.
	\eeq
Thus the Bernoulli constant $K = 8\rho_0 u_{\rm max}^{2} (\la_0/a)^2 +p_{0}$. Substitution in the Bernoulli integral leads to the density profile:
      \beq
        \frac{ \rho}{\rho_0} = 1 + 32 \left(\frac{\rho_0 u_{\rm max}^{2}}{p_{0}}\right) \left(\frac{\la_0}{a} \right)^2 \left(\frac{y}{a}\right) \left(1-\frac{y}{a} \right).
      \eeq
The resulting density profile is also parabolic. The density increases from $\rho(0)=\rho_0$ at the bottom of the channel to a maximum value half way up the channel and decreases symmetrically back to $\rho_0$ at the top. Thus, we have,
	\beq
	\frac{\rho_{\rm max}}{\rho_{0}}= 1 + 8 \left(\frac{\rho_{0} u_{\rm max}^{2}}{p_{0}} \right) \left(\frac{\la_0}{a} \right)^{2}.
	\eeq
We note that in isothermal conditions, we may write, $c_{s}^{2} = p_{0}/\rho_{0}$, where $c_s$ is the isothermal sound-speed. Since the Mach number of the flow along the centre is, $M^{2}= u_{\rm max}^2/c_s^2$, we have the relation:
	\beq
	\frac{\rho_{\rm max}}{\rho_{0}}= 1 + 8 M^{2} \left(\frac{\la_0}{a} \right)^{2}.
	\eeq
$M$ can take any value in principle. The second factor, $(\la_0/a)^2$, is by assumption a very small number. For moderate Mach numbers, the density increase is rather small. The flow superficially resembles Poiseuille flow and satisfies the same boundary conditions, but is strictly non-dissipative. It should be noted that Poiseuille flow involves a constant pressure {\em gradient along the flow} driving the latter against viscosity, whereas in the present case, there is no variation of any quantity along the flow.

\subsubsection{Adiabatic channel flow}
\label{s:channel-adibatic}

For adiabatic channel flow $(p/p_0) = (\rho/\rho_0)^\gamma$ and $h = \frac{\gamma}{\gamma-1} \frac{p}{\rho}$. We employ the same parabolic velocity profile as in the isothermal case. Since $\la^2 \rho$ is a constant, the R-Euler equation (\ref{e:reg-euler-channel-flow}) becomes a ``twirl force'' Bernoulli's equation:
	\beq
	\frac{\la^2 \rho}{2} \dd{u_y^2}{y} = - \rho \dd{h}{y} \quad
	\text{or} \quad
	\pdr_y \left(\half \la^2 \rho u_y^2 + p \right) = 0.
	\eeq
As before we obtain the exact first integral $\half \la^2 \rho u_y^2 + p = K$. Making use of the constitutive relation we evaluate the Bernoulli constant at $y =0$ by choosing $p_0 = p(0)$ and $\rho_{0}=\rho(0)$:
	\beq
	K = \half \la_{0}^2 \rho_{0} u_y(0)^2 +  p_{0}.
	\eeq	
Substitution in the exact integral to eliminate $K$, we obtain the pressure (and density) distributions
	\beq
	\frac{p}{p_0} = \left( \frac{\rho}{\rho_0} \right)^\gamma = 1 + 32 \left(\frac{\rho_{0} u_{\rm max}^2}{p_0} \right) \left(\frac{\la_0}{a} \right)^2 \left( \frac{y}{a} \right) \left(1 - \frac{y}{a} \right).
	\eeq
For adiabatic flow, $p/p_0$ varies with height in exactly the same way as $\rho/\rho_0$ in the isothermal case!

\subsection{Isothermal plane vortex sheet}

As a typical illustrative example, we consider a steady plane vortex sheet under isothermal conditions. The vortex sheet is assumed to lie in the $x$-$z$ plane and to have a thickness $\tht$ in the $y$-direction. We assume the velocity points in the $x$-direction $\bfv = (u(y),0,0)$ and approaches {\em different} asymptotic values $u_\pm$ as $y \to \pm \infty$. The density $\rho$ is also assumed to vary only with height $y$. Exactly as in channel flow, we obtain the equation for time-independent flows (\ref{e:R-euler-isotherm-channel-flow}):
	\beq
	\pdr_y \left(\half \la^2 \rho u_y^2 + p_{0} \frac{\rho}{\rho_{0}}\right) = 0.
	\eeq
The steady state is not unique and this equation can be used to find the density profile for any given vorticity profile. To model a vortex sheet of thickness $\tht$ we take the vorticity profile in $y$ to be given by
	\beq
	u_{y} = \Delta u \: \left(\frac{\theta}{\pi} \right) \left[\frac{1}{\theta^{2}+y^{2}} \right] \quad \text{where} \quad \bfw = - u_y(y) \: \hat z
	\eeq
Here $\Delta u = u_{+} - u_{-}$ and $w_{0} = -\Delta u/\pi \theta$ is the $z$-component of vorticity on the sheet. We obtain, as usual, the first integral,
	\beq
	\half \la_{0}^2 \rho_{0} u_y^2 + \frac{p_{0} \rho}{\rho_0} = K.
	\eeq
The suffix in this instance refers to quantities on the sheet ($y=0$). The Bernoulli constant $K = p_0 + \half \rho_0 (\Delta u)^2 \, \left( \frac{\la_0^2}{\pi^2 \tht^2} \right).$
We obtain the velocity profile by integration:
	\beq
	u(y) = u_- + (\Delta u) \left(\frac{\theta}{\pi} \right) \int_{-\infty}^{y} \frac{d\mu}{\theta^{2}+\mu^{2}} = u_{-} + (\Delta u) \left[\half + \frac{1}{\pi}\arctan\left(\frac{y}{\theta} \right) \right].
	\eeq
Assuming $u_+ > u_-$, the velocity monotonically increases from $u_-$ to $u_+$ with increasing height $y$. Moreover, the velocity on the sheet $u(0) = \half(u_- + u_+)$ is the average of its asymptotic values. The density profile follows from the first integral:
	\beq
	 \frac{\rho}{\rho_0} = 1 + \half \left(\frac{\la_0 }{\pi \theta} \right)^2 \left[\frac{\rho_0 (\Delta u)^{2}}{p_0} \right] \left(1- \left[\frac{\theta^{2}}{\theta^{2}+y^{2}} \right]^{2} \right).
	\eeq
In particular, the asymptotic densities are
	\beq
	\frac{\rho_{\pm \infty}}{\rho_0} = 1 + \half \left( \frac{\la_0}{\pi \theta} \right)^2 \left[\frac{\rho_0 (\Delta u)^2}{p_0} \right].
	\eeq
Thus, the density is decreased at the sheet relative to the values at $\pm \infty$. If the sheet thickness $\theta \gg \la_0/\pi$, the decrease is not significant. If the thickness is comparable to the regularizing length $\la_0$, the density decrease at the sheet can be considerable, depending upon the `relative flow Mach number' defined as, $(\Delta M)^{2} = (\rho_0/p_0)(\Delta u)^2$. Unlike velocity, the density increases from the sheet to the same asymptotic values on either side of the sheet ($y = \pm \infty$), reflecting the symmetry of the assumed vorticity profile. This is similar to the rotating vortex/tornado model (\ref{s:modeling-vortex}) where an increase in density outwards from the core of the vortex is balanced by a corresponding decrease in vorticity.

\subsection{Regularized plane flow}
\label{s:plane-flow}

It is interesting to consider the R-Euler equations for flow on the $x$-$y$ plane with $\bfv = (u(x,y),v(x,y),0)$. First consider incompressible flow $\grad \cdot \bfv = 0$ with constant $\rho$, and hence constant $\la$. The condition $u_x + v_y = 0$ is solved in terms of a stream function $u = - \psi_y$ and $v = \psi_x$ (subscripts denote partial derivatives). Vorticity points vertically $\bfw = w \hat z$ with $w = v_x - u_y = \Delta \psi$. The twirl acceleration is proportional to the gradient of $w^2$:
	\beq
	\bfw \times (\grad \times \bfw) = w \hat z \times (w_y \hat x - w_x \hat y) = w \grad w = (1/2) \grad w^2.
	\eeq
So for constant $\la$, the incompressible 2d R-Euler equation becomes 
	\beq
	\pdr_t \bfv + \bfw \times \bfv = - \grad \left( \sigma + (1/2) \la^2 w^2 \right).
	\eeq
The twirl acceleration term may be absorbed into a redefinition of stagnation enthalpy $\sigma$. In particular, the regularization drops out of the evolution equation for vorticity $\bfw_t + \grad \times (\bfw \times \bfv) = 0$, which states that $\bfw$ is frozen into the incompressible flow field $\bf v$. In other words, for incompressible plane flow, the regularization plays no role in vortical dynamics. This is to be expected: enstrophy $\int w^2 \: dx \, dy$ is bounded in incompressible 2d flows (indeed it is conserved) and there is no vortex stretching. 

By contrast, compressible flow on a plane is richer. For simplicity, consider steady flow with ${\bf v} = u \hat x + v \hat{y}$, ${\bf w} = w(x,y) \hat{z}$ and $\nabla \times {\bf w}=w_{y}\hat{x}-w_{x}\hat{y}$. The continuity equation $\grad \cdot (\rho \bfv) = 0$ is solved using a stream function: $\rho u = - \psi_{y}$, $\rho v = \psi_{x}$. The R-Euler equation becomes
	\beq
	 w u =-\sigma_y - \la^2 ww_y \quad \text{and} \quad
	 -w v = -\sigma_x -\la^2 w w_x.
	\eeq	
Using the relation, $\sigma = h + \half \bfv^2 = h + (\grad \psi)^2/2 \rho^2$, we obtain the equivalent equations:
	\beq
	 w \psi_{y} = \rho \left[ h + \frac{1}{2\rho^2} (\grad \psi)^2 \right]_y + \la^2 \rho ww_y \quad \text{and} \quad
	 w \psi_{x} = \rho \left[ h + \frac{1}{2\rho^2} (\grad \psi)^2 \right]_x + \la^2 \rho ww_x.
	\label{e:plane-flow-R-euler-stream-fn}
	\eeq
From the constitutive relation $\la^2 \rho = \la_0^2 \rho_0$ is a constant. Assuming $w$ is not zero, we get
	\beq
	\psi_y = \frac{\rho}{w} \left[ h + \frac{1}{2\rho^2}(\grad \psi)^2 \right]_y + \la_0^2 \rho_0 w_y
	\quad \text{and} \quad
	 \psi_x = \frac{\rho}{w} \left[ h + \frac{1}{2\rho^{2}}(\grad \psi)^2 \right]_x + \la_0^2 \rho_0 w_x.
	\eeq
Differentiating the first equation in $x$, the second in $y$ and subtracting, we see that, $\rho/w$ has a vanishing Jacobian with $\sig = h + (\grad \psi)^2/2\rho^2$. Thus the equations say that $\sig$ is an arbitrary function $\Sigma$ of $\rho/w$. Setting $\rho/w = \Theta$, we get
	\beq
	\psi_y = \Theta \: \Sigma'(\Theta) \: \Theta_y + \la_0^2 \rho_0 w_y \quad \text{and} \quad
	\psi_x = \Theta \: \Sigma'(\Theta) \: \Theta_x + \la_0^2 \rho_0 w_x.
	\eeq
It follows that we may integrate the equations to get $\psi = \la_0^2 \rho_0 \, w + H(w/\rho)$. Here $H$ is an arbitrary function related to $\Sigma$ through a quadrature $H = \int \Sigma'(\Theta) \Theta d \Theta$. Since $w = (\psi_x/\rho)_x + (\psi_y/\rho)_y$, a specification of $H$ reduces this to a non-linear PDE for the two unknowns $\psi$ and $\rho$. The under-determinacy of this system is a common feature of the {\em steady} compressible R-Euler equations.

Alternatively, suppose we do not divide the R-Euler equation by $w$ [which could vanish in a region] but simply note that differentiating the first equation of (\ref{e:plane-flow-R-euler-stream-fn}) in $x$ and the second in $y$, and subtracting, we get an equation involving two Jacobians:
	\beq
\frac{\partial (w,\psi)}{\partial (x,y)}=\frac{\partial (\rho,\sig)}{\partial (x,y)}
	\eeq
We may consider the ansatz $w = J(\psi)$ where $J$ is an arbitrary function so that the LHS vanishes. For the RHS to vanish, $\sigma$ must be a function of $\rho$, say $\sigma = Z(\rho)$. Thus the `compatibility condition' on (\ref{e:plane-flow-R-euler-stream-fn}) can be satisfied by introducing two arbitrary functions $J$ and $Z$. There may be many other, much more complicated solutions of (\ref{e:plane-flow-R-euler-stream-fn}) but we do not investigate them here. Given, $J$ and $Z$ and the equation of state $p = p(\rho)$ we can eliminate $w$ and $p$ to reduce (\ref{e:plane-flow-R-euler-stream-fn}) to two nonlinear PDEs for $\psi$ and $\rho$. The simplest case could be for example, $Z(\rho) = Z_0$, a constant in which case (\ref{e:plane-flow-R-euler-stream-fn}) becomes $\psi = \la_0^2 \rho_0 J(\psi)$ upon absorbing a constant into $\psi$. Once $J$ is specified and $\psi$ determined, $\rho$ is obtained from $\sigma(\rho) = Z_0$ given an equation of state. 

Another possible solvable case occurs for subsonic flows at relatively low Mach numbers. In $0^{\rm th}$ order, we may take $w=0$. There then exists a velocity potential $\phi(x,y)$ such that $\psi$ is its conjugate function. In zeroth order, $\rho$ is constant and hence $\phi$ is clearly the standard incompressible Euler velocity potential. The pressure variations are then determined by the constancy of $\sigma$. Evidently, they must be of order the square of the Mach number. The full nonlinear equation must then be linearised about this basic irrotational flow to calculate the vorticity in the next order. We do not pursue this here.

\subsection{Incompressible 3-d axisymmetric vortex flow}

We consider the steady, incompressible R-Euler equations in an axi-symmetric geometry. We have in mind applications to typical exterior flows where a spherical or cylindrical vortex capsule moves along the axis (e.g., Hill's spherical vortex). For simplicity, we consider incompressible flow $\grad \cdot \bfv = 0$ so both $\rho$ and $\la$ are a constant. We choose the axis to point along $\hat z$ and use cylindrical coordinates $(r,\phi,z)$. Axi-symmetry here means $\bfv$ does not have an azimuthal component ($v_\phi = 0$) and that pressure, $v_r$ and $v_z$ are independent of $\phi$. This is to be contrasted with the rotating vortex of \S \ref{s:modeling-vortex}, where the velocity was purely azimuthal. The continuity equation $\grad \cdot \bfv = r^{-1}\pdr_r (r v_r) + \pdr_z v_z = 0$ can be solved in terms of a stream function\footnote{Beware! Subscripts on $\psi, w$ denote partial derivatives, while those on $v$ denote components.}
	\beq
	\bfv = - \grad \times \left( r^{-1} \psi(r,z) \hat \phi \right) \quad
	\text{or} \quad
	v_r =  \psi_z/r \quad \text{and} \quad
	v_z = - \psi_r/r.
	\eeq
The vorticity is purely azimuthal ($\bfw = w \hat \phi$) while the pressure gradient, vortex and twirl accelerations have no azimuthal components:
	\beq
	w = \left( \pdr_z v_r - \pdr_r v_z \right) = \ov{r} \psi_{zz} + \pdr_r \left(\ov{r} \psi_r \right) = \grad^2 \left( \frac{\psi}{r} \right) - \frac{\psi}{r^3}
	\quad \text{and} \quad \bfw \times \bfv =  w v_z \hat r - w v_r \hat z \quad
	\text{and} \quad
	\bfT = \frac{w}{r} (r w)_r \hat r + w w_z \hat z.
	\label{e:vorticity-azimuthal-3d-vortex}
	\eeq
Thus the steady R-Euler equations $\bfw \times \bfv = - \grad \sigma - \la^2 \bfT$ reduce to two component equations:
	\beq
	w v_z = - \sigma_r - \la^2 \frac{w}{r} (r w)_r \quad \text{and} \quad
	- w v_r = - \sigma_z - \la^2 w w_z.
	\eeq
Taking the curl of the R-Euler equation we may eliminate pressure. Expressing $\bfv$ in terms of its stream function $\psi$, we obtain
	\beq
	\dd{(w/r, \psi)}{(r,z)} = - \frac{\la^2}{r} \left( w^2 \right)_z.
	\label{e:jacobian-condition-for-R-euler}
	\eeq
This Jacobian condition can be simplified by working with $\bfv_*$ rather than $\bfv$. Recall that the steady R-Euler equation is $\bfw \times \bfv_* = - \grad \sigma$ and the R-vorticity equation [steady freezing-in of $\bfw$ into $\bfv_*$] is 
	\beq
	\grad \times (\bfw \times \bfv_*) = \grad \times (w v^*_z \hat r - w v^*_r \hat z) = \left[ (w v^*_z)_z + (w v^*_r)_r \right] \hat \phi = 0.
	\eeq
Since $\bfv_*$ is divergence-free, we may express it in terms of a stream function $\psi^*$
	\beq
	v^* = - \grad \times \left(\frac{\psi^*}{r} \hat \phi \right) \quad \text{or} \quad v^*_r = \ov{r} \psi^*_z \quad \text{and} \quad
	v^*_z = - \ov{r} \psi^*_r.
	\eeq
In terms of $\psi^*$, the R-vorticity equation reduces to a vanishing Jacobian condition:
	\beq
	\left(- \frac{w}{r} \psi^*_r \right)_z + \left( \frac{w}{r} \psi^*_z \right)_r  = 0 \quad \text{or} \quad \dd{(w/r,\psi^*)}{(r,z)} = 0.
	\label{e:jacobian-condition-psi*}
	\eeq
Thus $\psi^*$ can be an arbitrary function of $w/r$ or $w \equiv 0$. To see what this means for $\psi$ we write $\bfv = \bfv_* - \la^2 (\grad \times \bfw)$ in components and read off the relation $\psi = \psi^* + \la^2 r w$ (upto an additive constant). Thus (\ref{e:jacobian-condition-psi*}) implies a vanishing Jacobian condition on $\psi$
	\beq
\dd{(w/r, \psi - \la^2 r w)}{(r,z)} = 0.
	\label{e:vanishing-jacobian-psi}
	\eeq
One checks that this is equivalent to (\ref{e:jacobian-condition-for-R-euler}). Thus $w/r$ must be an arbitrary function of $\psi - \la^2 r w$ or $w \equiv 0$. In the latter case (irrotational incompressible flow) the regularization plays no role and $\psi$ must satisfy\footnote{In the case of irrotational flow, we could work in terms of a velocity potential which is harmonic, unlike the stream function.}
	\beq
	w = \ov{r} \psi_{zz} + \left(\ov{r} \psi_r \right)_r = 0 \quad \text{or} \quad
	\grad^2 \left( \frac{\psi}{r} \right) = \frac{\psi}{r^3}.
	\label{e:w-eq-0-in-terms-of-psi}
	\eeq
Alternatively, $w/r$ must be constant on level surfaces of $\psi - \la^2 r w$, i.e. $w/r = H(\psi - \la^2 r w)$ where $H$ is an arbitrary function. This is an exact generalisation of Lamb's Eq.(13), Art. 165, p. 245 \cite{lamb} when $\lambda=0$. The appearance of an arbitrary function is another instance of the steady underdeterminacy of the R-Euler equation. Writing $w = r^{-1} \psi_{zz} + (r^{-1} \psi_r)_r$ we get a (generally non-linear) 2nd order PDE for $\psi$. Consider the simplest case where $H(g) = A - B g$ is a linear function ($[B] = 1/L^4$ and $[A] = 1/LT$). Then $\psi(r,z)$ must satisfy a 2nd order inhomogeneous linear PDE
	\beq
	\frac{w}{r} =  A - B \left[ \psi - \la^2 r w \right] \quad 
	\imply \quad
	\left(1 - \la^2 B r^2 \right) \left[ \ov{r^2} \psi_{zz} + \ov{r} \left( \ov{r} \psi_r \right)_r  \right] = A - B \psi.
	\eeq
The differential operator may be expressed in a more `invariant' manner in terms of the Laplacian of $\psi/r$:
	\beq
	\left[ \grad^2 - \ov{r^2} + \frac{B r^2}{1 - \la^2 B r^2} \right] \left( \frac{\psi}{r} \right) =  \frac{Ar}{1 - \la^2 B r^2}.
	\label{e:inhom-SE-for-psibyr}
	\eeq
When $A = 0$, (\ref{e:inhom-SE-for-psibyr}) becomes homogeneous and resembles the time-independent Schr\"odinger equation for a zero energy particle with wave function $f = \psi/r$ in a cylindrically symmetric non-central potential $V = r^{-2} - B r^2/(1 - \la^2 B r^2)$. If $B < 0$, then the potential is strictly positive and we would not expect any zero energy eigenstate. So when $A = 0$, we take $B > 0$.

\subsubsection{Spherical vortex}

The above equations may be used to model a spherical vortex of radius $a$ moving along the axis of symmetry in an irrotational exterior flow. An example of such irrotational flow occurs in the exterior of Hill's spherical vortex where
	\beq
	\psi = \half \: V_\infty r^2 \left[1- a^3/R^3 \right] \quad \text{for} \quad
	R^2 \equiv r^2 + z^2 > a^2.
	\eeq
This describes uniform flow far from the sphere, i.e. $v_r \to 0$ and $v_z \to - V_\infty$ as $R \to \infty$ (we go to the vortex frame and allow the fluid flow at infinity to be uniform). Furthermore, $\psi=0$ is a stream surface and hence the flow is tangential to the surface $R = a$. Within the sphere, if we choose $B=0$ in (\ref{e:inhom-SE-for-psibyr}), the regularization plays no role and we have to solve
	\beq
	\psi_{zz} + r \left(\psi_r/r \right)_r = A r^2.
	\eeq
This has a polynomial solution $\psi=\half Ar^{2} [a^{2}-r^{2}-z^{2}]$ vanishing on $R=a$. Continuity of velocity across $R =a$ implies
$A = -(3/2a^{2})V_{\infty}$. This constitutes Hill's famous ``spherical vortex'' solution. However, this makes $w_{\phi}$ {\it discontinuous} on $R=a$.
       
On the other hand, we could have chosen $A=0$ and left $B$ arbitrary in the interior. Then in spherical polar coordinates ($r = R \sin \tht, z = R \cos \tht$) (\ref{e:inhom-SE-for-psibyr}) becomes a Schrodinger equation for $f = \psi/r$
	\beq
	- \ov{R^2} \pdr_R \left(R^2 \dd{f}{R} \right) - \ov{R^2 \sin \tht} \pdr_\tht \left(\sin \tht \dd{f}{\tht} \right) + \left[ \ov{R^2 \sin^2 \tht} - \frac{B R^2 \sin^2 \tht}{1 - \la^2 B R^2 \sin^2 \tht} \right] f = 0.
	\label{e:SE-spherical-geom}
	\eeq
We must solve (\ref{e:SE-spherical-geom}) requiring $\psi=0$ on $R=a$ and regularity of $\psi$ at $R=0$. $B$ must then be chosen to match the outer solution. Note that since $w/r = -B(\psi - \la^2 r w)$, in this solution $w$ vanishes where $\psi$ does, and is therefore {\it rendered continuous} at the boundary (even for $\la = 0$), unlike in Hill's solution. We do not pursue here an explicit solution of (\ref{e:SE-spherical-geom}) for the regularized version of Hill's spherical vortex\footnote{Separation of variables does not work in (\ref{e:SE-spherical-geom}) since the `potential' $V$ depends on both $R$ and $\tht$.} but instead consider a cylindrical geometry where an explicit solution illustrating key features is easily found.

\subsubsection{Cylindrical vortex}

As the simplest special case of the above equations (\ref{e:vanishing-jacobian-psi}), we consider a cylindrical vortex (pipe-like flow). We imagine a flow with $v_\phi = 0$ as above, that is irrotational outside an infinite circular cylinder with axis along $z$ and with radius $a$. Vorticity is purely azimuthal inside the cylinder. We require the stream function, its normal derivative and $w$ to be continuous across the cylindrical surface $r = a$. The simplest irrotational flow in the region $r > a$ is a uniform flow with speed $c$ in the $-\hat z$ direction:
	\beq
	\psi = \frac{c}{2} \left(r^2 -a^2 \right) \quad \text{with} \quad v_r = v_\phi = 0 \quad \text{and} \quad v_z = - c.
	\eeq
The additive constant is chosen so that $r=a$ is a stream surface on which $\psi$ vanishes.

For $r \leq a$, $\frac{w}{r} = H(\psi - \la^2 r w)$ where $H$ is an arbitrary function. If $H = A$ is a non-zero constant, then $w(r=a^-) = a A$ cannot match the value $w = 0$ for $r > a$, so $H$ cannot be a constant. The next simplest possibility is a linear $H(g) = A - B g$. Choosing $A=0$ ensures that $w$ is continuous across the cylindrical surface $r=a$:
	\beq
	w(r,z) = - \frac{B r }{1 - \la^2 B r^2} \psi \quad \imply \quad
	w(r=a) = 0.
	\eeq
We get a zero energy Schr\"odinger eigenvalue equation for the `wave function' $f = \psi/r$ for $r \leq a$:
	\beq
	(- \grad^2 + V(r)) f = 0 \quad 
	\text{or} \quad
	- f_{zz} - \ov{r} (r f_r)_r + V(r) f(r) = 0
	\quad \text{where} \quad V(r) = \ov{r^2} - \frac{B r^2}{1 - \la^2 B r^2}.
	\eeq	
Unlike in the spherical vortex, the potential $V(r)$ is independent of both $\phi$ and $z$, so we may separate variables. $f$ could diverge at $r = 0$ in such a way that the stream function (or more importantly the velocity) is finite at $r = 0$. The BCs at $r=a$ are continuity of $\psi$ i.e. $\psi(r=a) = 0$ (which guarantees continuity of $w$) and its normal derivative $\psi_r$. The simplest interior solution is obtained by assuming that $\psi$ depends only on $r$ so that velocity is purely longitudinal $v_z = -r^{-1} \psi_r$. In this case the above Schrodinger-like equation reduces to a 2nd order linear ODE $- r^{-1} \left( r f_r \right)_r + V(r) f(r) = 0$ on the interval $0 \leq r \leq a$ with the BCs $\psi(r=a) = 0$ and $\psi'(r=a) = c a$.

\begin{figure}[h]       
    \begin{center}
    \includegraphics[width=4.5cm]{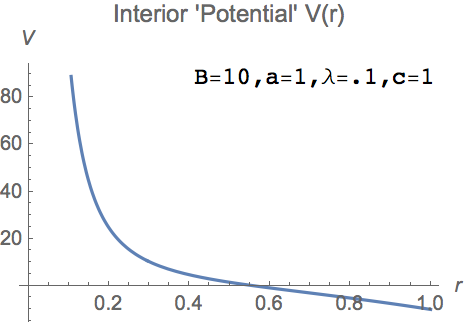}
    \hspace{0.5cm}
    \includegraphics[width=4.5cm]{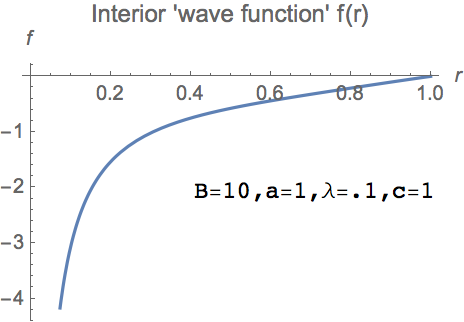}
    \hspace{0.5cm}
    \includegraphics[width=4.5cm]{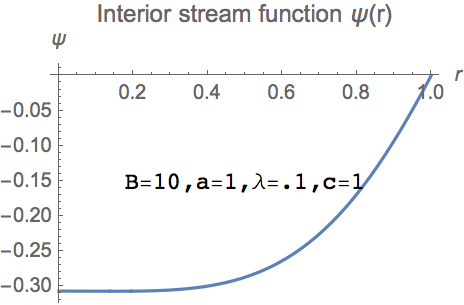}
    \caption{\footnotesize Interior potential $V(r)$ and wave function $f(r)$ for the Schrodinger-like equation for an infinite propagating axisymmetric cylindrical vortex of radius $a$ in a uniform external flow $-c \hat z$. The interior stream function $\psi(r) = r f(r)$ agrees with the exterior $\psi = (c/2) (r^2 - a^2)$ and its gradient at $r = a$.}
    \label{f:cyl-vortex-potn-wfn-stream-fn}
    \end{center}
\end{figure}
\begin{figure}[h]       
    \begin{center}
  \includegraphics[width=4.5cm]{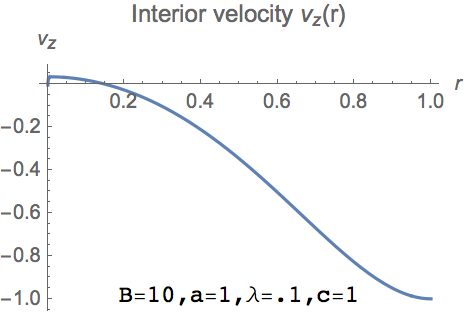}
    \hspace{0.5cm}
  	\includegraphics[width=4.5cm]{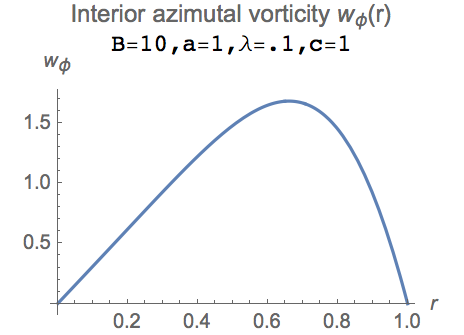}
    \caption{\footnotesize Velocity and vorticity profiles for cylindrical vortex of radius $a$. Velocity is longitudinal and increases in magnitude with increasing distance from the axis of the vortex and reaches the exterior flow value $-c \hat z$ at $r = a$. Voriticity is azimuthal for $r<a$ and matches the irrotational exterior flow at $r=a$. Radial derivative of vorticity is discontinuous across $r=a$ in our simple model.}
	\label{f:cyl-vortex-vel-vorticity}
    \end{center}
\end{figure}
This is a homogeneous second order ODE that may be put in the standard form
	\beq
	f'' + p(r) f' + q(r) f = 0, \quad \text{with} \quad p(r) = \ov{r} \quad \text{and} \quad q(r) = -V(r) = -\ov{r^2} + B r^2 + \la^2 B^2 r^4 + \ldots.
	\eeq
$p$ and $q$ have simple and double poles at $r=0$ and $q$ has simple poles at $r = \pm 1/\sqrt{\la^2 B}$, so the equation has $3$ regular singular points and could be transformed into the Hypergeometric equation. For sufficiently small $\la$, $r = 0$ is the only singular point in the physical region $0 \leq r \leq a$, around which the Frobenius method yields solutions. Making the ansatz $f(r) = r^\al \left[c_0 + c_1 \al + c_2 \al^2 + \ldots \right]$ with $c_0 \ne 0$ and comparing coefficients of $r^{\al - 2}$, we get the indicial equation $c_0 \al (\al -1) + \al c_0 - c_0 = 0$. Its roots $\al_1 = 1$ and $\al_2 = -1$ differ by an integer. Thus we have $2$ linearly independent solutions around $r=0$:
	\beqs
	f_1(r) &=& r \left[ c^{(1)}_0 + c^{(1)}_1 r + c^{(1)}_2 r^2 + \ldots \right] \quad \text{with} \quad c_0^{(1)} \ne 0  \quad \text{and} \cr
	f_2(r) &=& \ov{r} \left[ c_0^{(2)} + c^{(2)}_1 r + c^{(2)}_2 r^2 + \ldots \right] + c^{(2)} f_1(r) \, \log r \quad \text{with} \quad c_0^{(2)}, c^{(2)} \ne 0.
	\eeqs
Comparing coefficients of higher powers $r^{\al -2 + n}$ leads to recursion relations for $c_j^{(1)}$ and $c_k^{(2)}$. $f_1$ has a simple zero at $r=0$ while $f_2$ has a simple pole at $r=0$ in addition to a logarithmic branch cut ending at $r=0$. The solution of our boundary value problem with prescribed BC at $r=a$ is a linear combination of $f_1$ and $f_2$. A generic linear combination $f$ will diverge at $r=0$ like $c^{(2)}_0/r$. Thus we should expect the stream function $\psi = r f(r)$ to linearly approach a non-zero limit $c_0^{(2)}$ as $r \to 0$.

We have solved this ODE with the given BCs numerically. The results are illustrated Fig.\ref{f:cyl-vortex-potn-wfn-stream-fn},\ref{f:cyl-vortex-vel-vorticity} for cylinder radius $a = 1$, regularization length $\la = 1/10$, constant $B = 10$ and exterior flow speed $c = 1$. It is clear that $\psi(r), v_z(r)$ and $w(r)$ are all continuous at $r=a$. However the radial derivative of vorticity $w_r$ is discontinuous at $r =a$. A more careful treatment of a layer of thickness $\sim \la$ around $r=a$ should render $\pdr w/\pdr r$ continuous. Despite this discontinuity, the twirl acceleration $\bfT = \bfw \times (\grad \times \bfw) = r^{-1}w (r w)_r \hat r$ is continuous across the cylindrical surface since $w(a) = 0$. On the axis of the vortex, $\psi$ is divergent, though $v_z$ has a finite value, while $w$ vanishes there. In addition, we see that the radial derivative of $v_z$ is zero at $r=0$, as one expects from axial symmetry and smoothness of $v_z(r)$.

\section{Discussion and Conclusions}
\label{s:discussion}
The motivation for regularizing conservative, continuum systems like Eulerian ``ideal'' fluid mechanics
and ideal MHD were explained with some examples in \cite{thyagaraja}. A brief discussion of two famous examples should suffice here as a recapitulation of the arguments advanced earlier: the well-known example of Dirac-Pauli-Heisenberg Quantum Electrodynamics with its unpleasantly divergent (and practically useless) behaviour beyond the first order was ``tamed'' by Feynman-Schwinger-Tomonaga renormalizations and shown to work at all orders of the covariant perturbation theory by Dyson. The result of this profound set of ideas was a powerful tool which provided agreement between theory and experiment to unheard-of accuracy. The modern recognition that gauge theories which may even be non-Abelian, share this remarkable ``renormalizability'' has rightly focused researchers into constructing such theories. 

A more mundane but still deep example is provided by the KdV equation which is the regularized one dimensional ``kinematic wave'' equation (KWE).  The KWE is limited in its utility as a research tool due to its well-known failure to have single-valued solutions with finite gradients for all time. It might be argued that it should be the underlying physics which should provide the regularizing recipes for macro dynamical fluid systems: indeed, such systems can be regarded as suitable limits of fundamentally kinetic/particle systems, and as such must have dissipation terms like viscosity and thermal conduction which provide regularization. In fact, the simplest regularization of KWE is provided by Burgers' equation, which is indeed dissipative and even exactly soluble by the Cole-Hopf transformation into the heat equation. 

In spite of the general validity of dissipative, i.e. entropy producing regularizations of ideal fluid dynamics and MHD arising from underlying kinetic theories, experience has shown that the studies of purely conservative physical models often provide indispensable physical insight. Sure enough, planetary motion involves many dissipative processes which render singular phenomena such as simultaneous three-body collisions perfectly regular. Yet, most of Newtonian mechanics of point particles and rigid bodies profits enormously from using tools like Hamiltonian mechanics and action principles which are the hallmark of conservative dynamics\cite{arnold}. 

We note that kinetic approaches such as the Chapman-Enskog method based on, for example the Fokker-Planck equation of plasma theory, typically lead in higher orders in the mean-free-path asymptotic expansion to both ``entropy conserving reactive'' terms and to dissipative (entropy producing) terms in the stress tensor and the heat-flux vector (see  \cite{Braginskii,LifPit} and the more recent work \cite{karlin-gorban,struchtrup-torrilhon}). It is possible that terms like the ``twirl-acceleration'' (introduced here essentially as a formal conservative regularizing effect) could arise in higher order asymptotics (like the Burnett expansion) of kinetic equations. Somewhat analogous formal regularizers are commonly  encountered in effective field theory (e.g. the short-range repulsive Skyrme term with $4$ derivatives is believed to stabilize the singularity in the soliton solution of the QCD effective chiral Lagrangian \cite{bal-book}). String theory attempts to provide a relativistically acceptable short-range cut-off to the divergences encountered in the Einstein-Hilbert formulation of classical General Relativity in a manner which resembles in spirit the regularizers we have advocated for compressible fluid flow.

We have therefore adopted the principle that singularities such as unbounded enstrophies in Eulerian fluid dynamics and finite time failure of the models should be removed, if possible, by suitable {\it local} regularizing terms in the governing equations, in the spirit of Landau's mean field theory. These terms are required to satisfy certain strict physical criteria: i) The symmetries (usually global) obeyed by the original, unregularized system must be obeyed by the regularized system.  ii) The added terms must be ``minimal'' and ``small'' in some sense and should not alter the macro or meso-scale behaviour of the original system, although short-wavelength or ``ultra-violet-like catastrophes'' will have been significantly modified. iii) One should be able to derive appropriate conservation laws for the regularized equations and posit adequate boundary and initial data. These should be extended versions of the same laws for the original singular system. iv) The system dynamics must admit time-asymptotic behaviour corresponding to time-reversibility, Lagrange stability\cite{Nem} (but not necessarily integrability or Lyapunov stability), ergodicity and a valid statistical mechanics as in the case of 2D vortex systems considered by Onsager, London and Feynman in fluids [starting with the work of Kirchoff and Lamb] and successfully applied to 2D MHD by Edwards and Taylor \cite{edwards-taylor} and many others. It has been shown by several researchers (cf. \cite{Miura,Drazin,arnold}) that KdV has all of the above and many other interesting properties, like a Hamiltonian formulation, infinite set of non-trivial conservation laws, soliton scattering behaviour and exact solubility via the inverse-scattering transform. Apart from the incompressible systems considered in \cite{thyagaraja,AT1} we are not aware of any 3D continuum dynamical models with the far from trivial characteristics we have demonstrated for the R-Euler system.

We shall, in loose analogy with Dyson's concept of renormalizable field theories, introduce the idea of ``regularizable conservative continuum field theories''. Such theories must satisfy the criteria enumerated above. In the case of the Navier-Stokes equations or the visco-resistive MHD equations, and the Fokker-Planck kinetic equation of plasma theory (a regularized form of the Vlasov collisionless kinetics), we have dissipative regularizations. It is not yet fully clear to us if the Navier-Stokes equations admit continuous, unique solutions to initial-boundary value problems for reasonable data. We note that it has been shown in \cite{ladyzhenskaya-1} that NS can be ``regularized'' by adding a species of ``hyper-viscosity''. We conjecture that it may be possible that R-NS systems incorporating the twirl acceleration terms will, by definition, lead to a dynamical system with bounded enstrophy and one could demonstrate unique, classical solutions to such systems for the initial-boundary value problem for small, but non-zero values of $\lambda,c_{n}; n\geq 0$ (see \S \ref{s:other-const-laws-and-regs}). However, we do not attempt any proof of the existence of classical continuous solutions to the initial-boundary-value problems in this work. A 1D analogue of the twirl regularized viscous fluid and visco-resistive MHD models is the KdV-Burgers equation investigated by Grad and Hu \cite{Grad-Hu,Hu-KdV-Burgers} in the context of weak plasma shocks propagative perpendicular to a magnetic field. Mathematical examples of divergent series being ``summed'' to give perfectly well-defined and finite answers in Fourier analysis using summability methods of Abel and C\'{e}saro exemplify our approach to regularizability and its utility. Unlike the above dissipative regularizations, we focus here on the more difficult question of ``conservative regularizability'' of continuum fluid models.   
 
In the present paper, we have obtained the R-Euler equations which constitute the regularized Euler equations of compressible flow. These equations have a positive-definite energy that includes contributions from kinetic energy, compressional potential energy and the square of vorticity. We have shown that this nonlinear energy functional is a constant of the motion for typical conditions and thus prevents the unboundedness of enstrophy. These R-Euler equations are the natural generalizations of our earlier results for the incompressible fluid equations. 
 
The system motion takes place in the function space of $\rho(\bfx),\bfv(\bfx)$ which is ``foliated'' by the closed, nested surfaces formed by the constant energy functional. The R-Euler system is shown to be time reversible and to satisfy the symmetries of the Euler equations and have conservation laws corresponding with and generalizing those of the Euler system. There are even Kelvin-Helmholtz and Alfv\'en type freezing-in theorems and associated integral invariants. Furthermore, we have employed the elegant non-canonical Poisson Brackets (PBs) developed by Landau\cite{landau}(in quantum hydrodynamics), Morrison, Greene \cite{morrison-and-greene} and others to show that the R-Euler equations can be derived from the energy functional using  these PBs. This fact is remarkable in that we have demonstrated the existence and properties of a regularization of the Euler equations which ``conserves'' the Poisson structure, the conservation laws and global symmetries and guarantees the boundedness of enstrophy all at the same time. The Poisson bracket formalism implies that the system evolves on the intersection of the level hypersurfaces of energy and any other prevalent constants of motion, through a Hamiltonian, PB-mediated, infinitesimal 1-parameter group of time translations. 

Dissipative systems like NS are only associated with semi-groups and the system motion does not take place on a fixed manifold reversibly in time and typically involves ``strange attractors'' with complicated fractal properties. The regularization in the R-Euler dynamics is provided by the ``twirl acceleration'' $- \la^2 \bfT = - \la^2 \bfw \times (\grad \times \bfw)$. The size of this is determined by a parameter $\la$ with dimensions of length. The twirl term is expected to be important in high speed flows with vorticity or flows with large vorticity and its curl. At any given Reynolds number, it should dominate the viscous term for sufficiently high vorticity. The parameter $\la$ was a constant ``micro-length scale'' in incompressible R-hydrodynamics. For compressible flow, we have found that the simplest version satisfies a physically meaningful constitutive relation: $\lambda^{2}\rho$ is a constant. More generally, we have shown that a much wider class of constitutive relations are possible, some of which lead to bounded higher moments of the square of vorticity.

It is useful to note that a possible approach to the statistical mechanics of the R-Euler system is through the approach pioneered by E Hopf (see the extensive discussion by Stanisic, \cite{stan}). This was originally conceived as a method of investigating the statistical theory of hydrodynamic turbulence governed by the NS equations. However, it would seem that the ideas relating to the Hopf functional can certainly be of value in R-Euler statistical mechanics. Our PBs allow us to formulate Hopf's equation (analogue of the Liouville equation) $F_t + \{F, H \} = 0$ for the functional $F[\rho, \bfv, t]$. The Hamiltonian structure of the flow on the constant energy hyper-surface leads to micro-canonical statistical mechanics, and more generally to a canonical distribution (cf. Boltzmann-Gibbs or Fermi-Dirac, \cite{LanLif}). A statistical mechanics of entangled 3D regularized vortex tubes with bounded enstrophy and energy in dissipationless motion would be a significant extension of the 2D Onsager theory of line vortices, quantized or otherwise.

The ideas due to Koopman and von Neumann (cf. see the account given in \cite{RieszNagy}) in ergodic theory are also directly relevant provided a suitable measure can be developed for the constant energy surface on which the system motion takes place. The possibility of mapping the nonlinear evolution of the R-Euler flow on to unitary transformations in a function space of effectively finite number of degrees of freedom could have many practical applications. In numerical simulations, using the R-Euler system would enable one to avoid finite-time singularities in the enstrophy distribution and control the number of effective modes used depending on the initial data. A careful evaluation of our conserved swirl energy and other integral invariants should help to monitor the quality of simulations. Furthermore, we believe that the numerical study of plasma and fluid turbulence at very low collisionality (i.e. very high experimentally relevant Reynolds, Mach and Lundquist numbers) will be greatly facilitated by the use of our regularization.

Our examples show that as in standard Euler theory, steady solutions of the R-Euler equations are not unique, unlike the case with a dissipative regularization. Our plane vortex sheet and rotating vortex examples show that the regularization can effectively remove effects arising from discontinuities in velocity derivatives. The vortex sheet suggests that the density near the sheet is always lowered relative to asymptotic, far-field values, just as the density in the core of our rotating tornado is lower than outside. However in the corresponding R-MHD case, we find that the magnetic field tends to increase the core density due to the pinch effect. The stability theory of such regularized vortex sheets and rotating vortices would be of considerable interest. The a priori bound on enstrophy and kinetic energy demands a purely conservative non-linear saturation of any linearly growing mode. The behavior of such non-linear dynamics could provide insight into the statistics and kinematics of turbulent motions in the inertial range.

We have barely touched on higher dimensional solutions aside from obtaining the equations generalizing Hill's spherical vortex. Incidentally, all continuous potential flows of standard Euler theory in which ${\bf w} \equiv 0$ are also solutions of R-Euler. In otherwise irrotational flow, it is only when vortical singularities are encountered, that our theory differs by regularizing the solutions. However, it must be stressed that the twirl force cannot resolve all singularities of inviscid gas dynamics. A simple example is provided by the plane normal shock. Taking $\rho(x),u(x)$ and $p(x)$ as the basic variables in 1D, clearly at the shock front, these quantities change rapidly. However, no vorticity is associated with the flow and the twirl force is absent. It is well-known that collisional shock fronts involve entropy rises. Thus, to regularize them, one could add in the usual NS dissipative terms. On the other hand, both twirl and viscous stresses are likely to play a role in oblique shocks. To deal with collisionless shocks, one could extend the swirl Hamiltonian to include $(\grad \rho)^2$-type terms. 

A very fundamental problem would appear to be the following: given a periodic cubic domain and an arbitrary initial distribution of velocity [and associated vorticity] and some density distribution, how does the system evolve under R-Euler flow? In particular, it would be interesting to determine the long-time behaviour of the local energy and enstrophy distributions.


{\noindent \bf Acknowledgements:} We thank M Birkinshaw, S G Rajeev, J Samuel, A Sen and A Young for many interesting discussions. R Nityananda kindly informed us about recent work on regularization theory of kinetic theory. AT gratefully acknowledges the support and hospitality of CMI. This work was supported in part by the Infosys Foundation and a Ramanujan grant of the Department of Science \& Technology, Govt. of India.

\appendix

\section{Some properties of the Poisson brackets}
\label{a:PB-properties}
\subsection{Poisson Brackets in terms of scalar and vector potentials} 

We express the PBs among $\rho$ and $\bfv$ in terms of scalar and vector potentials. For irrotational flows these non canonical PBs may be expressed in terms of canonical Bose fields. To begin with, the Helmholtz theorem allows us to write $\bfv$ as a sum of curl-free and divergence-free fields $\bfv^{\rm irrot}$  and $\bfv^{\rm sol}$. The irrotational and incompressible fields admit scalar and vector potentials:
	\beq
	\bfv = \bfv^{\rm irrot} + \bfv^{\rm sol} =  - \grad C + \grad \times \bfQ.
	\eeq
Note that $C$ and $\bfQ$ are non-local in $\bfv$. If the flow domain is $\mathbb{R}^3$ and $\bfv$ falls off faster than $1/r$, then
	\beq
	C(\bfr) = \ov{4\pi} \int \frac{\grad_s \cdot \bfv(s)}{|\bfr- \bfs|} d\bfs \quad \text{and} \quad
	\bfQ(\bfr) = \ov{4\pi} \int \frac{\bfw(s)}{|\bfr- \bfs|} d\bfs \quad \text{with} \quad \grad \cdot \bfQ = 0.
	\label{e:scalar-and-vector-potentials-for-v}
	\eeq
We may treat $\rho, C$ and $\bfQ$ as dynamical variables in place of $\rho$ and $\bfv$. It is interesting to identify their PBs. Now, $\bfQ$ commutes with $\rho$ since $\bfw$ does. On the other hand $\{ C(x), \rho(y) \} = \del(x-y)$ since $\{ \bfv(s) , \rho(y) \} = - \grad_s \del(y-s)$ and $\grad^2_s (1/|\bfr - \bfs|) = -4\pi \del(\bfr -\bfs)$. The PBs of $\bfQ$ and $C$ are more involved:
\small
	\beqs
	16 \pi^2 \{ C(\bfx) , Q_j(\bfy)\}
	&=& \int  \left [\frac{(\bfx - \bfr) \cdot (\bfy - \bfr) w_j - (\bfx - \bfr) \cdot \bfw (y_j - r_j)}{\rho(\bfr) \: |\bfx - \bfr|^3 |\bfy - \bfr|^3} \right] \; d\bfr  \cr
	\text{and} \quad 16\pi^2 \{ Q_i(\bfx), Q_j(\bfy) \} &=& \int \left[ \frac{\eps_{ijk} (y_k - r_k) (\bfx-\bfr) \cdot \bfw - (x_i - r_i) ((\bfy - \bfr) \times \bfw)_j}{\rho(\bfr) \: |\bfx - \bfr|^3 |\bfy - \bfr|^3} \right] d \bfr.
	\label{e:AA-Aphi-PB}
	\eeqs \normalsize	
Since $\{ \rho(x), \rho(y) \} = 0$ it is natural to ask whether $\{ C, C \} = 0$ so that $C$ and $\rho$ would be canonically conjugate. We find \small
	\beq
	16 \pi^2 \{ C(\bfa) , C(\bfb)\} = \int \frac{\pdr_{r^i}\pdr_{s^j} \{v_i (r), v_j(s)\}}{|\bfa - \bfr||\bfb - \bfs|} d\bfr \: d\bfs
	= \int \frac{(a_i - r_i)(b_j - r_j)}{|\bfa - \bfr|^3 |\bfb - \bfr|^3} \frac{\om_{ij}}{\rho(\bfr)} d\bfr 
	= \int \frac{(\bfa - \bfr) \times (\bfb - \bfr) \cdot \bfw}{ \rho(\bfr) |\bfa - \bfr|^3 |\bfb - \bfr|^3} d\bfr.
	\label{e:phiphi-PB}
	\eeq \normalsize
The integrals in (\ref{e:scalar-and-vector-potentials-for-v}, \ref{e:AA-Aphi-PB},\ref{e:phiphi-PB}) are finite as may be seen in spherical coordinates centered at $\bfr = \bfa$. Defining $\tl \bfr = \bfr - \bfa$, the double pole at $\tl \bfr = 0$ is cancelled by the double zero in the volume element $\tl r^2 d\tl r d\Om$. The same applies to a neighborhood of $\bfr = \bfb$.

Note that, $\{ C(\bfa) , C(\bfa)\} = 0$, consistent with anti-symmetry. But $C$ at distinct locations don't generally commute. It suffices to show this in a special case. We take $\bfa = (0,0,0)$, $\bfb = (0,1,0)$, asymptotically constant $\rho = z/(z^2 + 1)$ and rapidly decaying $\bfv = x^2 (y - 1)^4 e^{-r^2}\: \hat z$. This ensures (\ref{e:phiphi-PB}) is manifestly convergent, $\bfw$ has zeros at $\bfa$ and $\bfb$ to cancel the apparent triple poles: \small
	\beqs
	\bfw &=& e^{-r^2}\left[ 2x^2 (y-1)^3 (2 -y(y-1)) \; \hat x + 2 x (y-1)^4 (x^2 - 1) \: \hat y\right] \quad \text{and} \cr
	 \{C (\bfa) , C(\bfb)\} &=&  \ov{16\pi^2} \int^{\infty}_{-\infty} \frac{\left(z^2 + 1\right)e^{-r^2}\left( 2 x^2 (y-1)^3 [2 - y(y-1)] \right)}{r^3 \: \left({x^2 + (y - 1)^2 + z^2}\right)^{3/2}} \, dx\, dy\,dz \approx -0.026 \ne 0.
	 \eeqs \normalsize
Thus $\rho$ and $C$ are {\it not} canonically conjugate in general. But in irrotational flow, $\bfw = \bfQ = 0$ so $\{ C(a), C(b) \} \equiv 0$ and $\rho, C$ are canonically conjugate. This is reminiscent of the number density-phase PB and suggests the introduction of the complex field $\psi = \sqrt{\rho} e^{i C/\kappa}$ where $\kappa$ is a constant with dimensions of diffusivity\footnote{A natural choice is $\kappa = c_s L$ where $c_s$ is a sound speed and $L$ a macroscopic length associated with the flow. In quantum theory $\kappa = \hbar/m$.}. The $C$-$\rho$ PB (for $\bfw = 0$) then imply that $\psi$ and $\psi^*$ satisfy canonical Bose PB: $\{ \psi, \psi \} = \{ \psi^*, \psi^* \} = 0$ and $\{ \psi(x), \psi^*(y) \} = (i/\kappa) \del(x-y)$. The evolution equation for $\psi$ in the irrotational case is reminiscent of the 3d Gross-Pitaevskii or nonlinear Schr\"odinger equation (especially for $\gamma = 2$ where$U'(\rho) \propto \rho = |\psi|^2$)
\beq
i \kappa \{ \psi , H\} = i \kappa \frac{\pdr \psi}{\pdr t}  = - \frac{\bfv^2}{2} \psi - U'(\rho) \: \psi - \frac{i \kappa}{2 \rho} \grad \cdot (\rho \bfv) \psi \quad \text{where} \quad \bfv = -\frac{\kappa}{2i}\left(\frac{\psi^* \grad \psi - \psi \grad \psi^*}{|\psi|^2}\right) \quad \text{and} \quad \rho = |\psi|^2.
\eeq
However, the above calculation implies that $\psi$ and $\psi^*$ are not canonical Bose fields for flows with vorticity. For flows with vorticity, Clebsch potentials give a way of identifying canonically conjugate variables (see Ref. \cite{ecg-nm}).

\subsection{Poisson brackets of mass current and swirl velocity}
\label{a:pb-mass-curr-and-v*}

The PB of mass current ${\bf M} = \rho \bfv$ are of particular interest. Suppose $\bf a$ and $\bf b$ are a pair of constant vectors, then using (\ref{e:PB-among-basic-var}),
	\beqs
	(a) && \{ \bfM(\bfx) , \rho(\bfy) \} = - \rho(\bfx) \grad_\bfx \del(\bfx-\bfy), 
	\cr
	(b) && \{ {\bf a} \cdot {\bf M}(\bfx) , {\bf b} \cdot \bfv(\bfy) \} = \left[ ({\bf a} \times {\bf b}) \cdot \bfw(\bfx) - ({\bf a} \cdot \bfv)(\bfx) \, {\bf b} \cdot \grad_\bfx \right] \del(\bfx-\bfy),
	\cr
	(c) &&  \{ \bfa \cdot \bfM(\bfx) , \bfb \cdot \bfM(\bfy) \} = \left[\rho(\bfy) (\bfa \cdot \bfv(\bfx)) (\bfb \cdot \grad_\bfy) - (\bfa,\bfx \leftrightarrow \bfb, \bfy) + \rho \, (\bfa \times \bfb) \cdot \bfw  \right] \del(\bfx-\bfy),
	\cr
	(d) && \{ \bfa \cdot \bfM(\bfx) , \bfb \cdot \bfw(\bfy) \} = \rho(\bfx) \, \sum_i \left( \bfb \times \grad_\bfy \left( \rho^{-1} (\bfw \times \bfa)_i \del(\bfx-\bfy) \right) \right)_i.
	\eeqs
Given the important dynamical role that the swirl velocity $\bfv_* = \bfv + \la^2 \grad \times \bfw$ plays, we mention some of its PBs. For e.g. the PB of $\bfv_*$ with $\rho$ is the same as that of $\bfv$ with $\rho$ (as $\la$ and $\bfw$ commute with $\rho$):
	\beq
	\{ \bfv_*(\bfx), \rho(\bfy) \} = \{ \bfv(\bfx) + \la(\bfx)^2 \grad \times \bfw, \rho(\bfy) \} = \{ \bfv(\bfx) , \rho(\bfy) \}.
	\eeq
Using $\grad \times (\grad \times \bfv) = \grad (\grad \cdot \bfv) - \grad^2 \bfv$, the swirl velocity may be got from $\bfv$ by the action of the tensor operator $T_{ik}(\bfx)$:
	\beq
	v_{*i} = \left[ \del_{ik} + \la^2 \left(\pdr_i \pdr_k - \del_{ik} \grad^2 \right) \right] v_k \equiv T_{ik} v_k.
	\eeq
The PB of $\bfv_*$ with other quantities can be conveniently expressed in terms $T_{ik}$\footnote{$T_{ik}$ commutes with $\bfw$ and $\rho$, but not $\bfv$. Moreover $\grad \times \bfw = \la^{-2} (T - I) \bfv$. The non-dynamical $\la^{-2} (T_{ik} - \del_{ik}) = (\pdr_i \pdr_k - \del_{ik} \grad^2)$ commutes with everything.}:
	\beqs
	(a) && \left\{ v_{*i}(\bfx) , w_j(\bfy) \right\} = T_{ik}(\bfx) \left\{ v_k(\bfx), w_j(\bfy) \right\}, \cr
	(b) && 	\left\{ v_{*i}(\bfx) , \frac{w_j(\bfy)}{\rho(\bfy)} \right\} = - \{ v_i(\bfx), \rho(\bfy) \} \frac{w_j(\bfy)}{\rho^2(\bfy)} + \ov{\rho(\bfy)} T_{ik}(\bfx) \left\{ v_k(\bfx), w_j(\bfy) \right\}, \cr
	(c) && \{ v_{*i}(\bfx), v_j(\bfy) \} = T_{ik}(\bfx) \{ v_k(\bfx) , v_j(\bfy) \} + (\grad \times \bfw)_i(\bfx) \left\{ \la^2(\bfx), v_j(\bfy) \right\}, \cr
	(d) && \{ v_{*i}(\bfx), v_{*j}(\bfy) \} = (\grad \times \bfw)_i(\bfx) \{ \la^2(\bfx) , v_j(\bfy) \} + (\grad \times \bfw)_j(\bfy) \{ v_i(\bfx), \la^2(\bfy) \} + T_{ik}(\bfx) T_{jl}(\bfy) \{ v_k(\bfx) , v_l(\bfy) \}, \cr
	(e) && 	\{ \rho(\bfx) v_{*i}(\bfx) , \rho(\bfy) \} = \rho(\bfx) \{ v_i(\bfx), \rho(\bfy) \}, \cr
	(f) && \{ \rho v_{*i}(\bfx), v_j(\bfy) \} = \rho(\bfx) T_{ik}(\bfx) \left\{ v_k(\bfx), v_j(\bfy) \right\} + v_i(\bfx) \{ \rho(\bfx) , v_j(\bfy) \}, \cr
	(g) && \{\rho v_{*i}(\bfx), w_j(\bfy) \} = \rho(\bfx) T_{ik}(\bfx) \{ v_k(\bfx), w_j(\bfy) \}.
	\eeqs

\subsection{Poisson brackets of solenoidal and irrotational linear functionals}
\label{a:solenoidal-irrot-pb}

The PB (\ref{e:pb-between-functionals-of-rho-v}) of (especially linear) functionals of $\bfv$ and $\rho$ have interesting properties. Suppose $F[\bfv] = \int \bff \cdot \bfv \, d\bfr$ and $G[\bfv] = \int \bfg \cdot \bfv \, d\bfr$ are two {\it linear} functionals of $\bfv$, with $\bff$ and $\bfg$ a pair of test vector fields vanishing sufficiently fast at infinity. Then
	\beq
	\{ F[\bfv] , G[\bfv] \} = \int (\bfw/\rho) \cdot (\bff \times \bfg) \: d\bfr.
	\eeq
An interesting sub-class of such linear functionals are the `solenoidal' ones $F_s[\bfv] = \int \bff \cdot \bfv \, d\bfr$ where $\bff$ is solenoidal $\grad \cdot \bff = 0$. Writing $\bff = \grad \times \bfA$\footnote{In terms of Clebsch potentials $\alpha, \beta$ for the solenoidal field $\bff = \grad \alpha \times \grad \beta = \grad \times (\alpha \grad \beta)$, we may take $\bfA = \alpha \grad \beta$.} and assuming $\bfA$ vanishes at infinity, $F$ can be written as a linear functional of vorticity:
	\beq
	F_s[\bfv] = \int (\grad \times \bfA) \cdot \bfv \, d\bfr 
	= \int \bfA \cdot (\grad \times \bfv) \, d\bfr + \int \grad \cdot (\bfA \times \bfv) \, d\bfr = \int \bfA \cdot \bfw \, d\bfr.
	\eeq
Since $\{\bfw, \rho \} = 0$, it follows that a solenoidal $F_s$ commutes with any functional of density $\{ F_s[\bfv], H[\rho] \} = 0$. Associated to a solenoidal $F_s$, we may define the functional $F^\rho_s = \int \bff \cdot \rho \bfv \, d\bfr$. Then one checks that $\{ F_s, F_s^\rho \} = 0$:
	\beq
	\{ F_s, F_s^\rho \} = \int \left[ \frac{\bfw}{\rho} \cdot (\bff \times \rho \bff) - \bff \cdot \grad (\bff \cdot \bfv) \right] d\bfr = \int (\bff \cdot \bfv) (\grad \cdot \bff) \, d\bfr = 0.
	\eeq
Similarly, if $\phi$ is any function (independent of $\rho$ and $\bfv$) then it follows that $F^\phi_s = \int \bff \cdot \phi \bfv d\bfr$ commutes with $F_s$ (but not with $F^\rho_s$ in general) if $\bff$ is solenoidal.

Similar to solenoidal linear functionals we may define irrotational linear functions $F_i[\bfv] = \int \bff \cdot \bfv \, d\bfr$ where $\bff = \grad \alpha$ is irrotational. Then $F_i[\bfv] = - \int \alpha(\bfr) (\grad \cdot \bfv) \, d\bfr$. The PB of two irrotational functionals is in general non-zero:
         \beq
         \{ F_i,G_i \} = \int ({\bf w}/{\rho}) \cdot [\nabla \alpha \times \nabla \beta] d\bfr \quad
         = \; -\int {\bf v}\cdot \grad \times \frac{ [\nabla \alpha \times \nabla \beta]}{\rho} d\bfr. 
         \eeq
They commute if the potentials $\alpha$ and $\beta$ are functionally dependent. The PB of an irrotational functional with a linear functional of density $H[\rho] = \int h \rho \, d\bfr$ is also non-zero in general
	\beq
	\{ F_i[\bfv], H[\rho] \} = - \int \bff \cdot \grad h \, d\bfr = - \int \grad \alpha \cdot \grad h \, d\bfr,
	\eeq
but vanishes if $\bff = \grad \alpha$ and $\grad h$ are orthogonal. 

\subsection{Proof of Jacobi identity for three linear functionals of velocity and density}
\label{a:jacobi}

Suppose $F, G$ and $H$ are three {\em linear} functionals of velocity and density \small
	\beq
	F = \int \left[ \bff(\bfr) \cdot \bfv(\bfr) + \tl f(\bfr) \rho(\bfr) \right] \: d\bfr, \;\;
	G = \int \left[ \bfg(\bfr) \cdot \bfv(\bfr) + \tl g(\bfr) \rho(\bfr) \right] \: d\bfr, \;\;
	H = \int \left[ \bfh(\bfr) \cdot \bfv(\bfr) + \tl h(\bfr) \rho(\bfr) \right] \: d\bfr,
	\eeq \normalsize
where $\bff, \bfg, \bfh$ are three smooth test vector fields and $\tl f, \tl g, \tl h$ are three test functions all vanishing sufficiently fast at infinity. We prove that the Jacobi expression $\{ \{ F, G \} , H \} + {\rm cyclic} = 0$. This is a non-trivial special case of the Jacobi identity. As a corollary, the Jacobi identity for three linear functionals of vorticity is also satisfied. For, we can write any linear functional of vorticity $F[\bfw] = \int \bfA \cdot \bfw \: d\bfr = \int (\grad \times \bfA) \cdot \bfv \, d\bfr$ as a solenoidal functional of velocity and use the previous result.

We will first obtain an interesting formula (\ref{e:jacobi-expr-3-lin-fnals-of-v}) for the Jacobi expression. Recall from (\ref{e:pb-between-functionals-of-rho-v}) that the PB of two linear functionals is $\{ F, G \}	= \int \left[ \rho^{-1} \bfw \cdot (\bff \times \bfg) - \bff \cdot \grad \tl g + \bfg \cdot \grad \tl f \right] \: d\bfr$. To find $\{ \{ F, G \} , H \}$ we need the functional derivatives
	\beqs
	\deldel{\{ F, G \}}{\rho} &=& - \ov{\rho^2} \bfw \cdot (\bff \times \bfg) \quad \text{and} \quad
	\deldel{\{ F, G \}}{\bfv} = \grad \times \left( \frac{\bff \times \bfg}{\rho} \right).
	\cr
	{\rm Thus} \quad J_1 &=& \{ \{ F, G \} , H \} = - \int \left[ \deldel{\{ F, G \}}{\rho} \grad \cdot \deldel{H}{\bfv} - \left( \grad \cdot \deldel{\{ F,G \}}{\bfv} \right) \deldel{H}{\rho} + \deldel{\{ F, G \}}{\bfv} \cdot \left( \frac{\bfw}{\rho} \times \deldel{H}{\bfv} \right)  \right] \: d\bfr \cr
	&=& \int \left[ \frac{\bfw \cdot (\bff \times \bfg)}{\rho^2} \grad \cdot \bfh - \grad \times \left( \frac{\bff \times \bfg}{\rho} \right) \cdot \frac{\bfw \times \bfh}{\rho} \right] \: d\bfr.
	\eeqs
Notice that $J_1$ is independent of the test functions $\tl f, \tl g$ and $\tl h$ so that the dependence of $F,G$ and $H$ on $\rho$ does not play any role in the Jacobi condition. We would like to separate the dependence on dynamical variables $\bfw$ and $\rho$ from the dependence on $\bff,\bfg$ and $\bfh$. Using the curl of a cross product we arrive at
	\beqs
	J_1 = \{ \{ F,G \}, H \} &=& \int \frac{\grad \rho}{\rho^3} \cdot \left[ \bfw \cdot (\bfh \times \bff) \: \bfg + \bfw \cdot (\bfg \times \bfh) \: \bff \right] \: d\bfr
	\cr && + \int \frac{\bfw}{\rho^2} \cdot \left[ \bfh \times [\bff, \bfg] + (\bff \times \bfg) (\grad \cdot \bfh) - (\bfg \times \bfh) (\grad \cdot \bff) - (\bfh \times \bff) (\grad \cdot \bfg) \right] d\bfr.
	\eeqs
Here $[\bff , \bfg] = (\bff \cdot \grad) \bfg - (\bfg \cdot \grad) \bff$ is the commutator of vector fields. Notice that the $1^{\rm st}$ term involves the gradient of $\rho$ while the $2^{\rm nd}$ does not. $J_2$ and $J_3$ are obtained by cyclic permutations of $\bff,\bfg,\bfh$. Adding $J_1 + J_2 + J_3 = J$, several terms cancel leaving
\small
	\beqs
	J &=& J^{\pdr \rho} + J^\rho = - \int \grad\left(\rho^{-2} \right) \cdot [(\bfw \cdot (\bff \times \bfg)) \bfh + (\bfw \cdot (\bfg \times \bfh)) \bff + (\bfw \cdot (\bfh \times \bff)) \bfg] \: d\bfr \cr
	&+& \int \frac{\bfw}{\rho^2} \cdot \left[ \left( \bff \times [\bfg, \bfh] + \bfg \times [\bfh, \bff] + \bfh \times [\bff, \bfg]  \right)  + \left\{ (\bfh \times \bfg) (\grad \cdot \bff) + (\bff \times \bfh) (\grad \cdot \bfg) + (\bfg \times \bff) (\grad \cdot \bfh)  \right\} \right] d\bfr.
	\label{e:jacobi-expr-3-lin-fnals-of-v}
	\eeqs \normalsize
For the Jacobi identity to be satisfied, this must vanish for arbitrary test vector fields $\bff,\bfg,\bfh$ and any fixed $\rho$ (asymptotically constant) and $\bfw$ (vanishing at infinity). The $1^{\rm st}$ term involves $\grad \rho$, so we call it $J^{\pdr \rho}$ while the second term is called $J^\rho$. In the integrand of $J^\rho$ the dependence on $\bfw,\rho$ is factorized from the dependence on $\bff,\bfg,\bfh$. This is not quite the case with $J^{\pdr \rho}$.

{\noindent \bf Proof that $J=0$:} We expand the test vector fields as a linear combination of fields along the coordinate directions $\hat x, \hat y, \hat z$ and write the linear functionals\footnote{As remarked above, the dependence on $\rho$ of the {\it linear} functionals $F,G$ and $H$ does not enter the Jacobi expression.} as a sum $F[\bfv] = \sum_i \int \bff_i \cdot \bfv = F_1[\bfv] + F_2[\bfv] + F_3[\bfv]$. Thus the Jacobi expression becomes
	\beq
	J = \{ \{ F, G \} , H \} + \text{cyclic} = \sum_{i,j,k=1}^3 \{ \{ F_i, G_j \} , H_k \} + \text{cyclic}.
	\eeq
There are 27 terms of the form $\{ \{ F_i, G_j \} , H_k \}$ plus their cyclic permutations. Consider any one of the $27$ terms. There are three possibilities: (1) $i,j,k$ all distinct  (mutually orthogonal test fields); (2) $i=j=k$, (collinear test fields) (3) two indices the same and one distinct. We show below that the Jacobi identity is satisfied for three linear functionals of velocity $F,G,H$ falling into any one of the above categories. Consequently $J=0$ for any three linear functionals of velocity and density.

\subsubsection{Jacobi identity for 3 orthogonal test fields}
Let the three linear functionals $F,G,H$ in the Jacobi expression (\ref{e:jacobi-expr-3-lin-fnals-of-v}) point along $\hat x, \hat y \:$and$\: \hat z$, i.e. $\bff = \al(\bfr) \hat x$, $\bfg = \beta(\bfr) \hat y$, $\bfh = \gamma(\bfr) \hat z$ where $\al, \beta, \gamma$ are three test functions. Beginning with $(\bfw \cdot (\bff \times \bfg)) \bfh = \al \beta \gamma \, (\bfw \cdot \hat z) \hat z$ and their cyclic permutations we get (subscripts on $\al, \beta, \gamma$ denote partial derivatives)
	\beqs
	(\bfw \cdot (\bff \times \bfg)) \bfh + \text{cyclic} = (\al \beta \gamma) \bfw \quad \imply \quad 
	J^{\pdr \rho} = - \int (\alpha \beta \gamma) \: \bfw \cdot\grad(\rho^{-2}) \; d\bfr.
	\eeqs
The quantity $\al \beta \gamma = (\bff \times \bfg) \cdot \bfh$ is the volume of the parallelepiped spanned by the test vector fields. On the other hand $J^\rho$ is evaluated using $\bff \times [\bfg , \bfh] + (\bfh \times \bfg) (\grad \cdot \bff) + \text{cyclic} = - \grad(\al \beta \gamma)$. By the divergence theorem and $\grad \cdot \bfw = 0$ we get
	\beq
	J^\rho = - \int \frac{\bfw}{\rho^2} \cdot \grad (\al \beta \gamma) \, d \bfr = \int (\al \beta \gamma) \grad \cdot (\bfw/\rho^2) \, d\bfr 
	= \int (\al \beta \gamma) \bfw \cdot \grad (\rho^{-2}) \, d\bfr.
	\eeq
So, $J = J^\rho + J^{\pdr \rho} = 0$ and the Jacobi identity (\ref{e:jacobi-expr-3-lin-fnals-of-v}) for three orthogonal test fields is proved.

\subsubsection{Jacobi identity for three test fields in the same direction}

Let all three vector fields be collinear, say:
	$\bff = \alpha \hat x, \: \bfg = \beta \hat x \; \text{and} \; \bfh = \gamma \hat x.$ Since their cross product vanishes, $J^{\pdr \rho} = 0$ and the Jacobi expression (\ref{e:jacobi-expr-3-lin-fnals-of-v}) reduces to
	\beq
	J = \int \frac{\bfw}{\rho^2} \cdot \left[ \bff \times [\bfg, \bfh] + \bfg \times [\bfh, \bff] + \bfh \times [\bff, \bfg]  \right] \, d\bfr.
	\eeq
All the commutators point along $\hat x$, e.g. $[\bfg,\bfh] = (\beta \gamma_x - \gamma \beta_x) \hat x.$
It follows that the cross product of the vector fields and the commutators is zero, so $J = 0$.

\subsubsection{Jacobi identity for two collinear test fields and one orthogonal to them}

Let $2$ of the test fields be collinear and the $3^{\rm rd}$ point orthogonally. Without loss of generality we take $\bff = \alpha \hat x, \; \bfg = \beta \hat x, \; \bfh = \gamma \hat y$. Now, $\bfw \cdot(\bff \times \bfg) \bfh = 0, \;\bfw \cdot(\bfg \times \bfh) \bff = \al \beta \gamma w_z \hat x \; \text{and} \; \bfw \cdot(\bfh \times \bff) \bfg = - \al \beta \gamma w_z \hat x$. Therefore $J^{\pdr \rho} = - \int \grad\left(\rho^{-2} \right) \cdot [(\bfw \cdot (\bff \times \bfg)) \bfh + (\bfw \cdot (\bfg \times \bfh)) \bff + (\bfw \cdot (\bfh \times \bff)) \bfg] \: d\bfr = 0$. To compute $J^{\rho}$ (\ref{e:jacobi-expr-3-lin-fnals-of-v}) we need $\bff \times [\bfg, \bfh] + (\bfh \times \bfg) (\grad \cdot \bff) + \text{cyclic}$. Now, 
	\beq
	(\bfh \times \bfg) (\grad \cdot \bff) = - \al_x \beta \gamma \hat z, \quad
	(\bff \times \bfh) (\grad \cdot \bfg) = \al \beta_x \gamma \hat z, \quad
	\text{and} \quad
	(\bfg \times \bff) (\grad \cdot \bfh) = 0
	\eeq
So $(\bfh \times \bfg) (\grad \cdot \bff) + \text{cyclic} = \left(\al \beta_x \gamma - \al_x \beta \gamma \right) \hat z$. On the other hand, $\bff \times [\bfg , \bfh] =\al \beta \gamma_x \hat z, \;\bfg \times [\bfh , \bff] = - \al \beta \gamma_x \hat z\;\text{and} \; \bfh \times [\bff , \bfg] = (- \al \beta_x \gamma + \al_x \beta \gamma) \hat z.$
Adding these 
        \beq
	\bff \times [\bfg, \bfh] + \text{cyclic} = (\al_x \beta \gamma - \al \beta_x \gamma) \hat z.
	\eeq
It follows that $J = 0$, so the Jacobi identity is satisfied if two of the test fields are collinear and the third points orthogonally.

\subsection{Proof of Jacobi identity for non-linear functionals}
\label{s:Jacobi-general-proof}

Consider exponentials of three linear functionals of $\rho$ and $\bfv$:
	\beq
	{\cal F}[\rho, \bfv] = \exp i F[\rho,\bfv] \quad \text{where} \quad F[\rho, \bfv] = \int \left( \bff \cdot \bfv + \tl f \rho \right) \: d\bfr.
	\eeq
Here $\bff, \tl f$ are test field and test function as in \ref{a:jacobi}. ${\cal G}$ and $\cal H$ are defined similarly. Then
	\beq
	\deldel{\cal F}{\rho} = i \, \tl f \: {\cal F} \quad \text{and} \quad \deldel{\cal F}{\bfv} = i \, \bff \: {\cal F}, \quad {\rm e.t.c.}
	\eeq
Using (\ref{e:pb-between-functionals-of-rho-v}) the PB between the exponential functional $\cal F$ and an arbitrary functional $K$ is
	\beq
	\{{\cal F},K\}
     = i{\cal F}\int \left[ \frac{\bfw}{\rho} \cdot \left( \deldel{F}{\bfv} \times \deldel{K}{\bfv} \right) - \deldel{F}{\bfv} \cdot \grad \deldel{K}{\rho} + \deldel{K}{\bfv} \cdot \grad F_{\rho} \right] \: d\bfr = i \: {\cal F} \: \{F,K\}.
	\eeq
Taking $K= {\cal G}$, we have $\{{\cal F}, {\cal G}\}=-{\cal F}{\cal G} \{F,G\}$. Thus, the first term in the Jacobi expression becomes
	\beq
	{\cal J}_1 = \{ {\cal F}, \{ {\cal G}, {\cal H} \} \} = {\cal F} {\cal G} {\cal H} \left[ -i \{ F, \{ G, H \} \} + \{ G, H \} \: \{ F, G + H \} \right].
	\eeq
The product of PBs cancels out upon adding cyclic permutations, resulting in the Jacobi expression
	\beq
	{\cal J} \equiv {\cal J}_1 + \text{cyclic} = -i {\cal F} {\cal G} {\cal H} \left[  \{ F, \{ G, H \} \} + \{ G, \{ H, F \} \} + \{ H, \{ F, G \} \} \right] = 0.
	\eeq
So remarkably, the Jacobi expression for three exponential functionals is proportional to the corresponding expression for three linear functionals, which was shown to vanish in \S \ref{a:jacobi}. Thus we have proved the Jacobi identity for non-linear functionals that are exponentials of linear functionals!

The Jacobi identity for finite linear combinations of exponential functionals follows from linearity of the PBs. Now we propose that an arbitrary nonlinear functional $P[\rho,\bfv]$ can be formed by the following {\it functional} Fourier transform:
	\beq
	P[\rho,\bfv]=\int D[\tl p,{\bf p}] \; \hat{P}[\tl p,{\bf p}] \exp\left[ i \int [\tl p\rho + {\bf p} \cdot \bfv] d\bfx \right]
	\eeq
where $\int D[\tl p,{\bf p}]$ denotes functional integration over the test fields and test functions. Now, suppose $\hat{P}[\tl p,{\bf p}],\hat{Q}[\tl q,{\bf q}]$ and $\hat{R}[\tl r,{\bf r}]$ are suitable [functionally integrable] functionals of the test functions and test fields. $P,Q,R$ are clearly linear combinations of the exponential functionals considered above. It is then clear, by the linearity of PBs in each argument, that the nonlinear functionals $P,Q,R$ must satisfy Jacobi's identity since any three exponential functionals do, as shown above. A rigorous treatment of the above functional Fourier transform is beyond the scope of this work. We observe that this type of functional calculus is freely used in the Hopf functional theory and in modern quantum field theories based on the Wiener measure and Feynman's path integrals. This approach may also be applied to proving the Jacobi identity for other PBs including the canonical $\{x,p \}$ Poisson brackets of particle mechanics.


\end{document}